\documentclass{aa}  
\DeclareUnicodeCharacter{0327}{}
\usepackage{scrextend}
\usepackage{natbib,twoopt}
\usepackage[colorlinks=true,citecolor=blue,linkcolor=blue,urlcolor=blue,breaklinks=true]{hyperref} %% to avoid \citeads line fills
\bibpunct{(}{)}{;}{a}{}{,}             %% natbib format for A&A and ApJ
\makeatletter
  \newcommandtwoopt{\citeads}[3][][]{\href{http://adsabs.harvard.edu/abs/#3}%
    {\def\hyper@linkstart##1##2{}%
     \let\hyper@linkend\@empty\citealp[#1][#2]{#3}}}
  \newcommandtwoopt{\citepads}[3][][]{\href{http://adsabs.harvard.edu/abs/#3}%
    {\def\hyper@linkstart##1##2{}%
     \let\hyper@linkend\@empty\citep[#1][#2]{#3}}}
  \newcommandtwoopt{\citetads}[3][][]{\href{http://adsabs.harvard.edu/abs/#3}%
    {\def\hyper@linkstart##1##2{}%
     \let\hyper@linkend\@empty\citet[#1][#2]{#3}}}
  \newcommandtwoopt{\citeyearads}[3][][]%
    {\href{http://adsabs.harvard.edu/abs/#3}
    {\def\hyper@linkstart##1##2{}%
     \let\hyper@linkend\@empty\citeyear[#1][#2]{#3}}}
\makeatother

\usepackage{multirow}
\usepackage{amsmath}
\usepackage{bm}
\usepackage{txfonts}
\newcommand{\met}{$12 + \log$(O/H)}
\newcommand{\nism}{$n_\text{ISM}$}

\newcommand{\msun}{M$_\odot$}

\newcommand{\vs}{$v_{\text{shock}}$}
\newcommand{\vlos}{$V_{\text{los}}$}

\newcommand{\fluxcgs}{ergs~s$^{-1}$~cm$^{-2}$}

\newcommand{\chandra}{\textit{Chandra}}

\newcommand{\hst}{\textit{HST}}

\newcommand{\mappings}{\textsc{MAPPINGS}}

\newcommand{\ulx}{NGC 1313~X--1}
\newcommand{\theulx}{NGC 1313~X--1}

\newcommand{\ha}{H$\alpha$}
\newcommand{\hb}{H$\beta$}

\begin{document} 

   \title{MUSE spectroscopy of the ULX NGC 1313 X-1: A shock-ionised bubble, an X-ray photoionised nebula, and two supernova remnants}
\titlerunning{MUSE spectroscopy of the NGC 1313 X-1 bubble}
        %   \subtitle{I. Overviewing the $\kappa$-mechanism}

   \author{A. G\'urpide
          \inst{1},  M.Parra\inst{1, 2}, O. Godet\inst{1}, T. Contini\inst{1} \and
          J.-F. Olive\inst{1}
      %    \fnmsep\thanks{Just to show the usage
       %   of the elements in the author field}
          }
          \authorrunning{A. G\'urpide et al.} 

   \institute{IRAP, Université de Toulouse III/CNRS/CNES, 9 av. du Colonel Roche 31028 Toulouse, Cedex 04, France
      \and
   Univ. Grenoble Alpes, CNRS, IPAG, 38000 Grenoble, France}
   \date{}

% \abstract{}{}{}{}{} 
% 5 {} token are mandatory
 
  \abstract
   {The presence of large ionised gaseous nebulae found around some ultraluminous X-ray sources (ULXs) provides the means to assess the mechanical and radiative feedback of the central source, and hence constrain the efficiency and impact on the surroundings of the super-Eddington regime powering most of these sources. \theulx\ is an archetypal ULX, reported to be surrounded by abnormally high [O~I]$\lambda$6300/H$\alpha >$ 0.1 ratios, and for which high-resolution spectroscopy in X-rays has hinted at the presence of powerful outflows.}
   {We report observations taken with the integral field unit Multi-Unit Spectroscopic Explorer (MUSE) mounted at the Very Large Telescope of NGC~1313~X--1 in order to confirm the presence of a nebula inflated by the winds, investigate its main sources of ionisation and estimate the mechanical output of the source.}
   {We investigated the morphology, kinematics, and sources of ionisation of the bubble through the study of the main nebular lines. We compared the main line ratios with spatially resolved Baldwin–Phillips–Terlevich diagrams and with the prediction from radiative shock libraries, which allows us to differentiate regions excited by shocks from those excited by extreme ultraviolet and X-ray radiation.}
   {We detect a bubble of 452 $\times$ 266\,pc in size, roughly centred around the ULX, which shows clear evidence of shock ionisation in the outer edges. We estimate shock velocities to be in the $\approx160-180$\,km/s range based on the line ratios. This suggests that an average and continuous outflow power of $\sim(2-4.5) \times 10^{40}$~erg/s over a timescale of $(4.5-7.8) \times10^5$~yr is required to inflate the bubble. In the interior of the bubble and closer to the ULX we detect an extended ($\sim$140\,pc) X-ray ionised region. Additionally, we detect two supernova remnants coincidentally close to the ULX bubble of which we also report age and explosion energy estimates.}
   {The elongated morphology and the kinematics of the bubble strongly suggest that the bubble is being inflated by winds and/or jets emanating from the central source, supporting the presence of winds found through X-ray spectroscopy. The estimated mechanical power is comparable to or higher than the X-ray luminosity of the source, which provides additional evidence in support of NGC~1313~X--1 harbouring a super-Eddington accretor. }

   \maketitle
%
%-------------------------------------------------------------------

\section{Introduction} \label{sec:bubble_intro}
Ultraluminous X-ray sources (ULXs) are an enigmatic class of objects empirically defined as extragalactic, point-like, off-nuclear, and with an X-ray luminosity exceeding the classical Eddington limit for a $\sim$ 20\,\msun\ black hole (BH) \citep[see][for the most recent review]{fabrika_ultraluminous_2021}. While this empirical definition alone does not convey much information about the nature of these objects, there is now firm evidence suggesting that a fraction of them are powered by super-Eddington accretion onto stellar-compact objects. Evidence in support of this picture comes from their atypical X-ray spectral states compared to sub-Eddington X-ray binaries \citep[e.g.][]{gladstone_ultraluminous_2009, bachetti_ultraluminous_2013, gurpide_discovery_2021}, the discovery of X-ray pulsations in several ULXs with luminosities exceeding the classical Eddington limit for a neutron star (NS) by orders of magnitude \citep[e.g.][]{bachetti_ultraluminous_2014, furst_discovery_2016, israel_accreting_2017, carpano_discovery_2018, sathyaprakash_discovery_2019, rodriguez-castillo_discovery_2020, quintin_new_2021}, the detection of strongly Doppler shifted X-ray lines indicative of super-Eddington winds with speeds reaching fractions of the speed of light \citep[e.g.][]{pinto_resolved_2016, pinto_xmmnewton_2020, pinto_xmm-newton_2021}, the inflated or photoionised 40--400\,pc optical and radio bubbles found around them \citep{pakull_optical_2002, abolmasov_spectroscopy_2007, berghea_detection_2020}, and their similarities with the supercritically accreting Galactic source SS433 \citep{fabrika_jets_2004, liu_relativistic_2015, waisberg_collimated_2019, middleton_nustar_2021}. While these observational clues appear to match our understanding of the super-Eddington accretion regime to some extent \citep[e.g.][]{shakura_black_1973,poutanen_supercritically_2007, takeuchi_clumpy_2013, jiang_global_2014, middleton_spectral-timing_2015, narayan_spectra_2017}, many questions about the super-Eddington regime and its connection with ULXs remain open, such as the degree to which the emission is beamed \citep[e.g.][]{king_ultraluminous_2001, kaaret_high-resolution_2004,jiang_global_2014, mushtukov_pulsating_2021}, the fraction of energy carried in the outflows, the mechanical and radiative feedback induced by ULXs, and the exact accretion flow geometry allowing them to reach such high luminosities.  

Because the X-ray luminosity is the defining property of a ULX, it is natural that most studies attempting to tackle these issues have focused on the X-ray band \citep[e.g.][]{middleton_spectral-timing_2015, koliopanos_ulx_2017, walton_evidence_2018, gurpide_long-term_2021}. However, observations in this band are generally limited by our ability to obtain information restricted to our line of sight. For this reason, testing the degree of anisotropy of the X-ray emission remains challenging \citep[e.g.][]{middleton_nustar_2021}, yet strong anisotropy is predicted in many theoretical studies of the super-Eddington accretion regime \citep[e.g.][]{shakura_black_1973, poutanen_supercritically_2007, narayan_spectra_2017} and is a fundamental piece of information to determine the true radiative output in this poorly understood accretion regime.

On the other hand, the study of ULX bubbles offers the means to infer this energetic output by observing the impact of the ULX on the environment. Because the bubble nebula sees the full emission from the ULX, it essentially offers a 2D map of the ionising photon field emitted by the ULX, while also probing the presence and the impact of the super-Eddington winds or jets on the environment. Therefore, ULX nebulae can be used to constrain the energetics and overall properties of the super-Eddington accretion flows powering ULXs. A prime example is the study of the ULX nebula around the ULX NGC~6946~X--1 by \cite{abolmasov_optical_2008}, who showed that the presence of a UV source with $L$ $\sim$ 10$^{40}$\,erg/s was needed to explain the  high-excitation lines of the nebula. Later \hst\ observations in the far-UV by \cite{kaaret_direct_2010} confirmed the presence of the first ultraluminous UV source, in good agreement with the predictions made by \cite{abolmasov_optical_2008}. Observations of the HeII$\lambda$4686 recombination line of the X-ray photoionised nebula around Holmberg~II~X--1 have also been used to constrain the intrinsic X-ray flux of the source, finding that the degree of beaming of the X-ray emission must be small \citep{pakull_optical_2002, kaaret_high-resolution_2004, berghea_first_2010}. ULX nebulae are also of interest because they allow us to probe faint X-ray sources with luminosities below the ULX threshold, but for which most of the X-ray radiation might be simply not directed into our line of sight, or for which most of the energy might be released mechanically \citep{dubner_high-resolution_1998, pakull_300-parsec-long_2010, urquhart_newly_2019}. 

With integral field unit (IFU) instruments, the study of ULX nebulae becomes even more attractive as all the information needed to diagnose the physical conditions of the gas can be accessed with a single exposure. Indeed, several promising results can already be found in the literature \citep[e.g.][]{egorov_ultraluminous_2017, mcleod_optical_2019}. For this reason, we are undertaking a programme  to search for ULXs directly or serendipitously observed by the Multi-Unit Spectroscopic Explorer \citep[MUSE,][]{bacon_muse_2017}, an IFU instrument at the Very Large Telescope (VLT) covering the 480--930\,nm range, with a resolving power of $\sim$ 2484 at 650\,nm. 

In this paper we report the first of these analyses focusing on the well-known ULX \theulx\ \citep[e.g.][]{feng_spectral_2006, bachetti_ultraluminous_2013}. \theulx\ has a persistent unabsorbed luminosity in the 0.3--10\,keV band above $\sim$ 8 $\times$ 10 $^{39}$ erg/s, with a mean luminosity of 1.1 $\times$ 10$^{40}$ erg/s \citep{gurpide_long-term_2021}. It is among the few ULXs for which high-resolution spectroscopy has hinted the presence of powerful winds with velocities $\simeq$ 0.2\,$c$ (\citealt{pinto_resolved_2016, pinto_xmmnewton_2020}; see also \citealt{walton_iron_2016}), suggesting that the source harbours a supercritically accreting source, but whose nature remains unclear \citep{walton_unusual_2020}. It has been argued that the source should be seen at low inclinations, through the optically thin funnel formed by the wind around the rotational axis of the compact object \citep[e.g.][]{sutton_ultraluminous_2013, middleton_spectral-timing_2015, gurpide_long-term_2021}. While these winds are thought to be powerful enough to carve the aforementioned $\sim$100\,pc bubbles sometimes seen around ULXs \citep{pinto_thermal_2020}, to date this has not been directly confirmed. The presence of a nebula around NGC~1313~X--1 was hinted at in the seminal study of \cite{pakull_optical_2002}, who noted an exceptionally high [O~I]$\lambda$6300/\ha\ $>$ 0.1 ratio surrounding the ULX. Here we report the detection and analysis of a well-defined elongated 452 $\times$ 266\,pc shock-ionised bubble, providing the first link between wind detections in X-rays and their large-scale impact on the surroundings. In the interior of the bubble we find line ratios typical of an X-ray photoionised nebula, consistent with the results from \cite{pakull_optical_2002}.

Our focus here lies in investigating the main sources of ionisation of the bubble. Through different diagnostics and thanks to the MUSE data, we are able to spatially resolve and distinguish regions excited by shocks from those excited by extreme ultraviolet (EUV) and X-ray photoionisation from the ULX, providing important clues about the energy feedback of the source. We  show that while the bubble is likely to have been inflated by a wind, the inner regions show clear  evidence of hard photoionisation from the ULX. 

Coincidentally, we detect two supernova remnants (SNRs) at the edge of the bubble, of which we also provide line ratios and estimates of their kinetic explosion energy. An illustrative comparison between the energy required to inflate the ULX bubble invoking a SNR and these bona fide SNRs is provided. We show that the explosion energy required to inflate the ULX bubble is orders of magnitude higher than that of the SNRs, strongly disfavouring this hypothesis, in agreement with previous works. 

NGC 1313 ($z=0.001568$) is a face-on SB(s)d spiral galaxy at a distance of 4.25 Mpc \citep{tully_cosmicflows-3_2016} with an inclination $i$ = 48$^\circ$ \citep{ryder_new_1995}. \theulx\ is located in the northern part of the galaxy, within the inner 46" ($\sim$ 950\,pc) radius from the nucleus of NGC 1313, where \cite{ryder_new_1995} determined the rotational velocity to be 46.5\,km/s. This gives a line-of-sight velocity at the source position of $V_\text{rot} \sin i \cos \theta$ $\sim$ 30 km/s over the systemic velocity, where $\theta$ is the azimuth angle in the disk plane as defined by \citep{ryder_new_1995} and at the position of \theulx\ $\theta$ = 30$^\circ$.  

This paper is organised as follows. Section~\ref{sec:bubble_data_reduction} presents the observations considered and the data reduction. Section~\ref{sec:bubble_data_analysis} presents our analysis and results of the ULX bubble and the SNRs. In Section~\ref{sec:bubble_discussion} we discuss our results and implications. Finally we present our conclusion in Section~\ref{sec:bubble_conclusions}.

%--------------------------------------------------------------------
\section{Data reduction}\label{sec:bubble_data_reduction}

\theulx\ has been observed by MUSE in the Wide Field Mode (WFM) in three instances (PI F.P.A. Vogt) of 1.4 ks exposure time each. Two of these observations were taken in extended mode, which extends the observable bandpass in the blue part of the spectrum down to $\sim$ 465\,nm, compared to the nominal mode that reaches only to $\sim$ 480\,nm. The data cubes were downloaded from the ESO Science Portal\footnote{\url{http://archive.eso.org/scienceportal/home}}, which provides the data reduced by the automatic dedicated pipeline \citep{weilbacher_data_2020} and ready for scientific analysis. 

After inspection of the cubes, we discarded cube ADP.2019-12-20T09 12 08.117 from the analysis as $\sim$ 92\% of the spaxels had negative values, which rendered it unusable. We therefore considered only cubes ADP.2019-12-08T00:33:13.185 (cube 1) and ADP.2019-11-29T11:13:19.106 (cube 2) for the analysis (see Table~\ref{tab:data} for details). We found that both cubes had around 5\% of invalid pixels, which we interpolated using the 3$\times$3$\times$3 nearest pixels using the method available in the Zurich Atmosphere Purge (ZAP) tool \citep{soto_zap_2016}. 

\begin{figure}
    \centering
    \includegraphics[width=0.49\textwidth]{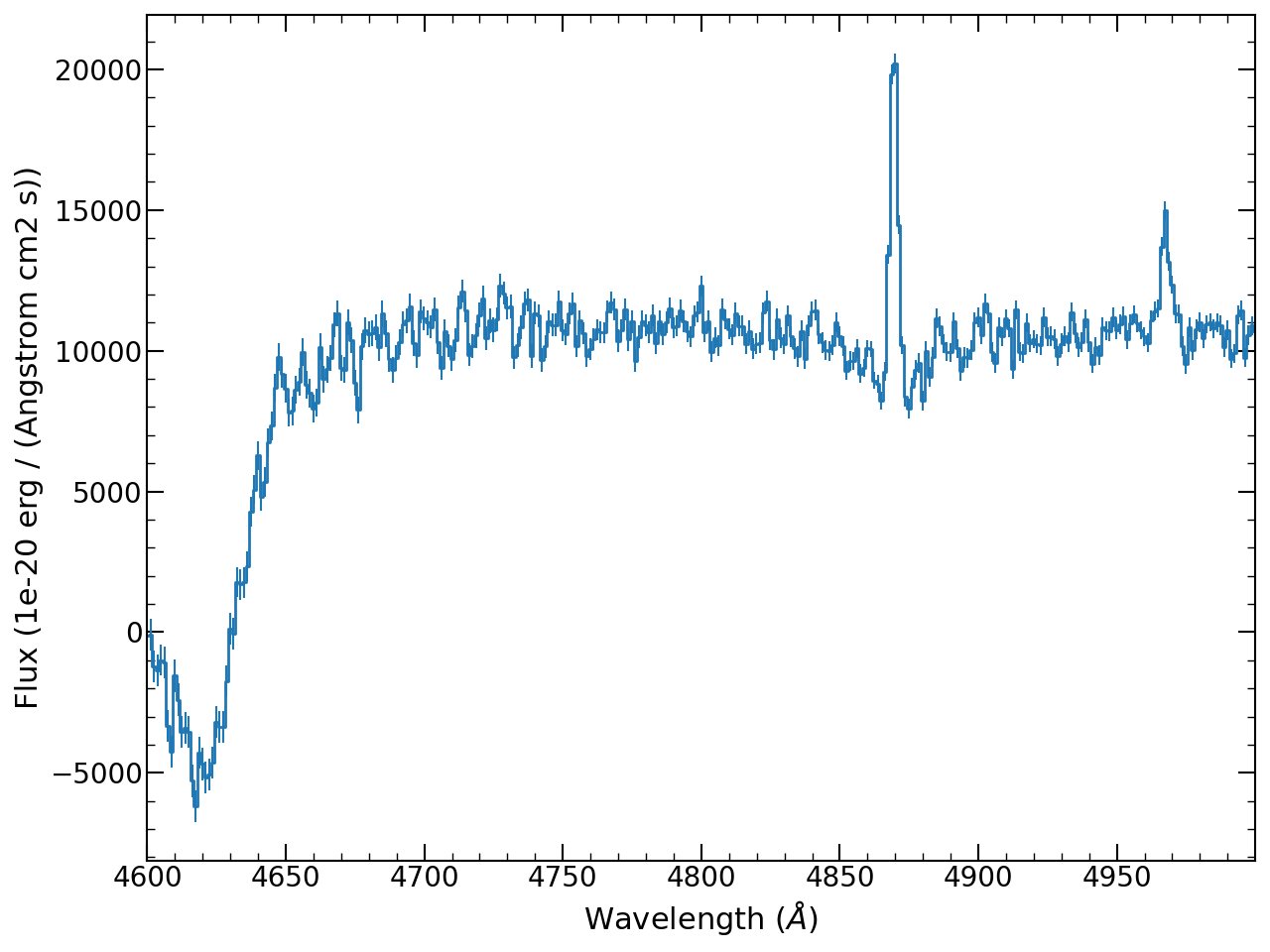}
    \caption{Spectrum extracted from a pixel of cube  ADP.2019-11-29T11:13:19.106 strongly affected by flux miscalibration effects at the blue edge of the wavelength range.}
    \label{fig:flux_cal}
\end{figure}

Cube 1 had slightly better spatial resolution (PSF FWHM $\sim$ 1" vs 1.2") and appeared of overall higher data quality as the sky subtraction had mostly affected the spectrum in the red part above $\sim$ 7000 \AA, whereas in cube 2 the sky residuals are also observed around the \ha-[N~II] complex. However, because the spatial distribution of the noise in each cube is different and the quality of each of the lines of interest depends strongly on the success of the sky correction, in some cases we could improve the results by combining the information from the two cubes (see Section~\ref{sec:camel}). For this reason, we decided to analyse  the two cubes independently and use them to cross-validate our results, although we mainly present the results from cube 1 for the reasons outlined above. We focused on the main nebular lines used for gas diagnostics covered by MUSE (i.e. \hb, [O~III]$\lambda$5007, [O~I]$\lambda$6300, the \ha-[N~II] complex, and the [S~II] doublet). While cube 2 was taken in extended mode, in principle allowing us to measure HeII$\lambda$4686, a relevant emission line in high-excitation nebulae \citep[e.g.][]{kaaret_high-resolution_2004}, edge effects, and uncertainties on the flux calibration seemed to be affecting the cube in the blue part of the spectrum below 4700 \AA\ (Figure \ref{fig:flux_cal}), and thus we were unable to use this line in our analysis. The MUSE/WFM pixel scale of 0.2" and the PSF FWHM of $\sim$ 1" represent respectively a physical distance of 4.2\,pc and $\sim$ 20\,pc at the distance of \theulx\ of 4.25\,Mpc. 

Along with the MUSE observations, we made use of archival \hst\ data to correct the astrometry of the MUSE cubes and inspect the field around \theulx. In particular, we retrieved calibrated \hst\ images from the Hubble Legacy Archive\footnote{\url{https://hla.stsci.edu/}} in the \textit{ACS/WFC/F555W} filter and in the \textit{WFC3/UVIS/F657N} and \textit{WFC3/UVIS/547M} filters. The first filter was chosen as it has substantial overlap with the MUSE wavelength range (Table \ref{tab:data}), which was required to perform the astrometric correction, whereas the other two filters allowed us to inspect the morphology of the nebular emission with a higher spatial resolution than provided by MUSE, in particular to aid in the identification of the SNRs (Section~\ref{sub:snrs}).

Finally, we also retrieved the longest available \chandra\ observation of \theulx\ in order to inspect for the presence of nearby X-ray sources (see Section~\ref{sub:snrs}) associated with the identified SNRs. We reprocessed these observations using the script \texttt{chandra\_repro} using CIAO version 4.12 and CALDB 4.9.3. Table~\ref{tab:data} provides a summary of all the data used in this work.
\begin{table*}
    \centering
    \caption{Log of the observations used in this work.}
    \label{tab:data}
    \begin{tabular}{cccccc}
    \hline
     \noalign{\smallskip}
        Telescope & Detector & ObsID & Date & Band & Exposure  \\ 
         &  &  &  &  & ks \\
         \hline
         \hline
         \noalign{\smallskip}
         \chandra\ & ACIS-S & 2950 & 2002-10-13 & 0.3--10 keV &19.9  \\ 
        % \chandra\ & ACIS-I & 3550 & 2002-11-09 & 0.3--10 keV & 14.76  \\
         %\chandra\ & ACIS-I & 3551 & 2003-10-02 & 0.3--10 keV &  15.02  \\
         \hline
          \noalign{\smallskip}
         \hst\ & ACS/WFC/F555W & hst\_9796\_01\_acs\_wfc\_f555w &  2003-11-17  &4777--6017 \AA  & 1.16 \\
         \hst\ & WFC3/UVIS/F657N & hst\_13773\_15\_wfc3\_uvis\_f657n & 2015-02-24 & 6503--6628    &  1.5 \\
         \hst\ & WFC3/UVIS/F547M & hst\_13773\_15\_wfc3\_uvis\_f547m & 2015-02-24 & 5108--5821    &  0.55 \\
          \hline
         \noalign{\smallskip}
         VLT & MUSE & ADP.2019-12-08T00:33:13.185 & 2019-10-17 &  4750-9350 \AA & 1.4   \\
         VLT & MUSE & ADP.2019-11-29T11:13:19.106 & 2019-10-16 &  4600--9350 \AA & 1.4   \\
          \noalign{\smallskip}
         \hline
         \hline
    \end{tabular}
\end{table*}

\begin{figure*}
    \centering
    \includegraphics[width=0.98\textwidth]{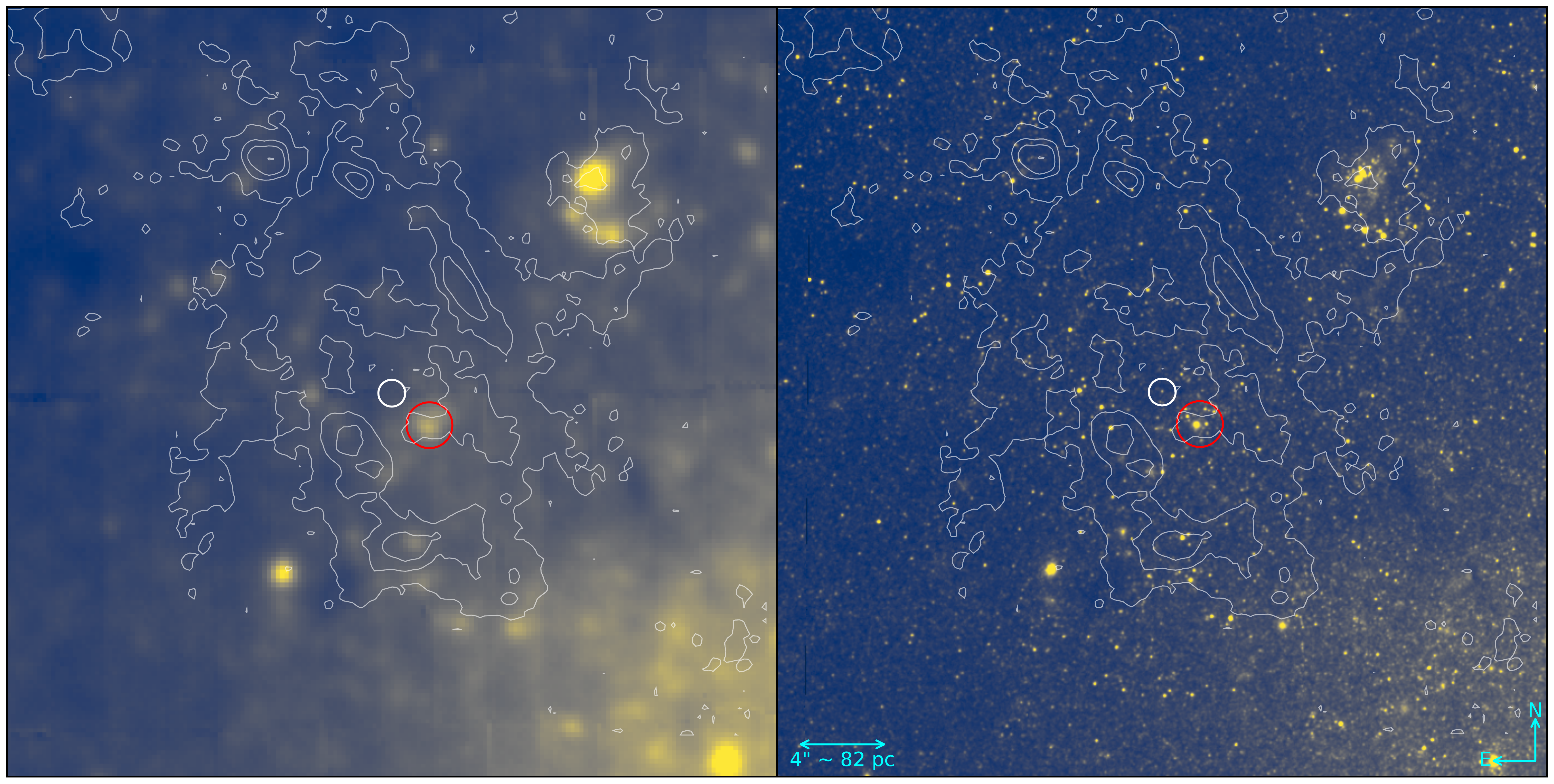}
    \caption{Comparison between the (left) MUSE equivalent \hst/ACS/F555W bandwidth and (right) \hst\ images of the field around NGC~1313~X--1, after astrometric correction of the MUSE cube. The white circle in both images is 0.61" in radius and shows the 3$\sigma$ uncertainty on the position of the ULX \theulx\ in the MUSE cube. The position of the optical counterpart of \theulx\ determined by \cite{yang_optical_2011} can be seen on the \hst\ image. The red circles indicate a bright star of a young cluster of massive stars \citep{ryon_effective_2017} to aid the comparison between the images. The contours show the flux of the [O I]$\lambda$6300 line (see Figure~\ref{fig:flux_maps}). Both images have the same orientation and show a region of 30"$\times$30".}
    \label{fig:astro}
\end{figure*}

The optical counterpart of \theulx\ was identified by \cite{yang_optical_2011}, which is depicted in the \hst\ image in Figure \ref{fig:astro}. To identify the position of the counterpart in the MUSE cube, we used the \hst\ image taken in the ACS/F555W filter to correct the astrometry of the cube and register it to the \hst\ coordinates (Figure \ref{fig:astro} left). To do so, we used the \texttt{python} package \texttt{mpdaf} \citep{piqueras_mpdaf_2017} to integrate the MUSE cube in wavelength space, weighting each wavelength by the ACS/F555W filter throughput in order to generate a MUSE image in the F555W filter (central wavelength of 5360 \AA\ with a bandwidth of 360 \AA). Then we used the task \texttt{adjust\_coordinates} from the same package to run an autocorrelation function between the F555W MUSE and the \hst\ images. The maximum of the autocorrelation function was used to shift the cube coordinates relative to the \hst\ image and register its coordinates to this reference image. The resulting alignment is shown in Figure \ref{fig:astro} and showed excellent agreement with the \hst\ image. We considered the uncertainty on the position of the counterpart in the cube to be the nominal \hst\ PSF added in quadrature to the pixel scale of MUSE (0.2"). We fitted a 2D Gaussian profile to the optical counterpart in the \hst\ image to determine the PSF FWHM = 0.110$\pm$0.003". Thus, the 3$\sigma$ uncertainty of the position of the optical counterpart in the MUSE cube was found to be $\sim$0.61". We note that the counterpart was not clearly detected in this MUSE image (but see Section~\ref{sub:optical_counterpart}). 

\section{Data analysis and results}\label{sec:bubble_data_analysis}

\subsection{Line maps} \label{sec:camel}
In order to derive physical quantities from the main nebular lines, we used the publicly available code CAMEL\footnote{\url{https://gitlab.lam.fr/bepinat/CAMEL}} which is described in \cite{epinat_massiv_2012}. CAMEL performs pixel-by-pixel Gaussian line fitting to derive flux and kinematic maps for each fitted line. We enhanced the signal-to-noise ratio (S/N) of the faint lines in the cubes by applying a 2D spatial Gaussian smoothing with FWHM = 2 pixels as the spatial resolution of the cube (PSF FWHM $\sim$ 1") was well above the pixel scale of MUSE (0.2"). A  single Gaussian component was chosen to fit the lines, and CAMEL was run in five independent spectral cuts so the continuum could be easily modelled with a one degree polynomial in each cut. The only exception to this was the \hb\ line that presented some Balmer absorption (see the dipping structure around the \hb\ in Figure~\ref{fig:flux_cal}) that we took into account using a Gaussian absorption line together with a zero-degree polynomial to model the continuum. We found that the relative differences in \hb\ flux between the approach where the continuum was modelled with a one degree polynomial and the approach taking into account the Balmer absorption were 1--50\% (10th and 90th percentiles). Cuts in the spectral space were performed around the \hb\ line, both [O~III] lines, the [O~I]$\lambda$6300 line, the [N~II]-\ha\ complex, and finally one for the [S~II] doublet. In each case we assumed the same line width for each pair of forbidden lines and the same line-of-sight velocity for the whole set of lines fitted in each cut.

\begin{figure}
    \centering
    \includegraphics[width=0.41\textwidth]{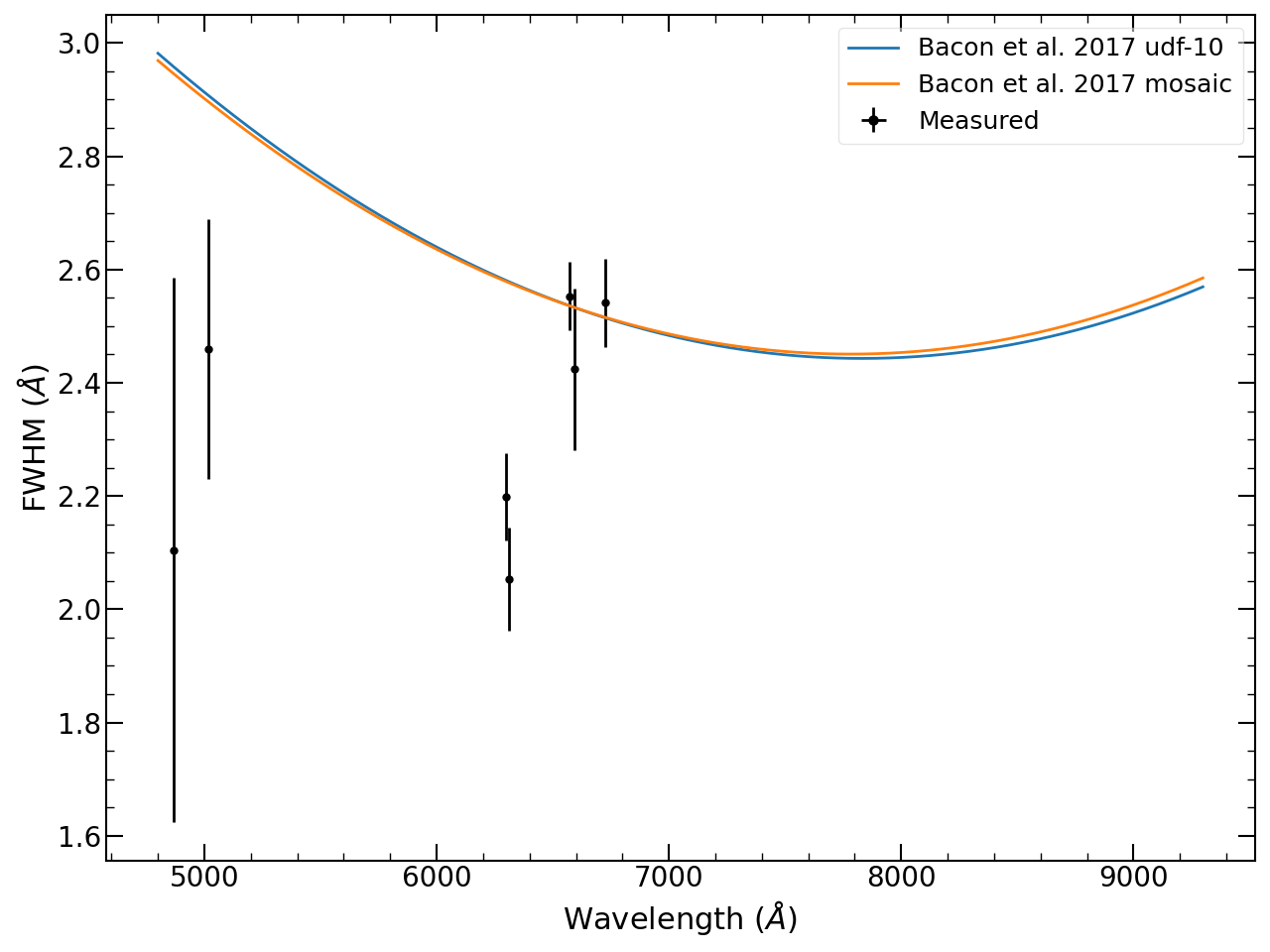}
    \caption{Estimated FWHM (see text for details) compared to the parametrisation of \cite{bacon_muse_2017}. The FWHM estimated from the data cube does not show a good agreement with the parametrisation of \cite{bacon_muse_2017} towards the blue part of the spectrum.} \label{fig:fwhm_est}
\end{figure}

The dispersion maps were corrected for instrumental broadening by subtracting the instrumental line spread function (LSF) FWHM in quadrature. In order to estimate the LSF FWHM, we computed the median and the standard deviation of the 10th percentile FWHM of the dispersion maps produced by CAMEL for all the lines we measured. We additionally included the sky line at [O~I]$\lambda$6300, which was present throughout the spatial extent of the cube at its rest-frame wavelength.  We only considered lines with  S/N $>$ 5 to avoid introducing noise in the calculation. Table~\ref{tab:fwhm_estimation} reports the determined FWHM for the different lines along with the number of pixels from the dispersion map used in the calculation whereas Figure~\ref{fig:fwhm_est} shows a comparison with the parametrisation from \cite{bacon_muse_2017} (see their Equations 7 and 8) for cube 1. The results for cube 2 were similar, and we do not report them for brevity. 

The discrepancy found in the \hb\ line might be due to the stellar absorption around this wavelength, affecting the estimation of its FWHM. The discrepancy between the other lines' FWHM values and the parametrisation of \cite{bacon_muse_2017} is unclear. The parametrisation from \cite{bacon_muse_2017} was derived by fitting the sky lines from a combined cube composed of 25-minute exposures totalling 10--37h with a specific observation pattern, where the field was rotated by 90$^\circ$ between exposures, an observing strategy far more complex than our single-exposure data cubes. It is possible that the recipe derived by \cite{bacon_muse_2017} might not be suited for our more simplistic case. We decided to focus only the FWHM of the [N~II]-\ha\ and [S~II] lines, which showed good agreement with the parametrisation proposed by \cite{bacon_muse_2017}. We corrected the observed FWHM using our own estimation of the LSF FWHM, although the results are unchanged if instead we used the values obtained from \cite{bacon_muse_2017}. 

\begin{table} 
 \centering 
 \caption{Results from the instrumental LSF FWHM estimation (see text for details). For the forbidden doublets only one line is shown as the same FWHM value was assumed.}  \label{tab:fwhm_estimation}
 \begin{tabular} {lcc} 
 \hline
  \noalign{\smallskip}
 Line & FWHM\tablefootmark{a} & N\tablefootmark{b} \\ 
 & \AA & \\
 \hline
 \hline 
 \noalign{\smallskip}
  \hb\ & 2.1$\pm$0.5 & 2289 \\  
  {[O~I]}$\lambda$6300\tablefootmark{$\dagger$}&  2.20$\pm$0.08 & 2880 \\   
  {[O~I]}$\lambda$6300 &  2.05$\pm$0.09 & 2734 \\ 
{[O~III]}$\lambda$5007 & 2.5$\pm$0.2 & 3101 \\ 
\ha\ & 2.55$\pm$0.06 & 3276 \\ 
{[N~II]}$\lambda$6583 & 2.4$\pm$0.1 & 3276 \\ 
{[S~II]}$\lambda$6716 & 2.54$\pm$0.08 & 3276 \\ 
\noalign{\smallskip}
\hline
\end{tabular}
\tablefoot{\tablefoottext{a}{Average and standard deviation of the pixels used.}\tablefoottext{b}{Number of pixels from the dispersion map used.}
 \tablefoottext{$\dagger$}{Sky line}}
 \end{table}
 %python ~/scripts/pythonscripts/muse/determine_lsf.py -s 5 SII6716 HALPHA HBETA OI6300 OI6300_sky OIII5007 NII6583 -p 10

The resulting line-of-sight and dispersion maps were more severely affected by the invalid pixels and noise present in the cubes in comparison with the flux maps. We therefore decided to use the information from cube 2 to improve them. As stated in Section~\ref{sec:bubble_data_reduction}, because the spatial distribution of the noise is unique to each cube, we found that we could improve the S/N of these maps by averaging the information obtained from cube 1 and cube 2. For the velocity maps, we further noted that the [N~II]-\ha\ and [S~II] were consistent with each other, both spatially and numerically (the median and the standard deviation of the velocity differences between the two maps were --0.09$\pm$5 km/s). We therefore decided to combine the information from the [N~II]-\ha\ and [S~II] maps from cube 1 and cube 2 in order to create an average map of these lines. The [O~III] and [O~I]$\lambda$6300 maps were also produced by averaging cubes 1 and 2. We similarly averaged the maps from the \ha, [S~II]$\lambda$6716, and [N~II]$\lambda$6583 lines for the dispersion maps to obtain a smoother map as the maps showed again good consistency  (the median and standard deviation of the differences between the \ha\ and [S~II]$\lambda$6716 maps were 13$\pm$12 km/s, with the same values found for the differences between the \ha\ and [N~II]$\lambda$6583 maps). In this case we did not use the information from cube 2 as it was considerably affected by noise and did not provide any significant improvement.

\begin{figure}
\centering
\includegraphics[width=0.85\linewidth]{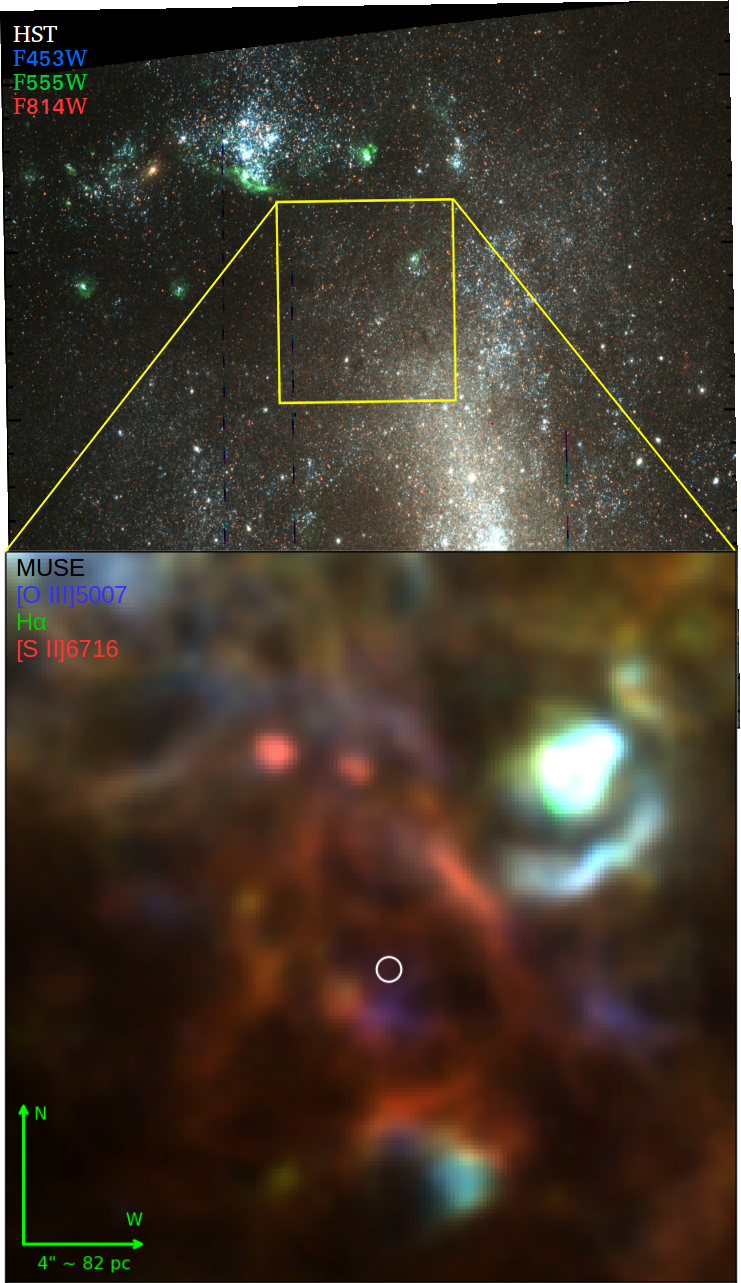}
\caption{(Top) Composite \hst\ image showing F435W (Blue), F555W (Green) and F814W (red) ACS/WFC images downloaded from the HLA archive (ObsID hst\_9796\_01\_acs\_wfc\_f814w\_f555w\_f435w). The yellow square shows the region shown in the MUSE image in the bottom panel. (Bottom) MUSE RGB image of the nebula around \theulx\ corresponding to the yellow square in the \hst\ image, showing fluxes in \ha\ (green channel), [O III]$\lambda$5007 (blue channel) and [S II]$\lambda$6717 (red channel) lines.}
    \label{fig:rgb_image}
\end{figure}

\begin{figure*}
    \centering
\includegraphics[width=0.49\textwidth]{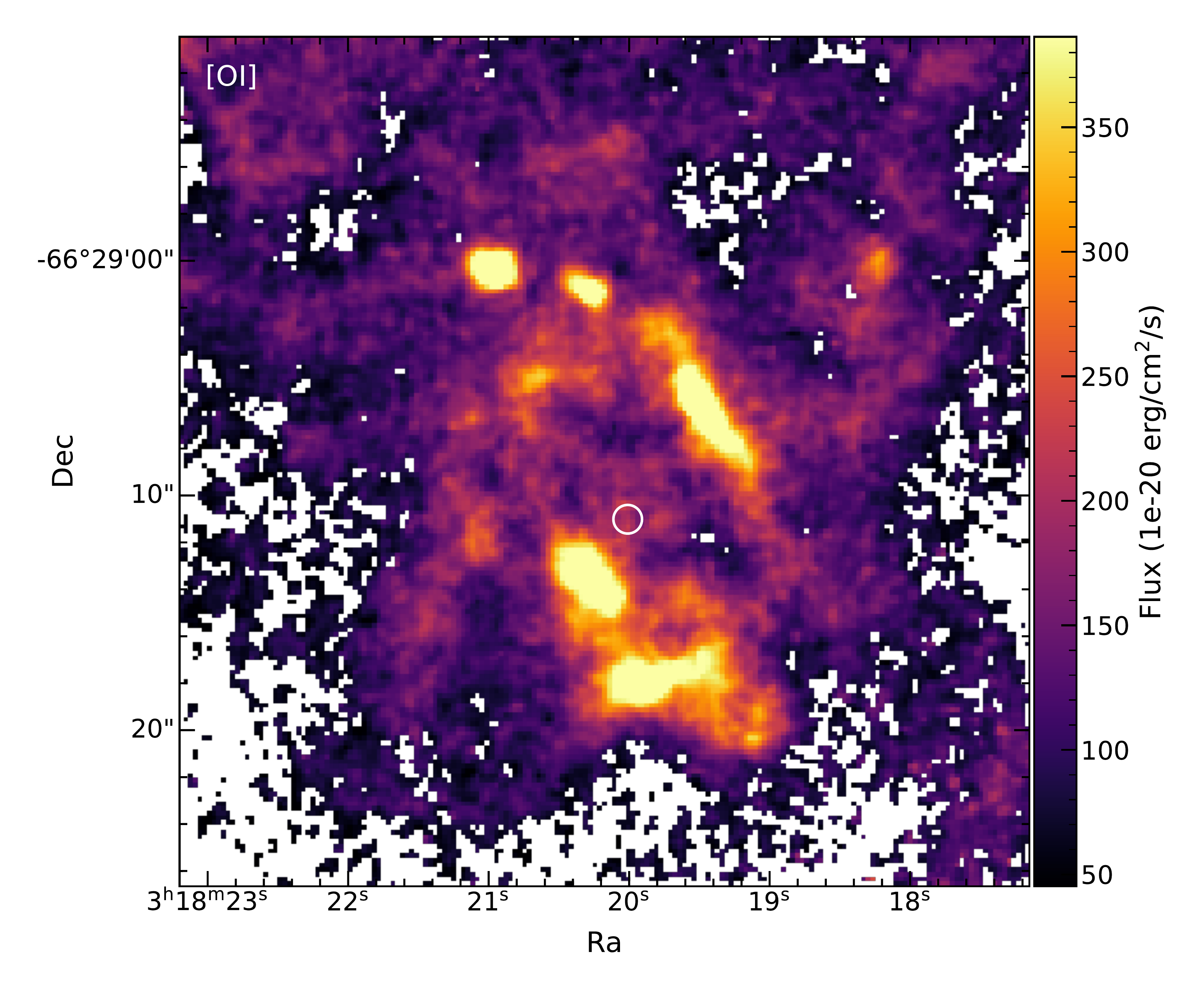}
 \includegraphics[width=0.49\textwidth]{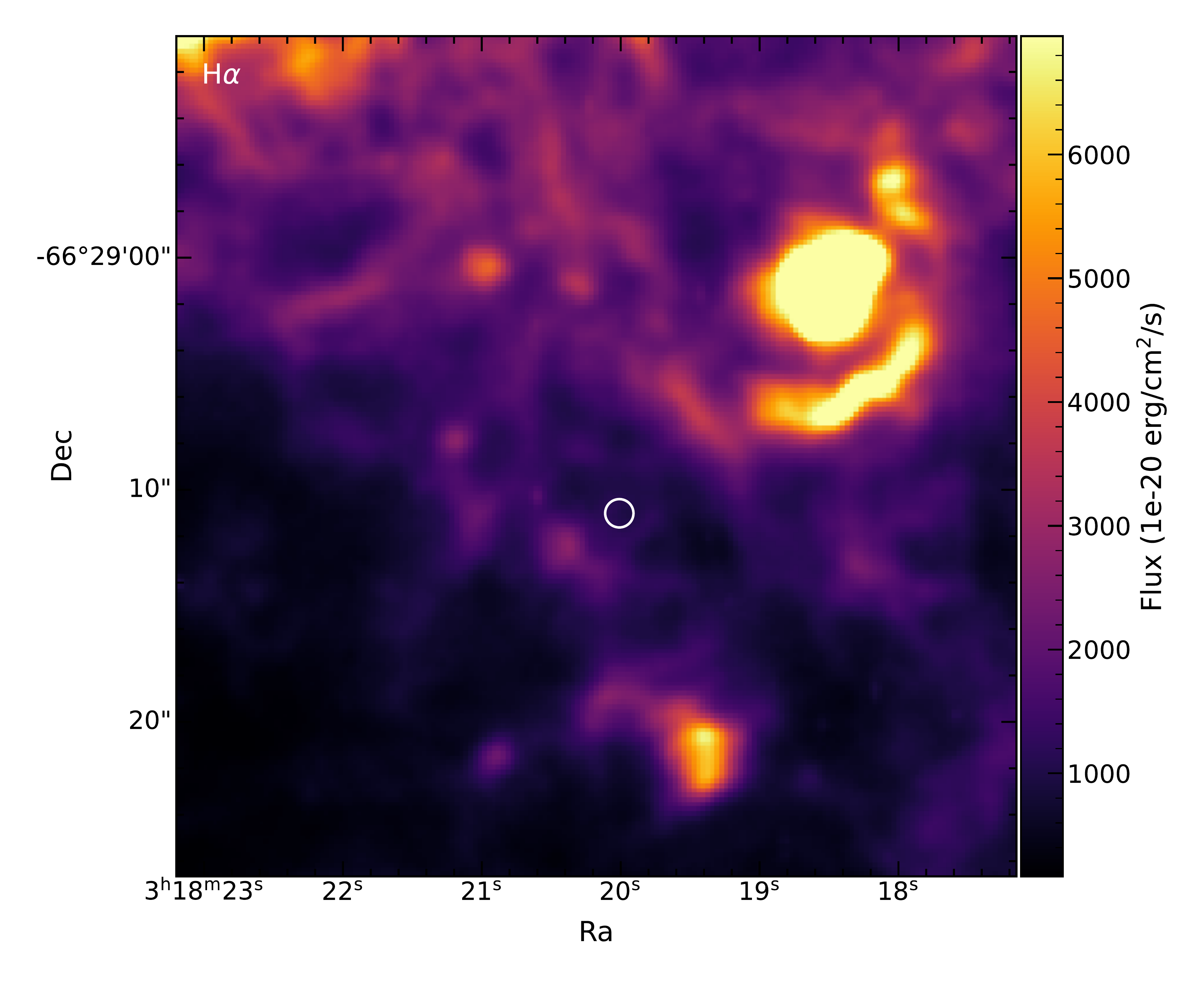}
\includegraphics[width=0.49\textwidth]{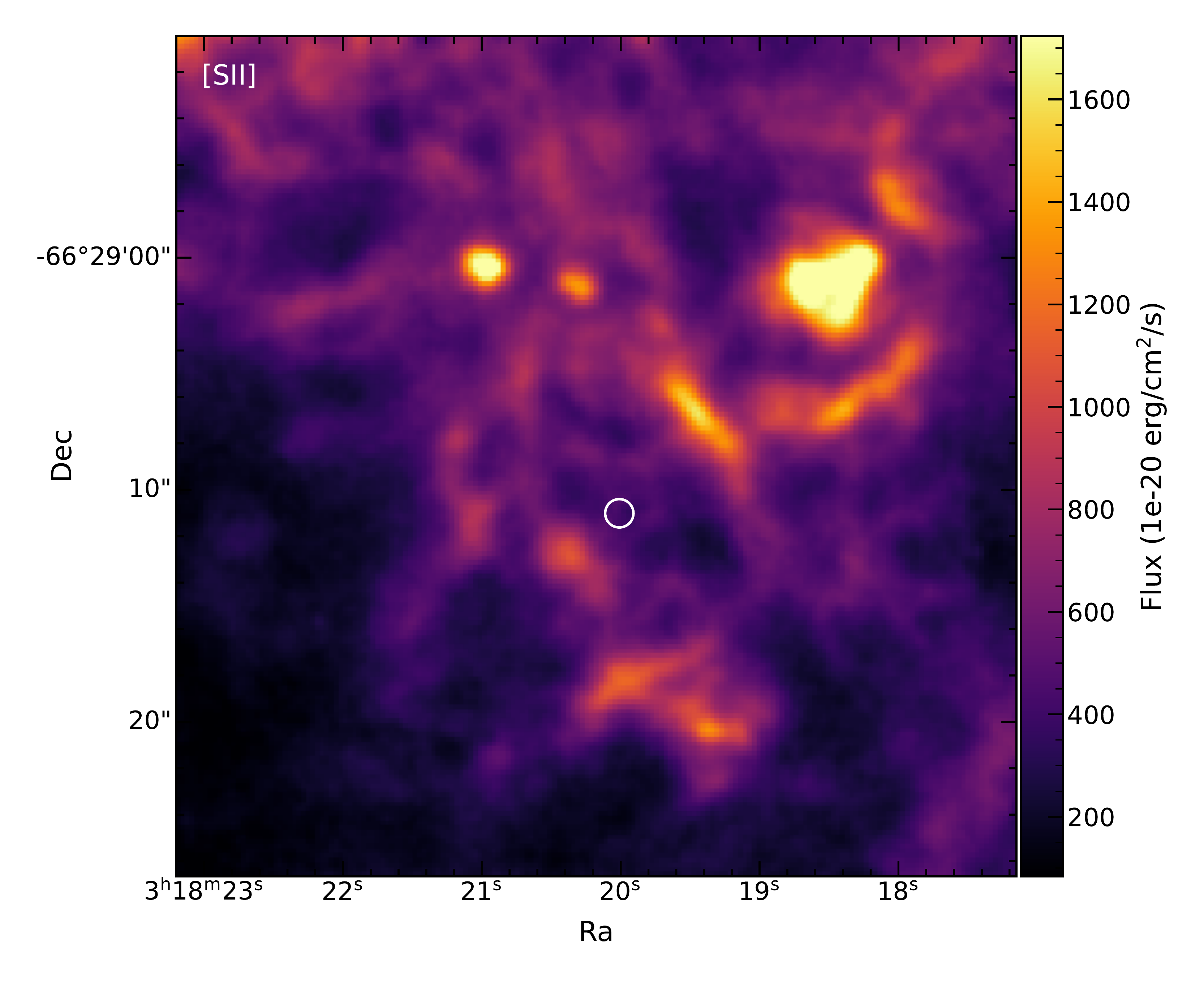}
\includegraphics[width=0.49\textwidth]{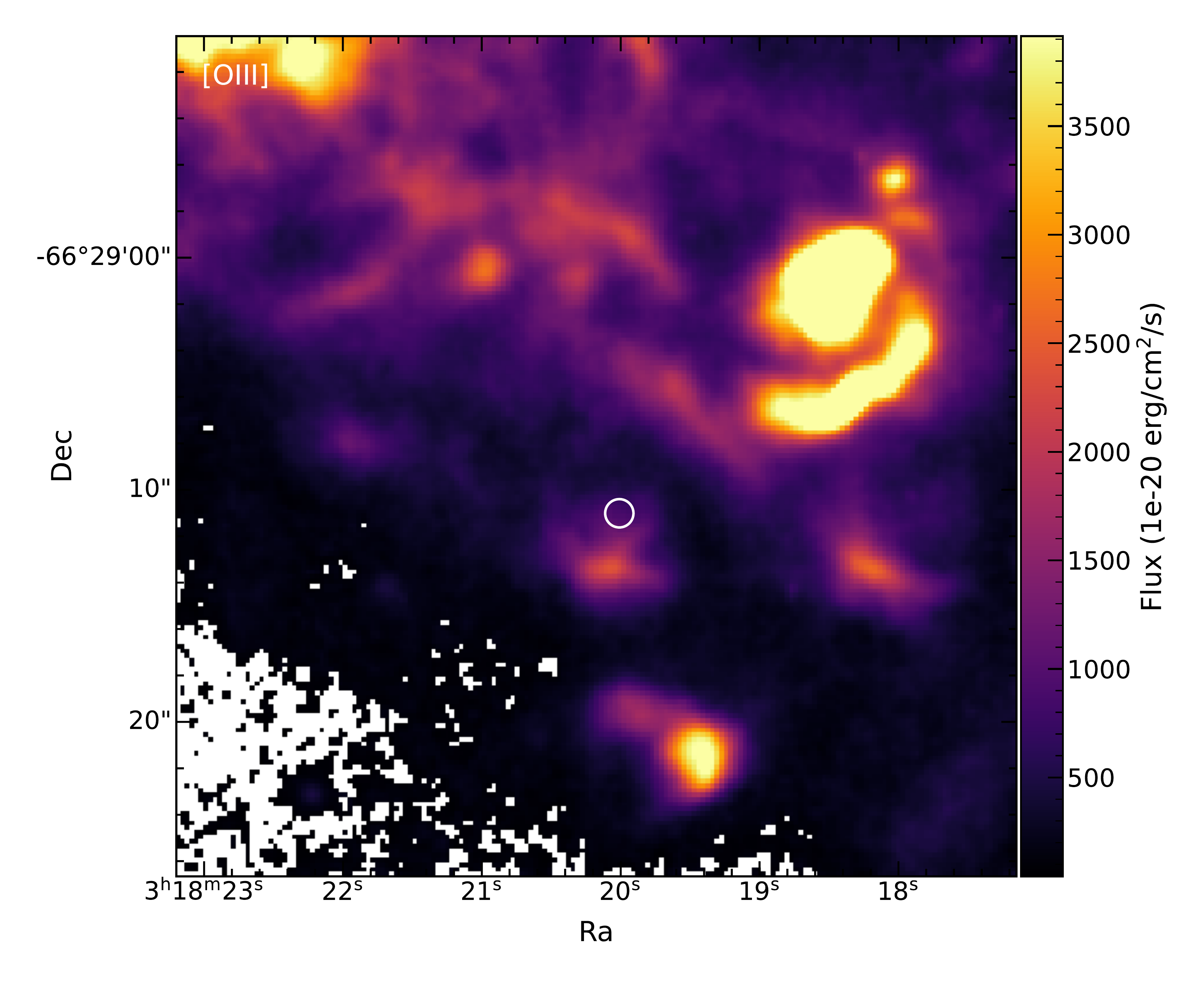}
    \caption{Flux maps for [O~I]$\lambda$6300 (top left), \ha\ (top right), [S~II]$\lambda$6716 (bottom left), and [O~III]$\lambda$5007 (bottom right). The small white circle indicates the position of \theulx\ and its uncertainty (see text for details). The scale and direction are the same as in Figure \ref{fig:rgb_image}.}
    \label{fig:flux_maps}
\end{figure*}

\begin{figure*}
    \centering
    \includegraphics[width=0.49\textwidth]{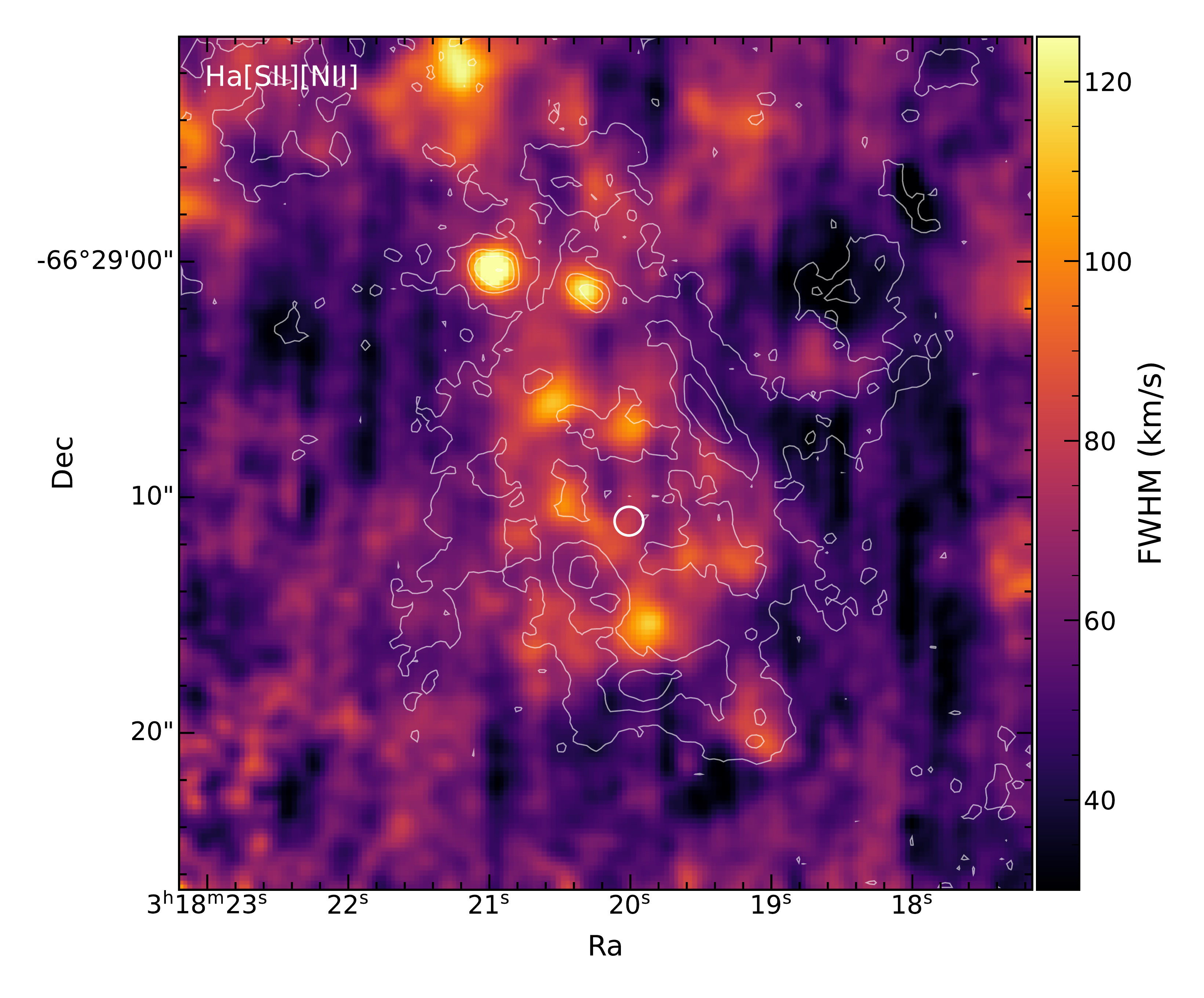}
      \includegraphics[width=0.49\textwidth]{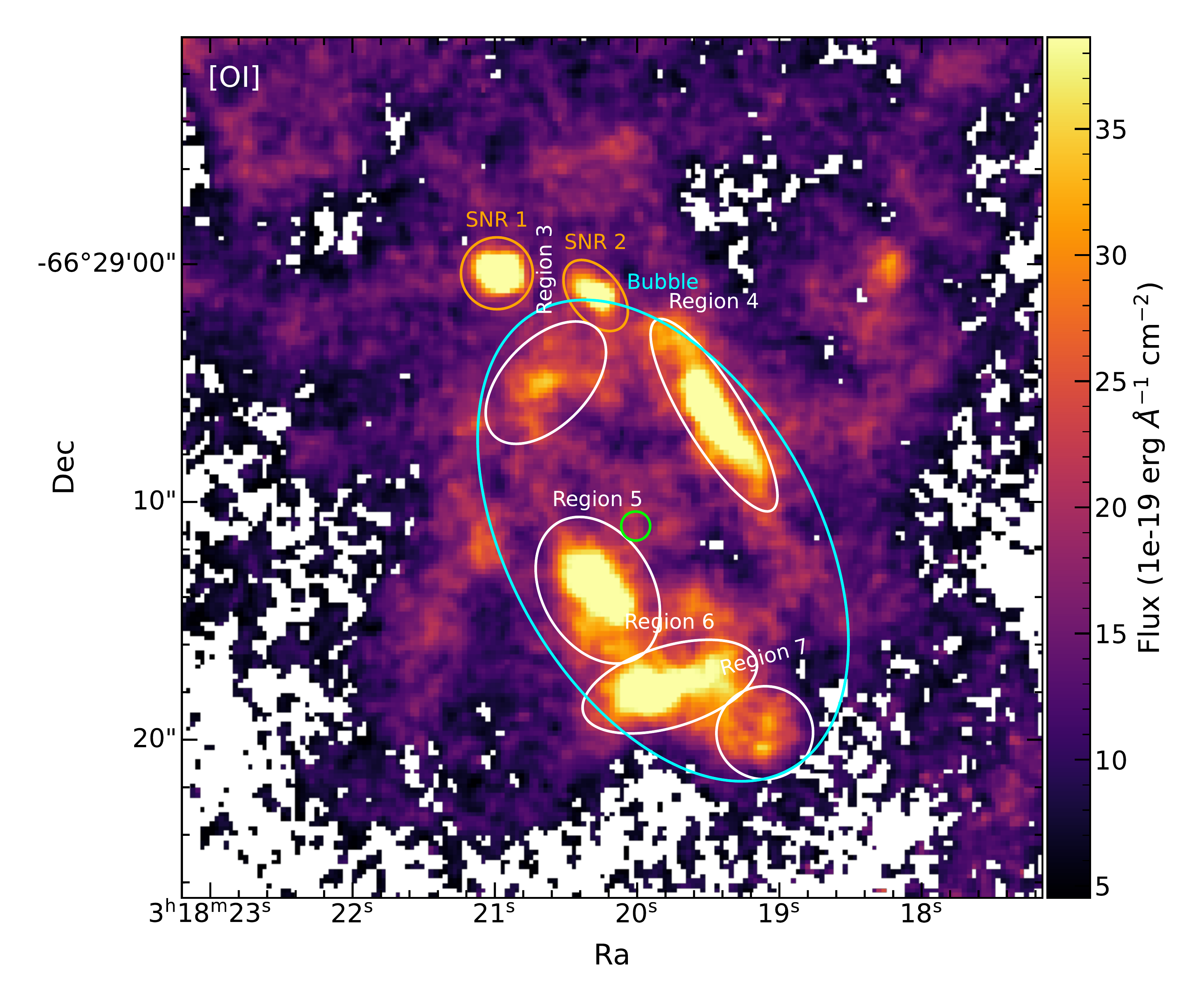}
    \caption{Dispersion and flux maps of the region around \theulx. (left) The dispersion map of the average of the \ha, [N~II]$\lambda$6853, and [S~II]$\lambda$6716 lines, corrected for instrumental broadening (contours as in Figure \ref{fig:astro}). The map has been smoothed with a Gaussian kernel with $\sigma$ = 1.5 pixels. (right) [O~I]$\lambda$6300 flux map showing some regions discussed throughout the paper, the ULX bubble (cyan), and the coincidentally close SNRs (orange). The position of the ULX is indicated with a green circle. }
    \label{fig:disp_maps}
\end{figure*}

\begin{figure*}
    \centering
    \includegraphics[width=0.99\textwidth]{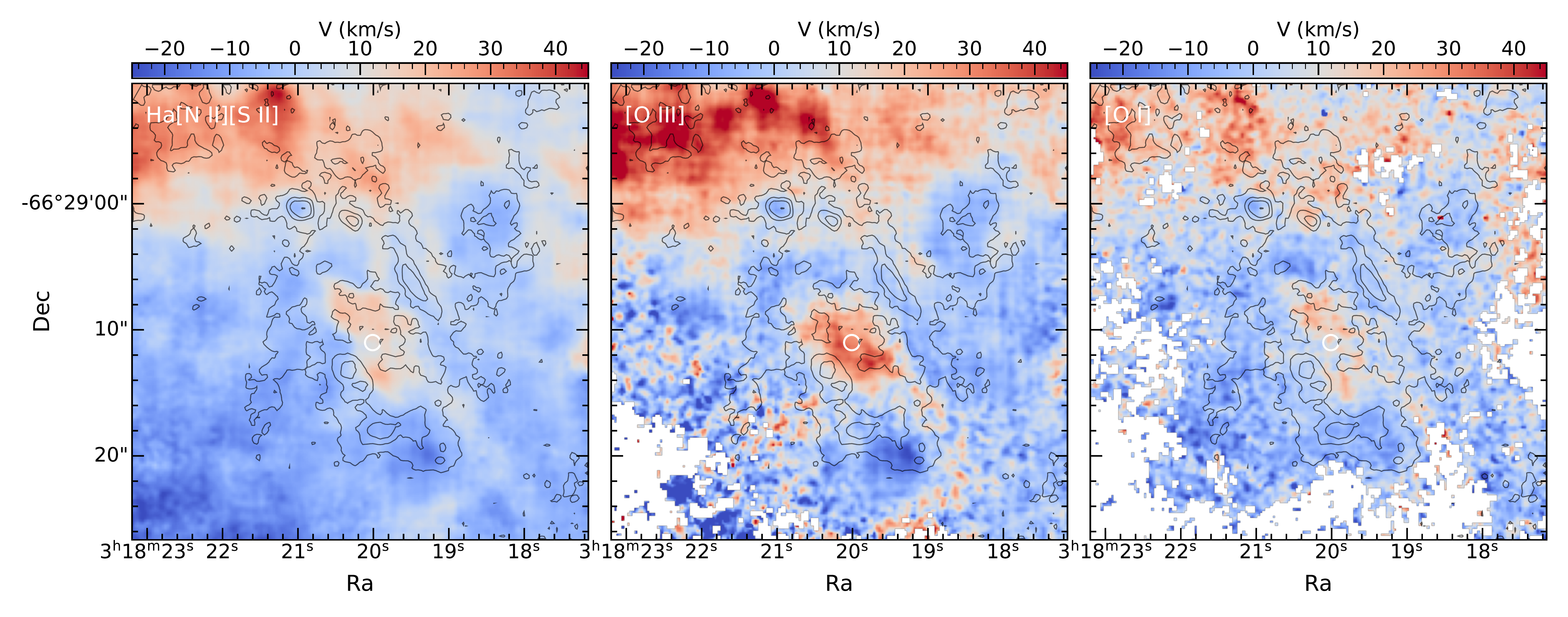}
    \caption{Line-of-sight velocity maps over the system velocity of NGC 1313 ($\sim$ 470 km/s) for the combined [N~II]-\ha\ and [S II] lines (left), [O III] (middle), and [O I]$\lambda$6300 (right). An offset of 25 km/s (the median value estimated from the whole map after excluding the central bubble) has been subtracted to reflect more closely the local velocities, which is in good agreement with the 30 km/s estimated at the position of \theulx\ based on the values given by \cite{ryder_new_1995} (see Introduction). The scale is the same in all panels, the contours as in Figure \ref{fig:disp_maps}. }
    \label{fig:vel_maps}
\end{figure*}

The resulting maps for the line flux, dispersion and line-of-sight velocity are shown in Figures~\ref{fig:flux_maps}, \ref{fig:disp_maps}, and \ref{fig:vel_maps} for some of the nebular lines with the best S/N. Pixels with a line detection S/N below 5 have been masked. Figure~\ref{fig:rgb_image} shows a composite image of the \ha, [O III]$\lambda$5007, and [S II]$\lambda$6716 lines. For the line-of-sight velocity maps, we subtracted the median value measured from the map after excluding the central parts of the bubble in order to reflect the local velocities more closely. This value was found to be 25 km/s, which is in good agreement with the estimated 30\,km/s based on the measurements from \cite{ryder_new_1995} in Section~\ref{sec:bubble_intro}. 

The right panel of Figure~\ref{fig:disp_maps} shows some regions of interest that will be discussed throughout the paper, and in particular in Sections~\ref{sub:density_est} and \ref{sub:mec_power}. These were obtained by fitting 2D Gaussian ellipsoidal profiles to the [O~I]$\lambda$6300 flux map, except for SNR 1 for which we used the \ha\ map instead (see Section~\ref{sec:mappings}) and a circular Gaussian given its apparently symmetric profile. The radii of the regions were set equal to their respective FWHMs obtained, with the exception of regions 5, 6, and 7. For the first two we reduced the sizes obtained to avoid overlapping between them, whereas region 7 was simply determined visually. The integrated fluxes of these regions and the bubble are quoted in Table~\ref{tab:line_fluxes}, while the sizes are given in Table~\ref{tab:pre_shock}. The two bright [S II] and [O I]$\lambda$6300 northernmost blobs are likely two SNRs coincidentally close to the ULX bubble, as we explain  in Section~\ref{sub:snrs}. We describe first the regions concerning the interaction of the ULX with the interstellar medium (ISM).

The \ha, [O~I]$\lambda$6300, and [S II]$\lambda$6716 flux maps clearly reveal an elongated bubble in the south-west--north-east direction,  452 $\times$ 266\,pc in size (highlighted in the right panel of Figure~\ref{fig:disp_maps} in cyan). The ULX is roughly at 1.2" from the centre of the ellipsoidal shape marked by the bright [O I]$\lambda$6300 regions, corresponding to a physical distance of 25\,pc.  

\begin{table*}
\caption{Observed fluxes of the different nebular lines relative to \hb, integrated in each of the regions highlighted in the right panel of Figure~\ref{fig:disp_maps}. Uncertainties are given at the 1$\sigma$ level by adding the uncertainty on each pixel in quadrature.}\label{tab:line_fluxes}
\centering
\resizebox{\textwidth}{!} {
\begin{tabular}{ccccccccc}
\hline \hline
Line & SNR 1& SNR 2& Region 3& Region 4& Region 5& Region 6& Region 7 & Bubble\\
\hline \hline 
\noalign{\smallskip}
\hb\ (10$^{-15}$\,\fluxcgs)   & 1.5$\pm$0.1  & 0.95$\pm$0.08  & 2.1$\pm$0.1  & 3.2$\pm$0.2  & 2.4$\pm$0.1  & 2.2$\pm$0.1  & 1.7$\pm$0.1  & 24.1$\pm$0.4  \\
\hline
\noalign{\smallskip}
[O III]$\lambda$4959 &  0.698$\pm$0.002  & 0.713$\pm$0.003  & 0.500$\pm$0.002  & 0.589$\pm$0.001  & 0.707$\pm$0.002  & 0.386$\pm$0.002  & 0.470$\pm$0.002  & 0.550$\pm$0.001 \\
$[$OIII]$\lambda$5007 & 2.175$\pm$0.010  & 2.156$\pm$0.011  & 1.543$\pm$0.007  & 1.799$\pm$0.006  & 2.143$\pm$0.008  & 1.154$\pm$0.006  & 1.449$\pm$0.008  & 1.674$\pm$0.002  \\ 
$[$O I]$\lambda$6300 &0.414$\pm$0.002  & 0.433$\pm$0.002  & 0.455$\pm$0.001  & 0.474$\pm$0.001  & 0.722$\pm$0.002  & 0.770$\pm$0.002  & 0.417$\pm$0.002  & 0.496$\pm$0.001 \\
$[$N II]$\lambda$6548 & 0.191$\pm$0.001  & 0.192$\pm$0.001  & 0.207$\pm$0.001  & 0.232$\pm$0.001  & 0.276$\pm$0.001  & 0.252$\pm$0.001  & 0.220$\pm$0.001  & 0.227$\pm$0.001\\
\ha\ & 3.614$\pm$0.005  & 3.718$\pm$0.004  & 3.490$\pm$0.002  & 3.688$\pm$0.002  & 3.555$\pm$0.002  & 3.741$\pm$0.002  & 3.687$\pm$0.003  & 3.580$\pm$0.001   \\ 
$[$N II]$\lambda$6583 & 0.610$\pm$0.006  & 0.610$\pm$0.005  & 0.646$\pm$0.003  & 0.732$\pm$0.003  & 0.850$\pm$0.004  & 0.794$\pm$0.003  & 0.691$\pm$0.004  & 0.711$\pm$0.001 \\
$[$S II]$\lambda$6716 & 1.304$\pm$0.003  & 1.284$\pm$0.004  & 1.374$\pm$0.002  & 1.387$\pm$0.002  & 1.522$\pm$0.002  & 1.804$\pm$0.002  & 1.305$\pm$0.003  & 1.388$\pm$0.001  \\
$[$S II]$\lambda$6731 & 0.954$\pm$0.002  & 0.920$\pm$0.002  & 0.962$\pm$0.001  & 0.976$\pm$0.001  & 1.079$\pm$0.001  & 1.263$\pm$0.001  & 0.924$\pm$0.001  & 0.980$\pm$0.001 \\
\hline\hline
\end{tabular}
}
\end{table*}

Figure~\ref{fig:disp_maps} shows that the lines are broadened just inside the edges marked by the [O~I]$\lambda$6300 and [S~II]$\lambda$6716 flux maps (see particularly the strong broadening FWHM $\sim$ 100 km/s towards the south near region 6, north-east near region 3, and east to the \ulx\ near region 5). This is a telltale sign that the gas is being shocked in these regions and defines the limit of the interaction of the ULX with the surrounding medium. The maximum FWHM measured in these regions (3, 4, 5, 6, and 7) were in the 90--130 km/s range (Table~\ref{tab:pre_shock} gives the maximum FWHM values measured in each of the regions of interest). Instead, the [O~III]$\lambda$5007 flux map shows slightly different morphology, with a brighter region just south of the ULX position, also clearly visible in Figure~\ref{fig:rgb_image} as a light blue blob.  

\begin{figure*}
    \centering
    \includegraphics[width=0.99\textwidth]{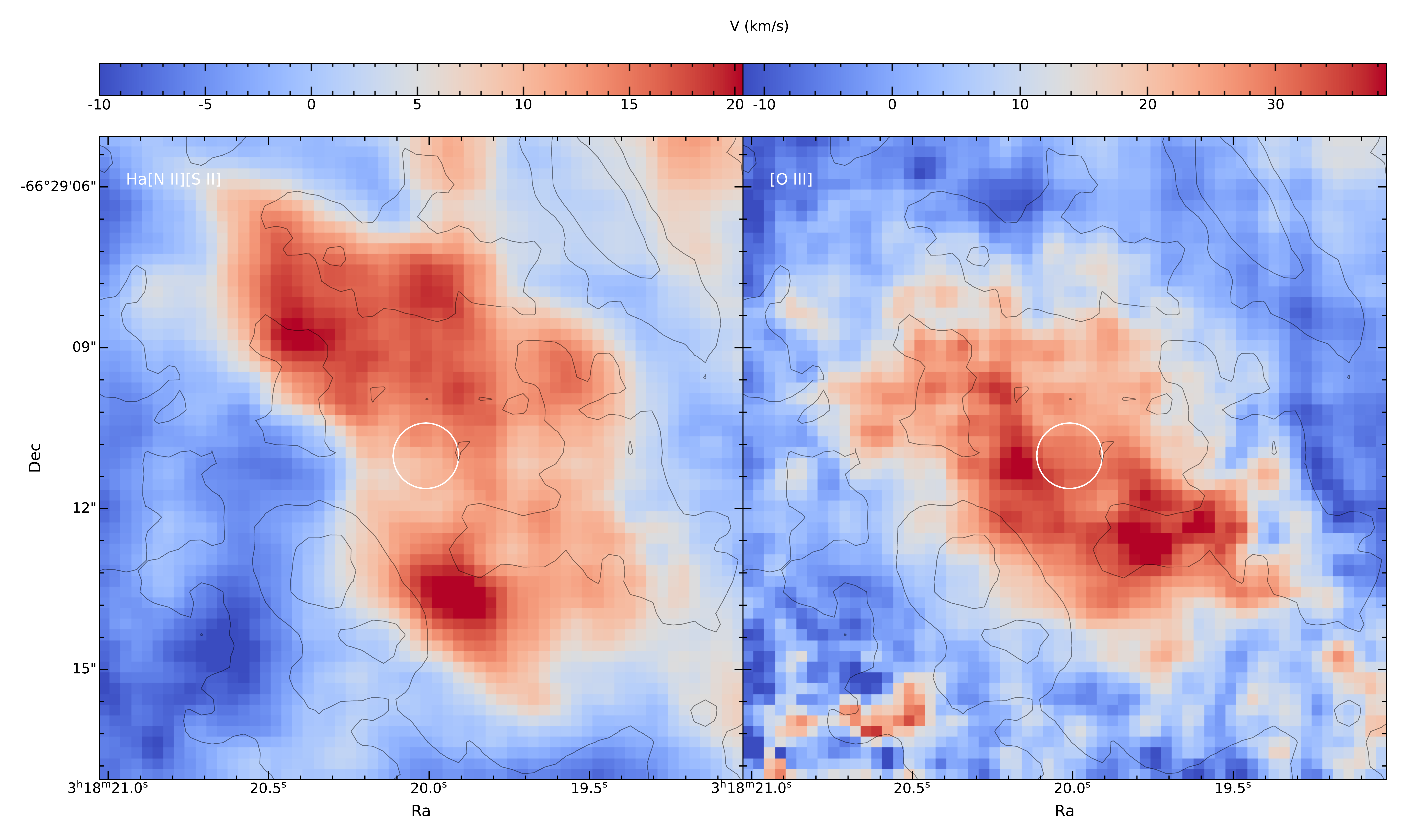}
    \caption{Close-up view of the velocity maps for the \ha[N~II][S~II] (\text{left}) and [O III] maps (\text{right}). Note the difference in scale between the two maps. The position of the ULX is indicated with a white circle. Contours as per Figure~\ref{fig:vel_maps}.}
    \label{fig:vel_profile} 
\end{figure*}

The line-of-sight velocity (\vlos) fields (Figures~\ref{fig:vel_maps}) show a structure roughly contained inside the [O I]$\lambda$6300 contours, with \vlos\ about 30--40 km/s higher than those of the surrounding gas, indicating that the gas inside the bubble is being dynamically perturbed by the ULX via outflows. This is clearly seen in the combined [N II]-\ha\ and [S II] and [O III]$\lambda$5007 \vlos\ fields. This is illustrated more clearly in Figure~\ref{fig:vel_profile}, which is a close-up of the two maps around the ULX position. We also note  the distinct structure of the two maps, which suggests that the [O III]$\lambda$5007 map is tracing different gas compared to the combined \ha[N II][S II] map. The combined [N II]-\ha\ and [S II] velocity fields show peak velocities of $\sim$ 20 km/s, whereas the surrounding gas shows slower velocities reaching at most $\sim$ 0 km/s with a typical 1$\sigma$ uncertainty of $\sim$ 1 km/s. The [O III]$\lambda$5007 shows higher velocities reaching up to $\sim$ 35 km/s, whereas the gas in the outer regions of the bubble has velocities of at most $\sim$ 5 km/s. The typical 1$\sigma$ uncertainty of the maps of the individual cubes here is slightly higher (by $\sim$ 4 km/s). The [O I]$\lambda$6300 map shows a somewhat similar structure to the \ha[N II][S II] maps, with two distinct regions north and south from the ULX, albeit with considerably higher noise.

Finally, we note that the dispersion map in the [O~I]$\lambda$6300 line reveals some strong broadening (uncorrected FWHM $\gtrsim$ 200 km/s) at the position of \theulx\ (see Figure~\ref{fig:extraction_region}). We note,  as stated in Section \ref{sec:bubble_data_reduction}, that the exact value of the instrumental-corrected FWHM is uncertain. Nevertheless, because this broadening is coincident with the source position, we used it in Section~\ref{sub:optical_counterpart} to extract a spectrum of the optical counterpart.

\subsection{Supernova remnants in disguise} \label{sub:snrs}

Figures~\ref{fig:rgb_image} and ~\ref{fig:flux_maps} reveal two compact spots bright in the [S~II] and [O~I]$\lambda$6300 lines, located north of the ULX (labelled SNR 1 and SNR 2 in the right panel of Figure~\ref{fig:disp_maps}). These are at $\sim$ 12" and 9.8" from the ULX position, corresponding to a physical distance of $\sim$ 250\,pc and 200\,pc, respectively. Here we find intrinsic FWHM values reaching up to $\sim$ 166\,km/s in the combined \ha, [N~II], and [S~II] dispersion map\footnote{The colour scale of the figure does not reflect such high values owing to the chosen scaling, which is meant to aid the visualisation of the other areas.}, the highest values in the whole nebula. The broadening and the strong [S~II] and [O~I]$\lambda$6300 lines implies that the gas is also being shocked in these regions. However, its association with the ULX is unclear as its distance is extreme even for ULX standards \citep[cf. 150~pc from S26~NGC~7793;][]{pakull_300-parsec-long_2010}, whereas its round morphology (notably for SNR 1) argues against a jet or wind excitation, for which we would likely expect an arc-shaped structure instead (cf. regions 3 and 7). A face-on view could still make an arc shape appear circular, but considering its already large projected distance this is highly unlikely. Recently, these regions have been identified as SNRs by \cite{kopsacheili_supernova_2021} based on their [S~II]/\ha $>$ 0.4 ratios, a common diagnostic used to identify SNRs. However, most of the gas inside the bubble also shows [S~II]/\ha $>$ 0.4 (see Section~\ref{sub:bpt_diagrams}) and in fact many ULX bubbles were initially misclassified as SNRs based on this diagnostic \citep[e.g.][]{1997ApJS..112...49M}. In this regard, \cite{urquhart_newly_2019} offers an interesting historical discussion on the early identification of several ULX bubbles as SNRs to which we refer the interested reader. For this reason we decided to investigate the nature of these regions to assess whether they were related to the ULX activity. Here we show that their identification as SNRs is nonetheless likely to be robust as it is  based on a more detailed imaging and spectral analysis.

\begin{figure*}
   \centering
    \includegraphics[width=0.9\textwidth]{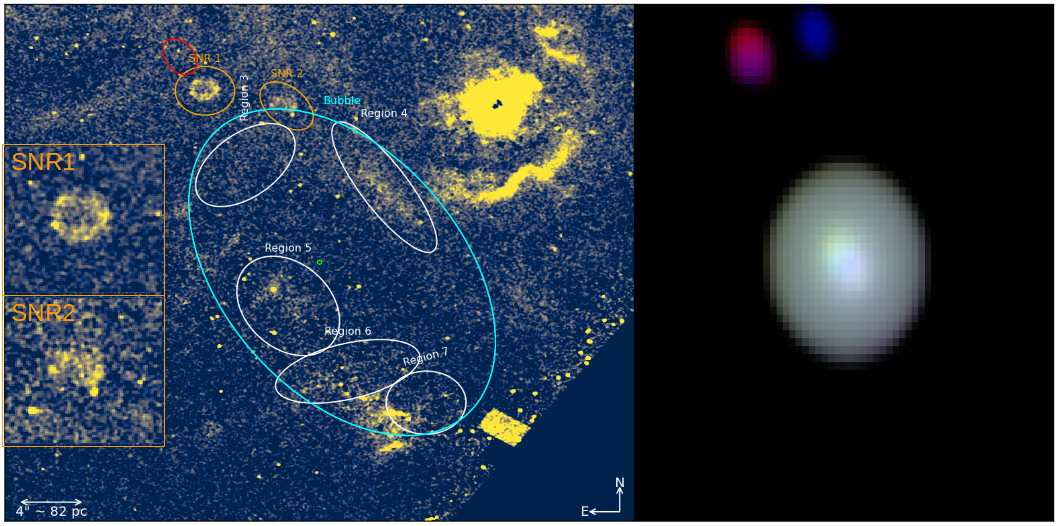}
    \caption{The field around NGC~1313~X--1 in optical and X-rays. The left panel shows the \hst\ continuum-subtracted WFC3/UVIS/F657N (\ha\ + [N II]) image. The WFC3/UVIS/F547M image was used for the continuum approximation and the position of NGC~1313~X--1 is indicated by a green circle. The red ellipse shows the 3$\sigma$ contour confidence level of the easternmost source with red-ish colours in the \chandra\ image. The same regions from Figure~\ref{fig:disp_maps} are also shown, and the insets show close-ups of the two SNRs, as indicated. The right panel shows a R,G,B = 0.5--1.2 keV, 1.2--2 keV, 2--7 keV \chandra\ image showing the same field. Both images show a of 32"$\times$32" field centred around the ULX and have been smoothed with 2 pixel Gaussian kernel.}
    \label{fig:F567N_chandra}
\end{figure*}

In order to inspect the morphology of the diffuse emission of these regions in more detail, we retrieved a \textit{WFC3/UVIS/F657N} image (which isolates the \ha\ + [N II] lines)  from the HLA archive, together with a \textit{WFC3/UVIS/F547M} image for the continuum subtraction. The details of these observations are reported in Table~\ref{tab:data}. In order to isolate more precisely the \ha\ + [N II] emission, we subtracted the continuum emission as measured from the F547M image. More specifically, we took into account the bandwidths of the  different filters  by measuring the ratio between the intensities in these two images in four different circular regions of 4--5" containing stars and no obvious diffuse emission. The median of these ratios was used as a rescale factor to the continuum image before subtracting it from the  narrowband image (prior to this, the images were aligned as explained below). The subtracted image (Figure~\ref{fig:F567N_chandra}) shows that the emission from SNR 1 is almost spherical, with an angular radius of $\sim$0.85" (17.5~pc), confirming what had been hinted at from the MUSE \ha\ image, and strongly supporting its identification as a SNR \citep{kopsacheili_supernova_2021}. Based on the morphology alone, the situation might be less clear for SNR 2 due to the low S/N of the diffuse emission, but its shape might also be suggestive of a SNR. We also note  the brighter emission from these two regions compared to the rest of the nebula.
% python ~/scripts/pythonscripts/hst/continuum_subtraction.py -n hst_13773_15_wfc3_uvis_f657n_drz.fits -c ../F547M/hst_13773_15_wfc3_uvis_f547m_drz.fits -r stars_renorm.reg

\begin{figure*}
    \centering
    \includegraphics[width=0.49\textwidth]{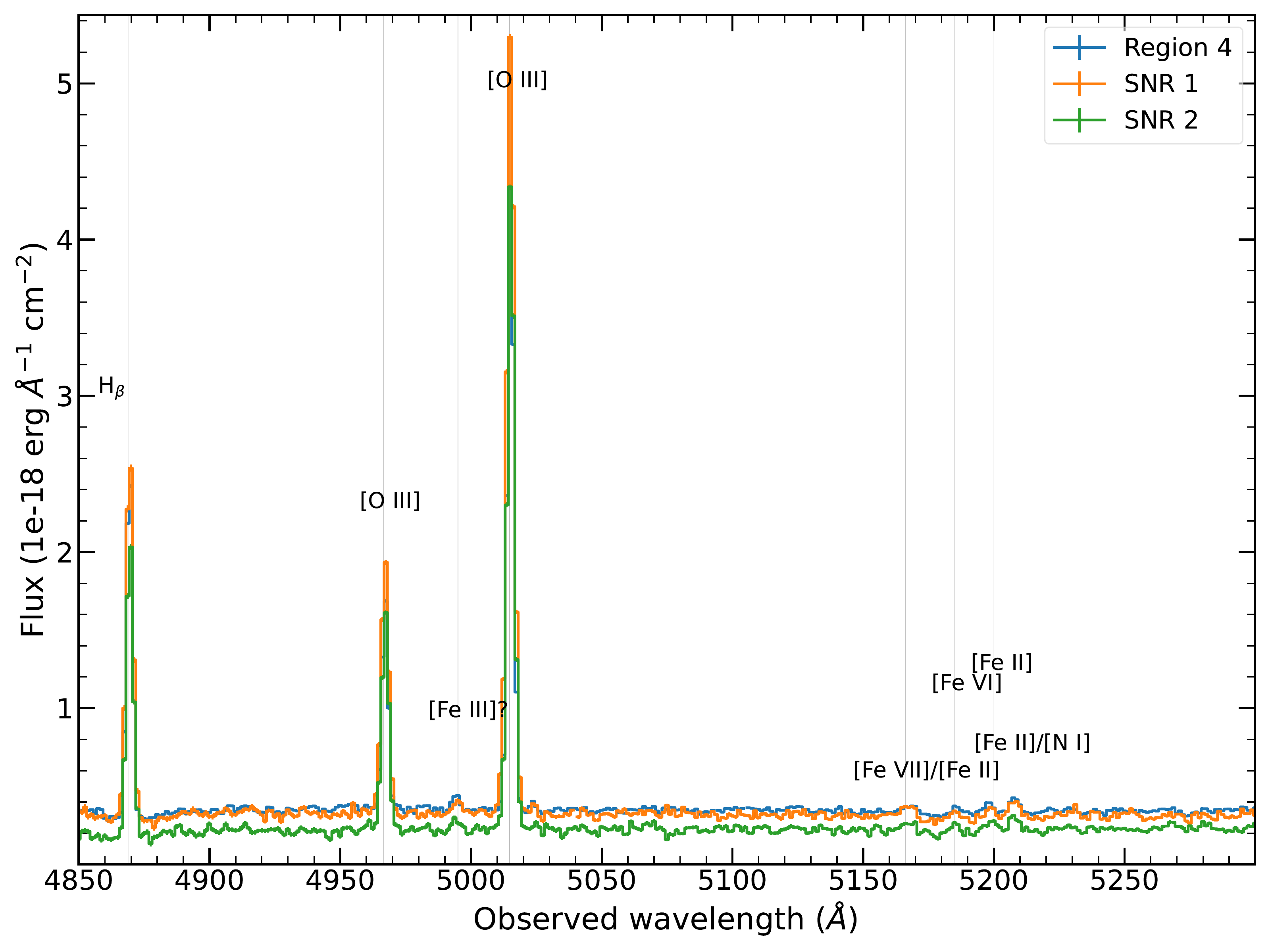}
    \includegraphics[width=0.49\textwidth]{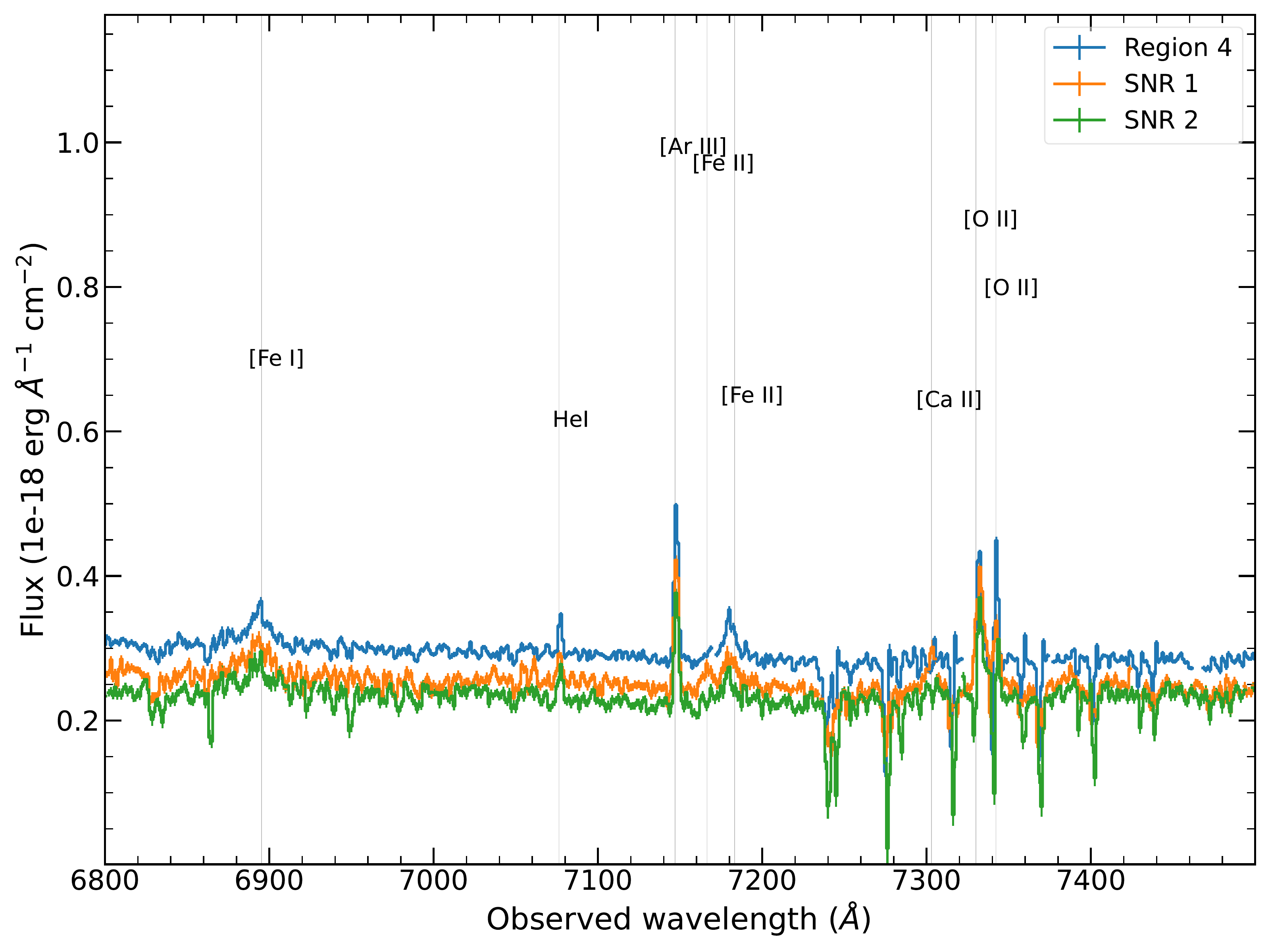}
    \caption{Average spectra extracted from regions SNR 1, SNR 2, and region 4 in Figure~\ref{fig:disp_maps} showing the spectral bands in the ranges 4850--5300 \AA\ (left) and 6800--7500 \AA\ (right). The thin vertical lines show the expected transitions at the redshift of NGC~1313 based on the line catalogue from \cite{fesen_catalog_1996}. The spectra have been smoothed with a Gaussian kernel of 2.5~\AA. Small gaps mask some pixels having negative values due to low S/N. }
    %create_snr_spectral_plots.txt
    \label{fig:snr_spectra}
\end{figure*}

We also present spectra of the two SNRs averaged over each of the regions. These are shown in Figure~\ref{fig:snr_spectra} along with the average spectrum from region 4 for comparison. The resemblance between the two SNR spectra and the presence of several emission lines typically seen in SNRs \citep{fesen_catalog_1996} strengthens the association of the SNR 2 as a SNR. A detailed analysis of these spectra is beyond the scope of the present paper, but based on the similarity between SNR 1 and SNR 2 we believe there is enough evidence to support the classification of SNR 2 as a SNR. Even if this is not the case, we note that our main conclusions remain unaffected. These spectra also illustrate the well-known difficulty in distinguishing ULX bubbles from SNRs \citep[e.g.][]{pakull_ultraluminous_2008}, and provide further evidence for region 4  being excited by shocks.

% python ~/scripts/pythonscripts/muse/plot_spectra.py corrnancleanngc1313x1_cubesubcubeRegion_4_sourcespec.fits ../snr1_mean/corrnancleanngc1313x1_cubesubcubeSNR_1_sourcespec.fits ../snr2_mean/corrnancleanngc1313x1_cubesubcubeSNR_2_sourcespec.fits --lines -z 0.001568 --lmin 6800 --lmax 7500
%python ~/scripts/pythonscripts/muse/plot_spectra.py corrnancleanngc1313x1_cubesubcubeRegion_4_sourcespec.fits ../snr1_mean/corrnancleanngc1313x1_cubesubcubeSNR_1_sourcespec.fits ../snr2_mean/corrnancleanngc1313x1_cubesubcubeSNR_2_sourcespec.fits --lines -z 0.001568 --lmin 4850 --lmax 5300
For completeness, we also decided to inspect the \chandra\ images for any evidence of associated X-ray emission which could strengthen their classification as SNRs. To this end, we used archival ObsID 2950 as it had the longest exposure time and the smallest PSF distortion around the ULX, owing to its off-axis angle of 2.317', compared to 6.884' for the other two  observations available in the archives (ObsIDs 3550 and 3551). The details of this observation is also reported in Table~\ref{tab:data}.

To precisely align the optical and X-ray images to a common reference frame, we matched the sources detected in both wavelengths to the \textit{Gaia} DR2 source catalogue \citep{gaia_collaboration_gaia_2018}. For the three \hst\ images (F555W, F547M, and F657N filters) we used the tool \texttt{tweakreg} from the \texttt{astrodrizzle} package\footnote{\url{https://drizzlepac.readthedocs.io/en/latest/index.html}}, which allows     both source detection and cross-matching. We tested different sets of parameters and decided to use a detection threshold of 75$\sigma$ and  to restrict the number of sources used in the matching stage to the 846 brightest sources (set to twice the amount of \textit{Gaia} sources in the F555W field of view). This was in order to restrict the matching stage to the brighter sources more readily detected by \textit{Gaia}. After cross-matching each of the images, we allowed the task to subsequently incorporate the new sources (not included in the original reference \textit{Gaia} catalogue) detected in the previous images to build a larger reference catalogue for the next image. Owing to its larger field of view the F555W image was used first, followed by the F547M image, and lastly the F657N due its narrower filter. We found 85, 33, and 307 matches for the F555W, F547M, and F657N filters respectively, resulting in rms residuals all below 0.007".

%python ~/scripts/pythonscripts/hst/astro_corr.py -m mask.txt --gaia -r 2.3 --update
For the \chandra\ images we used the \texttt{ciao} task \texttt{wavdetect} on the 0.5--7 keV band for source detection. A faint ($\sim$ 13 net counts) source north-east of the ULX is detected\footnote{This source is CXOU J031821.2--662858 in the work of \cite{mineo_x-ray_2012}, identified as a high-mass X-ray binary.} (Figure~\ref{fig:F567N_chandra}), with a 1$\sigma$ statistical uncertainty on the position of $\delta$ R.A. = 0.36" and $\delta$ DEC = 0.19". To examine whether this source could be the X-ray counterpart of SNR 1, we used the task \texttt{wcsmatch} with an initial search radius of 4" to match the X-ray sources to the \textit{Gaia} catalogue. After an initial transformation, an iterative process discards all matches with  radii above a certain value and continues to update the solution until no more matches are discarded. We visually inspected the resulting matches obtained for different radii and settled for a value of 0.6". This resulted in five matches: two background AGN, two foreground stars (one used in the work of \citealt{yang_optical_2011} for this same correction), and the supernova remnant SN1978K \citep[e.g.][]{kuncarayakti_evolving_2016}. The rms residuals were 0.16". Adding all the uncertainties in quadrature, the final 3$\sigma$ uncertainty on the position of the source is $\delta$ R.A. = 1.2" and $\delta$ DEC = 0.8". Based on the 3$\sigma$ confidence contour depicted in Figure~\ref{fig:F567N_chandra}, we rule out the association of the X-ray emission with SNR 1.

\subsection{Baldwin–Phillips–Terlevich diagrams} \label{sub:bpt_diagrams}
\begin{figure*}
    \centering
    \includegraphics[width=0.99\textwidth]{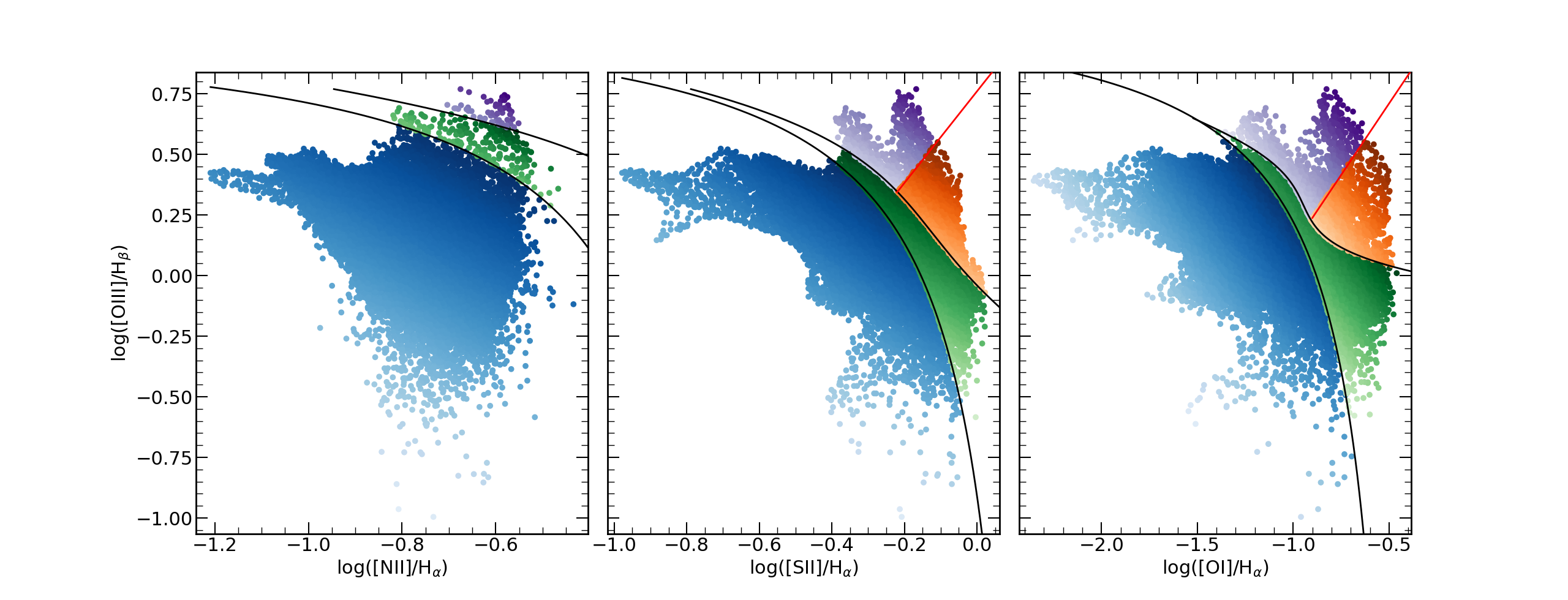}
    \includegraphics[width=0.98\textwidth]{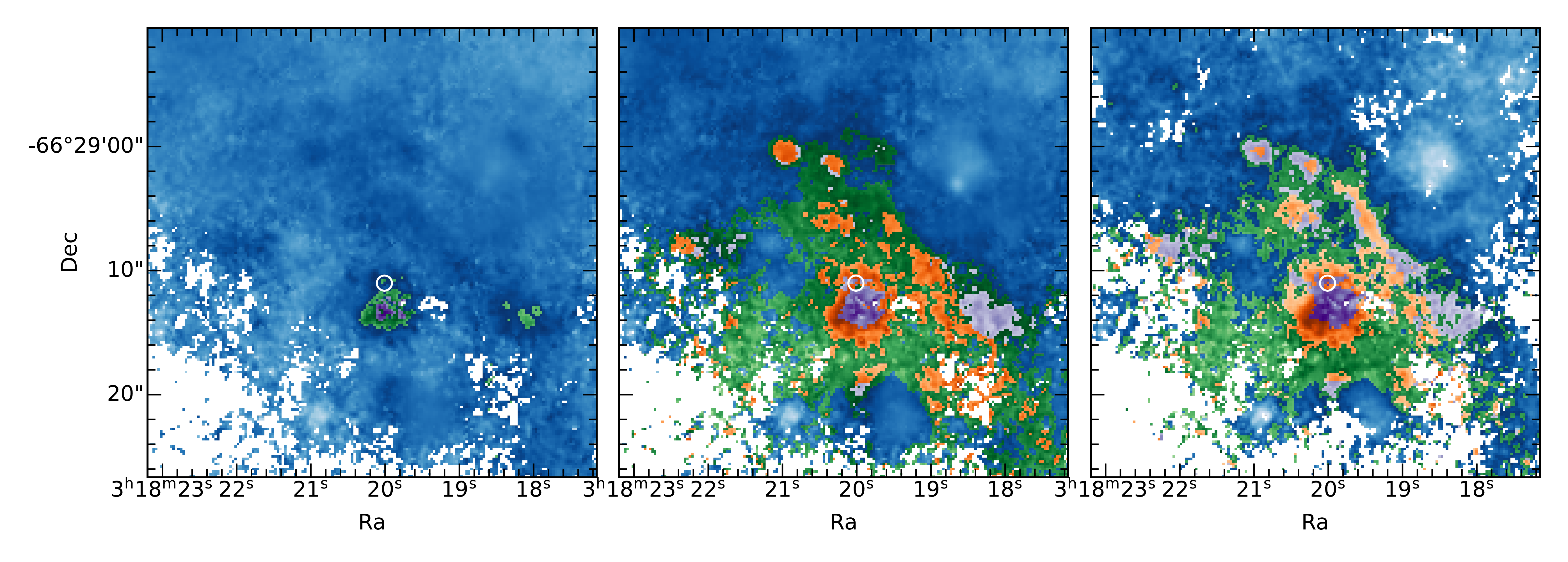}
    \caption{Results of the spatially-resolved BPT diagrams using the classification proposed by \cite{law_sdss-iv_2021}. Regions classified as HII, intermediate, AGN, and LINER are shown in blue, green, purple, and orange, respectively. The top panels show the classification of each pixel based on the line ratios, whereas the bottom panels show their corresponding location. Each pixel has been colour-coded proportionally to the sum of the x and y values in all panels to aid the visualisation and localisation of each pixel in the bottom panels.}
    \label{fig:bpt_diagrams}
\end{figure*}

In order to study the physical conditions of the gas around \ulx, we exploited MUSE access to the main nebular lines to derive spatially resolved [O~III]$\lambda$5007/\hb\ versus [N~II]/\ha, [O~III]$\lambda$5007/\hb\ versus [S~II]/\ha,\ and [O~III]$\lambda$5007/\hb\ versus [O~I]/\ha\ Baldwin–Phillips–Terlevich (BPT) diagrams \citep{baldwin_classification_1981, veilleux_spectral_1987}. These diagrams use the flux ratios of the Balmer lines to the forbidden lines to distinguish regions of gas ionised by stars (HII regions) from regions ionised by hard X-ray emission and/or shocks (AGN regions and LINER). To classify the regions based on these line ratios, a theoretical and empirical classification scheme is commonly used based on the works of \cite{kewley_theoretical_2001}, \cite{kauffmann_host_2003}, and \cite{kewley_host_2006}. For instance, \cite{kewley_theoretical_2001} derived the theoretical maximum ratios expected from regions photoionised solely by stars, using photoionisation and stellar synthesis population codes. This line is termed the maximum starburst line, and it is used to differentiate between HII regions from intermediate regions (where hard X-ray radiation and/or shocks start to contribute) and/or AGN-dominated regions (mainly ionised by shocks and/or hard X-ray radiation). However, these early works were based purely on line ratios of integrated spectra of galaxies, resulting in mixing of various components. By using IFU data from the MaNGA galaxy survey, \cite{law_sdss-iv_2021} recently refined these classification schemes by combining spatially resolved line ratios and the kinematic information provided by the dispersion along the line of sight of the \ha\ line. This classification results in a cleaner separation between the different components compared to early works and we therefore made use of it to derive the BPT diagrams shown in Figure~\ref{fig:bpt_diagrams}. We also note  that these diagrams are essentially independent of the exact reddening correction since the line ratios employed are sufficiently close in wavelength such that the correction can be safely ignored.

The diagrams clearly highlight that most of the pixels around the ULX are classified as AGN, LINER or intermediate, indicating that the ULX  contributes heavily to the ionisation budget of the surrounding environment. The [N II]/\ha\ BPT diagram is more sensitive to metallicity. For this reason, at low oxygen abundances (log(O/H) + 12 $\lesssim$ 8.4), AGN regions become difficult to distinguish from classical HII regions, mostly owing to the dependency of the [N~II]$\lambda$6583/\ha\ ratio with metallicity \citep{kewley_theoretical_2013}. We show in Section~\ref{sub:metalliticy} that the metallicity around \ulx\ is $\log$(O/H) + 12 $\approx$ 8.1--8.4. This likely explains why many of the regions classified as AGN--LINER in the other two BPT diagrams are found in the HII--intermediate region in the [N II]$\lambda$6583/\ha\ diagram. Nevertheless, this diagram is useful because the pixels classified as AGN likely trace the hard EUV/X-ray radiation from the source, as the hardness of the ionisation field tends to increase the [O~III]$\lambda$5007/\hb\ ratio \citep{kewley_theoretical_2013}, given the relatively high ionisation potential needed to doubly ionise oxygen. Thus,  the [N II]$\lambda$6583/\ha\ BPT diagram essentially offers a map of the regions strongly excited by EUV/X-rays from the ULX. 

The other two diagrams show a central roughly circular region slightly offset to the south from the position of \theulx, where the gas is classified as AGN and LINER. This classification is also observed in regions SNR 1 and SNR 2. Comparison of these diagrams with the kinematic data and the [O I]$\lambda$6300 and [S~II]$\lambda$6716 flux maps (see Section~\ref{sec:camel}) suggest that the lightly coloured regions found at the boundary of the AGN  and LINER regions mostly highlight regions excited by shocks. To further illustrate this point, we show the [O I]$\lambda$6300/\ha\ BPT diagram coloured based on the FWHM values from Figure~\ref{fig:disp_maps} in Figure~\ref{fig:bpt_fwhm}, whereas in Figure~\ref{fig:bpt_fwhm_regions} in the Appendix the same result is shown, but for each of the regions highlighted in the right panel of Figure~\ref{fig:disp_maps}. We can clearly see that pixels with largest FWHM values are concentrated close to the intermediate boundary,    between the AGN and LINER regions. These correspond mainly to regions SNR 1 and SNR 2 and regions close to 3 and 4 where   strong broadening is seen, which all have values $-1 <$ $\log$([O I]$\lambda$6300/H$\alpha$) $<-0.75$.

\begin{figure}
    \centering
    \includegraphics[width=0.49\textwidth]{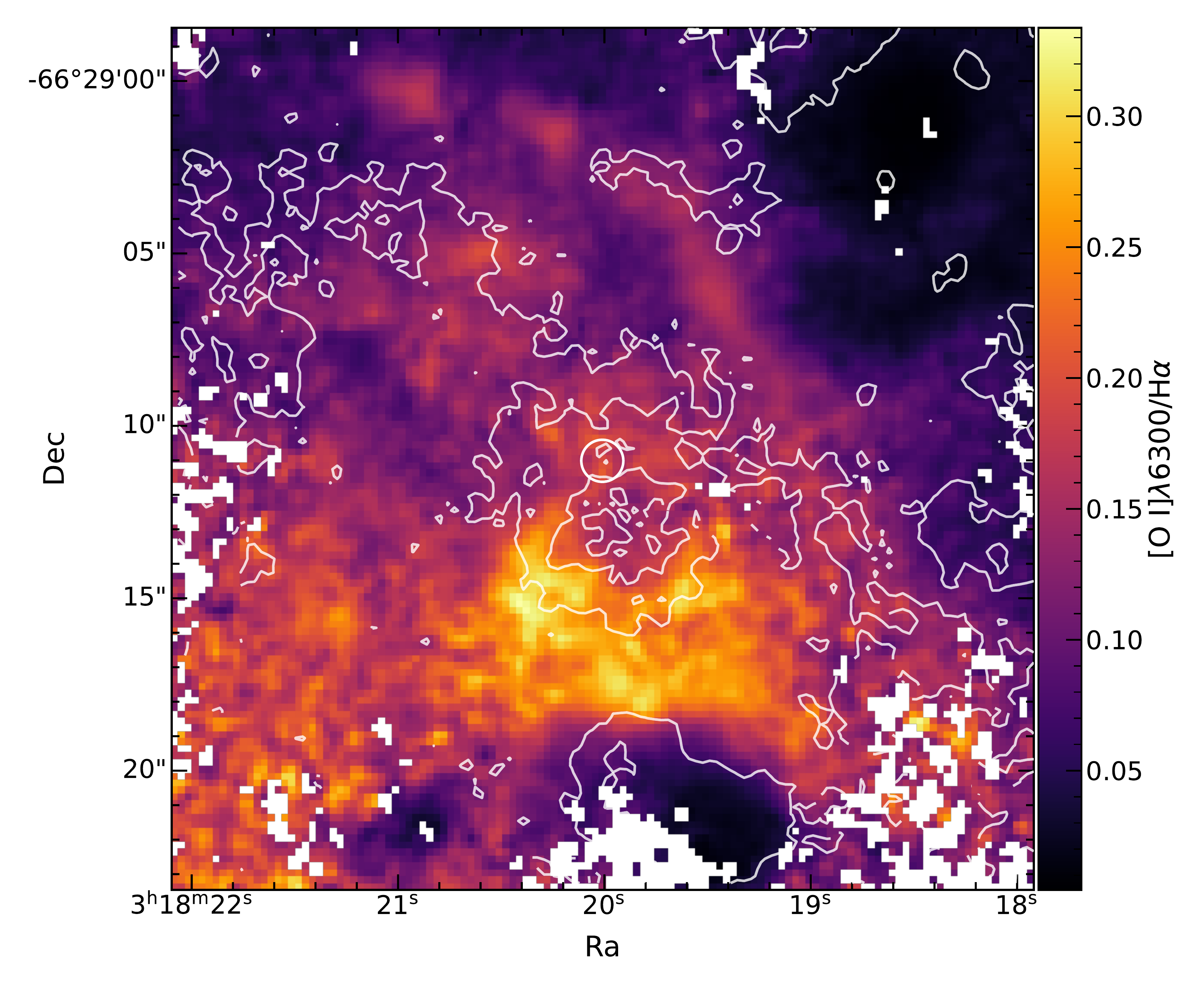}
    \caption{[O I]$\lambda$6300/\ha\ map with the [O III]$\lambda$5007/\hb\ contours overlaid in white showing a region of 25"$\times$25" around the ULX (white circle). Extended emission with high [O I]$\lambda$6300/\ha\ ratios is seen south to the peak of the [O III]$\lambda$5007/\hb\ indicating X-ray excitation.}
    \label{fig:xin}
\end{figure}

Regions 5 and 6 include some of the strong broadening seen in these areas, but the ratios where FWHM is strongest are instead pushed to higher [O I]$\lambda$6300/\ha\ ratios ([O I]$\lambda$6300/H$\alpha$ > 0.25 or $\log$([O I]$\lambda$6300/H$\alpha$) $>$ --0.6) compared to the other regions (Figure~\ref{fig:bpt_fwhm_regions}), most notably for region 6. These regions also show  a clear dependence with distance to the ULX, with the pixels closer to the ULX showing high $\log$([O III]$\lambda$5007/H$\beta$) $\gtrsim$ 0.25 as a result of the higher excitation by the ULX, whereas pixels further from the ULX are again found at the boundary between AGN, LINER and intermediate regions. We note in particular how the [O~III]$\lambda$5007/H$\beta$ ratio is much lower in region 6, whereas the [O I]$\lambda$6300/\ha\ value has increased. These pixels with [O I]$\lambda$6300/\ha\ are located to the south and east of the region classified as LINER in the [S II]/\ha\ or [O I]$\lambda$6300 diagrams. To illustrate this in more detail, we show in Figure~\ref{fig:xin} the [O III]$\lambda$5007/\hb\ contours over the [O I]$\lambda$6300/\ha\ image. This shows the presence of a highly ionised zone (in this case traced by the high [O III]$\lambda$5007/\hb\ ratio) surrounded by regions of bright forbidden transitions of neutral atoms (here traced by the [O I]$\lambda$6300/\ha\ $\gtrsim$ 0.25 ratio). This is consistent with the expectation of an X-ray ionised nebula \citep{halpern_x-ray_1980}, in very good agreement with the weakly ionised region reported by \cite{pakull_optical_2002}. Therefore in regions 5 and 6 we are likely seeing a mixture of shock and X-ray ionised gas.
In region 7 we note that  some of the pixels include the contribution from the stellar cluster located east of it, which explains the array of values found in the HII region, but again we observe some the pixels located at the AGN--LINER--intermediate boundary. In summary, we see that the environment of the ULX shows indications of being ionised by both EUV/X-ray radiation and shocks. While we do observe some pixels with FWHM $\sim$ 80--120\,km/s within the HII region in Figure\,\ref{fig:bpt_fwhm} (see the yellow coloured data points at --1.5 $<$ $\log($[O~I]$\lambda6300$/H$\alpha$) $<$ --1.25) associated with the northernmost region in Figure\,\ref{fig:bpt_fwhm}, we can rule out these as being due to shocks because of  their position in the BPT diagrams. 

\begin{figure}
    \centering
    \includegraphics[width=0.49\textwidth]{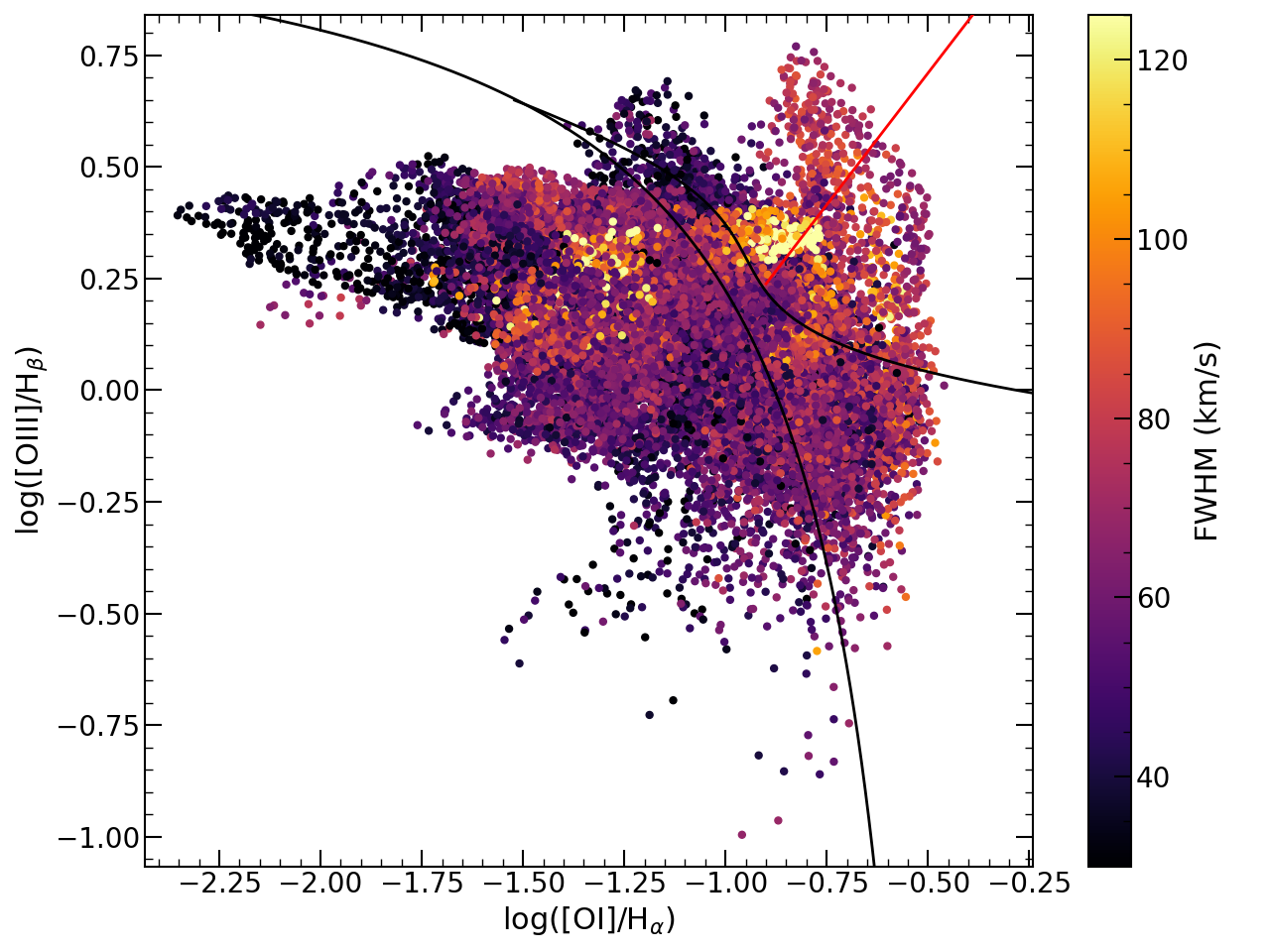}
    \caption{Spatially resolved [O~III]$\lambda$5007/\hb\ vs [O~I]$\lambda$6300/\ha\ BPT diagram coloured as a function of the intrinsic FWHM of the \ha\ line. The solid lines show the classification scheme proposed by \cite{law_sdss-iv_2021}, as in Figure \ref{fig:bpt_diagrams}.}
    \label{fig:bpt_fwhm}
\end{figure}

\subsection{Comparison with the \mappings\ libraries} \label{sec:mappings}
BPT diagrams are a powerful tool used to separate regions excited by stars from those excited by EUV/X-ray radiation and/or shocks (LINER and AGN regions). However, these diagrams cannot cleanly separate regions excited by shocks from EUV/X-ray radiation, although in Sections \ref{sec:camel} and \ref{sub:bpt_diagrams} we saw that the kinematic data can help in this discrimination. For this reason, it is often valuable to complement these diagrams with the predictions from radiative shock libraries. Here we chose to do so with the up-to-date set of pre-run shock-ionisation \textsc{MAPPINGS V} models \citep{2018ascl.soft07005S}  compiled by \cite{alarie_extensive_2019}, which is an extension of the \textsc{MAPPINGS III} models \citep{allen_mappings_2008}  to a wider range of abundances from the work of \cite{gutkin_modelling_2016}. With this analysis we can also attempt to provide an independent estimate of the shock velocities in the bubble (\vs), which will be relevant to evaluate the mechanical power needed to inflate the bubble (Section~\ref{sub:mec_power}) as well as the explosion energy of the SNRs (Section~\ref{sub:mec_snrs}).

\begin{figure*}
    \centering
    \includegraphics[width=0.99\textwidth]{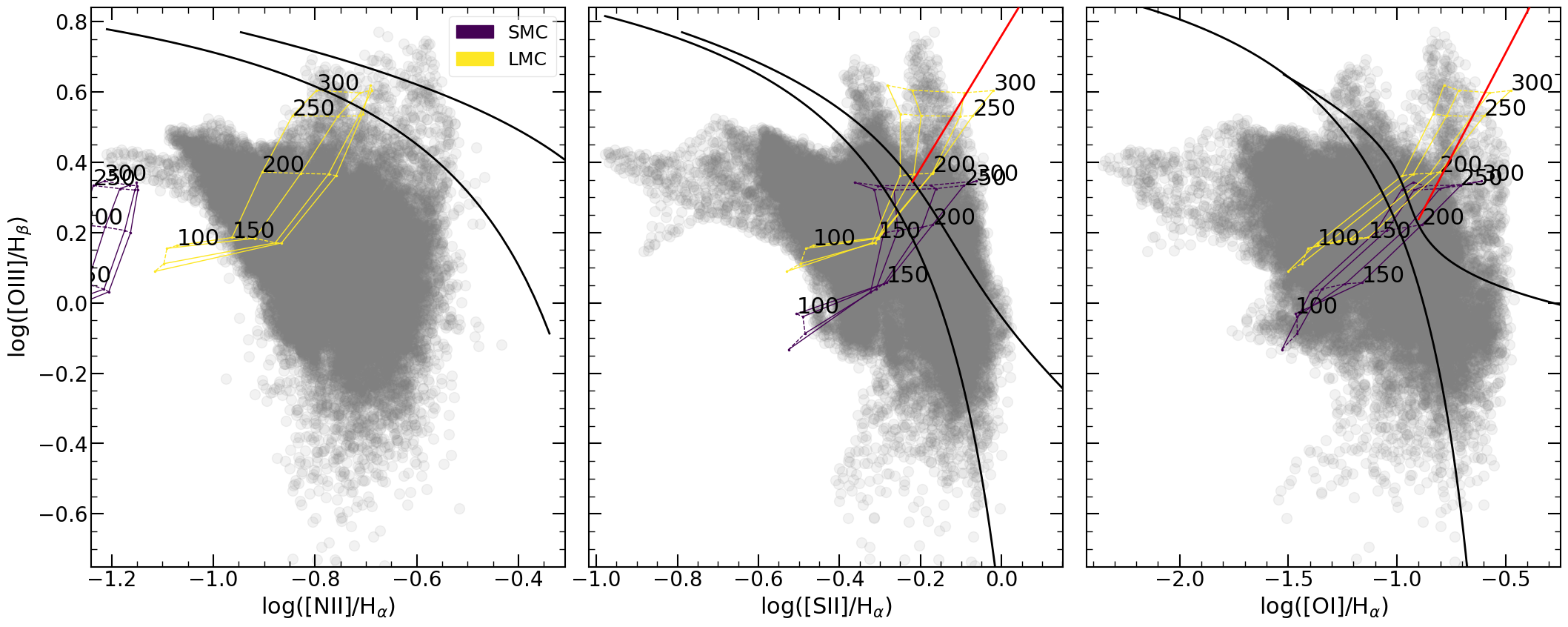}
    \includegraphics[width=0.99\textwidth]{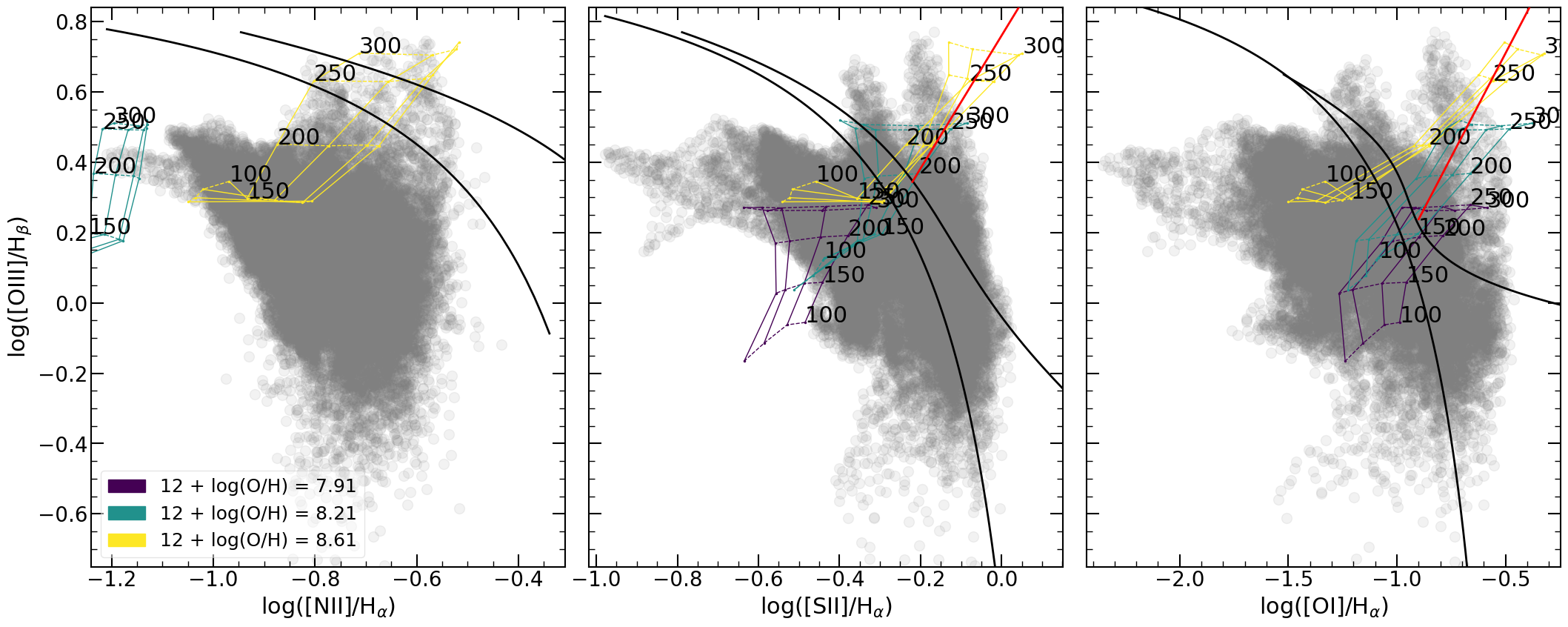}
    \caption{Comparison between the theoretical ratios predicted by the radiative shock libraries compiled by \cite{alarie_extensive_2019} for the precursor and shock models from abundances from \cite{allen_mappings_2008} (top) and abundances from \cite{gutkin_modelling_2016} (bottom) and the observed values. Each colour shows the predictions for a set of abundances (see text for details). Each line connects the predictions for shock velocities ranging from 100 to 300 km/s in steps of 50 km/s, as indicated.  The models for a given velocity are connected by a dashed line, indicating the effect of varying the pre-shock magnetic field for 10$^{-4}$, 1, 3.23, and 5 $\mu$G cm$^{3/2}$. The grey data points and the black solid lines correspond to the data and region delimitations from Figure~\ref{fig:bpt_diagrams}.}
    \label{fig:mappings}
    % python ~/scripts/pythonscripts/maxime_project/muse_project/read_mappingsV.py -a 0d002 0d004 0d010 -d 1 -m Gutkin16 -t "s+p"
    % python ~/scripts/pythonscripts/maxime_project/muse_project/read_mappingsV.py -a SMC LMC -d 1 -m Allen2008 -t "s+p"
\end{figure*}

We show in Section~\ref{sub:metalliticy} that the metallicity around NGC 1313 X--1 is \met\ $\approx$ 8.1--8.4. We therefore considered models for SMC (log(O/H) + 12 = 8.03$\pm$0.10) and LMC abundances (log(O/H) + 12 = 8.35$\pm$0.6) \citep{russell_abundances_1992} from the updated \cite{allen_mappings_2008} set, which have the closest set of abundances to our estimates. From the \cite{gutkin_modelling_2016} abundances, we considered four abundances to match roughly the observed range: $Z$ = 0.002, 0.004, and 0.01 which correspond approximately to \met\ = 7.91, 8.21, and 8.61, respectively\footnote{We have assumed a dust-to-metal mass ratio of 0.1 for the conversion \citep[see][]{gutkin_modelling_2016}.}. We also consider a representative range of pre-shock magnetic fields given in Figure~\ref{fig:mappings}. Using the \cite{gutkin_modelling_2016} models, for which the pre-shock density (\nism) can be varied logarithmically in the 1--10000 cm$^{-3}$ range, we found that the best match to the data was given for \nism\ = 1 cm$^{-3}$. Models with \nism\ = 10 cm$^{-3}$ were also acceptable, but mostly had the effect of making the line ratios less sensitive to variations in the pre-shock magnetic value than \vs. Therefore, the range of velocities found were roughly the same as those recovered for \nism\ = 1 cm$^{-3}$. We also note that our estimates from Section~\ref{sub:density_est} suggest \nism\ $\lesssim$ 5 cm$^{-3}$ even for velocities as low as $\sim$100\,km/s, and thus we considered only models with \nism\ = 1 cm$^{-3}$. The predicted line ratios from the libraries are compared with our line ratio estimates in Figure~\ref{fig:mappings}.

From the [S II]/\ha\ and [O I]$\lambda$6300/\ha\ diagrams, it is clear that velocities in excess of \vs\ $\gtrsim$ 150 km/s are needed to produce the observed line ratios close to the AGN, LINER, and intermediate boundaries, especially for the range of metallicities that  more closely match our estimates (Section~\ref{sub:metalliticy}). The LMC or \met\ = 8.24 set of abundances with velocities \vs\ $\sim$ 200 km/s or the SMC abundances with slightly higher velocities \vs\ $\sim$ 250 km/s, seem to provide a good match to our data (cf. Figure~\ref{fig:bpt_fwhm_regions}). It is also clear that values with log([O III]$\lambda$5007/\hb) $\gtrsim$ 0.45 cannot be explained by shocks alone as they either require shock velocities $\gtrsim$ 300 km/s, which can be ruled out by our kinematic data, or a set of abundances with abnormally high metallicity (cf. Figure~\ref{sub:metalliticy}), also ruled out by our estimates. Therefore, pixels at the uppermost right region of the BPT diagrams seem consistent with mainly EUV/X-ray excitation. A similar argument can be made about the pixels with the highest [O I]$\lambda$6300/\ha\ ratios seen in  regions 5 and 6. This is in line with the analysis presented in Sections~\ref{sec:camel} and \ref{sub:bpt_diagrams}, where it was shown that shocked regions are instead close to the AGN--LINER--intermediate boundary delimitation. Considering the [N II]/\ha\ diagram, it seems that the best match to the data would be given by the set of LMC abundances, in good agreement with our metallicity estimates (Section~\ref{sub:metalliticy}). The set of points with FWHM $\gtrsim$ 120 km/s seen in Figure~\ref{fig:bpt_fwhm} closer to the AGN, LINER and intermediate boundaries is clustered around --0.8 $<$ log([N~II]/\ha) $<$ --0.6 and 0.25 $<$ log([O III]$\lambda$5007/\hb) $<$ 0.5 in this diagram, which again would suggest 175 km/s $<$ \vs\ $<$ 200 km/s. In summary, the line ratios with values at rightmost part of the [O I]$\lambda$6300/\ha\ and [S II]/\ha\ diagrams or with very high ($\gtrsim$ 0.45) values of log([O III]$\lambda$5007/\hb) ratio cannot be explained by shocks, suggesting that these are regions excited by EUV/X-ray radiation. Instead, regions closer to the AGN--LINER--intermediate boundary line are closely matched by shocks with velocities in the $\sim$ 165 $<$ \vs\ $<$ 225 km/s range.  

It is often worth comparing these estimates with the kinematical data. In this regard, \cite{soria_ultraluminous_2021} provides an interesting discussion on how to relate the measured line FWHMs to the shock velocity in their Appendix, to which we refer the interested reader for more details. Briefly, if we are observing the projected central region of a thin spherically expanding bubble, we expect to observe a velocity distribution function $f_\text{c}$ = $\frac{1}{2}$[$\delta(v -v_\text{exp}) + \delta(v + v_\text{exp}$)], where $v_\text{exp}$ is the expanding velocity of the bubble. For radiative shocks, $v_\text{exp} = v_s$ \citep{dewey_kinematics_2010}, leading to \vs\ = 0.425\,FWHM. Considering that in reality we observe a finite region with a range of velocities instead of a delta function and that some gas lags behind the shock, \cite{soria_ultraluminous_2021} suggest that \vs\ = 0.47\,FWHM is a more appropriate relationship. Another possibility is to take a spectrum of the entire bubble instead of observing a finite region. In this case, assuming there is a single expansion velocity for the whole bubble, the observed velocity distribution is a uniform distribution $f_\text{T}$ = $\frac{1}{2}~v_\text{exp}$, which leads to \vs\ = 0.735\,FWHM. Given the morphology of the nebula, we considered only a circular region at the centre of the cyan ellipse in the right panel of Figure~\ref{fig:disp_maps} with radius 7.5" ($\sim$ 150 pc), roughly delimited by regions 4, 5, and 6. Here we measured a maximum FWHM of 127 km/s and a mean of 70 km/s from the mean \ha, [N~II], and [S~II] map. Assuming the case of a spherically thin expanding nebula, this would suggest \vs\ $\sim$ 55 km/s using either of the two estimates. However, such slow \vs\ values are at odds with the measured line ratios, as we  show in Section~\ref{sec:mappings}. We also note that a radial expansion does not appear to be supported by the \vlos\ maps, which shows instead some sort of outflowing gas emanating from the ULX (see the trough in the velocity field at the ULX position in Figure~\ref{fig:vel_profile}).

It is also possible in our case, given the high shock velocity inferred from the line ratios, that the FWHM we are measuring is a blend of the broadened line from the shock and the narrow component from the photoionised precursor, only broadened by thermal Doppler effect. We note that the presence of the shock precursor is strongly supported by the fact that this component is needed to match the line ratios in the LINER and AGN regions. Given the moderate resolution of MUSE (instrumental FWHM of $\sim$ 116\,km/s at 6560\,\AA), it is likely that we cannot resolve these two components. If the narrow component from the precursor dominates the emission, this can result in a narrower Gaussian fit to the blended line profile than if the shocked component could be resolved and modelled alone. We can see some evidence in support of this phenomenon in SNR 1 (Figure~\ref{fig:disp_maps}), where the FWHM measured with a single-Gaussian component was the strongest and where we found the strongest fit residuals, which could indicate the presence of a more complex line profile. We focused on the \ha\ line as it is the one with the strongest S/N, and extracted a spectrum from the region shown in Figure~\ref{fig:disp_maps}, determined by fitting a 2D Gaussian, profile as explained in Section~\ref{sec:camel}.
%on a region derived by fitting the \ha\ flux image around this region with a 2D circular profile. The resulting region is shown in Figure~\ref{fig:disp_maps} with a radius equal to the derived FWHM of 1.51$\pm$0.06" (1\,$\sigma$ uncertainty).

\begin{figure}
    \centering
    \includegraphics[width=0.49\textwidth]{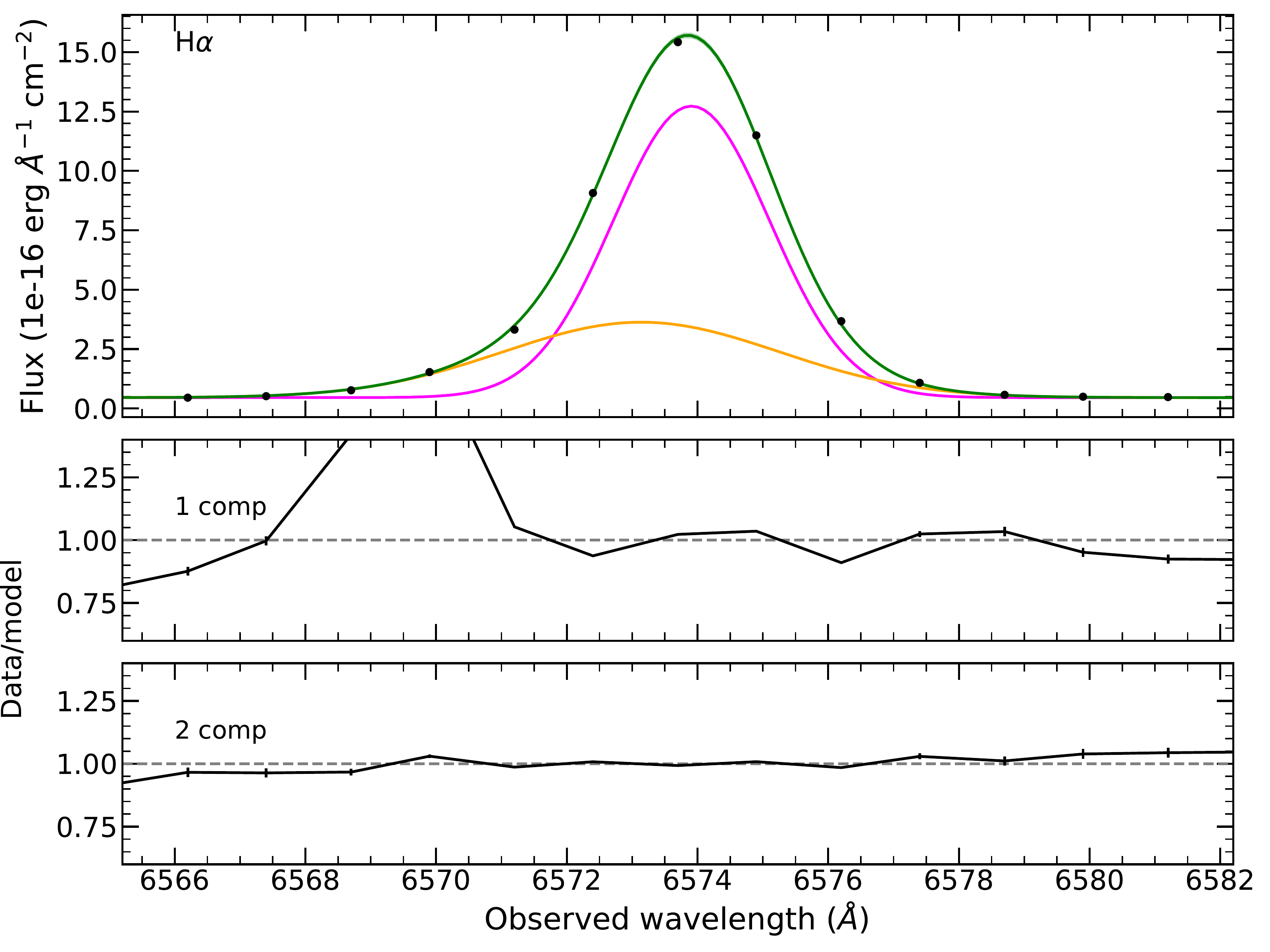}
    \caption{Results of the fitting process to the integrated \ha\ line of the SNR 1 region (Figure~\ref{fig:disp_maps}). (top) Best fit (green solid line) to the integrated \ha\ emission from region SNR 1 using two Gaussian components (shown in pink and yellow). The lower panels show the ratio plots of one- and two-component fits.}
    \label{fig:two_components}
\end{figure}

We performed Gaussian fits of the \ha\ line using the \texttt{python} package \texttt{lmfit}\footnote{\url{https://lmfit.github.io/lmfit-py/}} considering one or two Gaussian components. The continuum was modelled using a constant around the line centroid. The best-fit parameters for one and two components are shown in Table~\ref{tab:two_components}. Figure~\ref{fig:two_components} shows the best-fit two-Gaussian model component, with the corresponding residuals for one and two Gaussian components shown in the panels below. We also tried more complex profiles such as a Lorentzian or a Voigt profile, but they yielded worse fits than the single-Gaussian model. 

\begin{table*}
    \centering
    \caption{Best-fit parameters to the integrated \ha\ line of region SNR 1 (Figure~\ref{fig:disp_maps} right panel) for one and two Gaussian components (uncertainties at the 1\,$\sigma$ level).} \label{tab:two_components}
    \begin{tabular}{ccccc}
    \hline
    & Wavelength & $V_\text{los}$\tablefootmark{a}& FWHM & Flux \\
    & \AA & km/s & km/s & 10$^{-15}$\,erg cm$^{-2}$ s$^{-1}$ \\
    \hline\hline
            \noalign{\smallskip}
      1 Component   & 6573.78$\pm$0.07 & 31$\pm$3 & 98$\pm$6 & 5.2$\pm$0.1 \\
              \noalign{\smallskip}
      \hline
      \noalign{\smallskip}
    \multirow{2}{*}{2 Components} & 6573.92$\pm$0.03 & 38$\pm$1 & 56$\pm$5& 3.7$\pm$0.2  \\
    & 6573.15$\pm$0.09 & 3$\pm$4 & 196$\pm$8&1.7$\pm$0.2 \\
    \hline
    % results stored in bright_jet_sum_correct! bright_jet_sum has the old wrong extraction region
    \end{tabular}
    \tablefoot{\tablefoottext{a}{Line-of-sight velocity over the galaxy redshift ($z=0.001568$).}}
\end{table*}
The single-Gaussian component deviates strongly from the line profile at the blue wing of the line profile, suggesting the presence of a second broad component. In agreement with the above argumentation, the second broad component alone is broader (FWHM = 196$\pm$8\,km/s) than the single-Gaussian model component (FWHM = 98$\pm$6\,km/s). The former FWHM value is in turn much closer to the shock velocities inferred from the \mappings\ libraries. We conclude that it is possible that the narrow component of the precursor causes the FWHM of the lines to be underestimated. It must also be noted that a one-to-one correspondence between \vs\ and FWHM is not expected given that the FWHM might not necessarily sample the direction of maximum expansion. Considering the uncertainties associated with relating FWHM to the shock velocity, we decided to rely on the shock velocities inferred from \mappings\ rather than from the kinematic data.

Based on the different positions of the selected regions in the BPT diagrams (Figure~\ref{fig:bpt_fwhm_regions}) we assigned them a shock velocity depending on the typical values found for the SMC, LMC and \met\ = 7.91, 8.21 abundances. These are shown in Table~\ref{tab:pre_shock}. We also report in Table~\ref{tab:pre_shock} the maximum FWHM found in the selected regions. There is clear trend in the sense that regions with higher FWHM show higher shock velocity values as derived using the line ratios. This suggests that the estimated shock velocities based on the \mappings\ libraries are reasonable. The shock velocities found are typically a factor of $\sim$1.4 higher than the measured FWHMs. Similar factors were also found by \cite{2012MNRAS.427..956D} in the study of the microquasar NGC~7793~S26. Given that region 6 does not show much overlap with the line ratios predicted by the \mappings\ libraries, notably where FWHM is strongest (Figure~\ref{fig:bpt_fwhm_regions}) owing to the likely contribution from X-ray ionisation (Section~\ref{sub:bpt_diagrams}), we simply assigned it \vs = 1.4\,FWHM. We also note that the differences between the FWHM values and the shock velocities inferred from the line ratios we found are much smaller than in previous works such as \cite{urquhart_multiband_2018} or \cite{soria_ultraluminous_2021}, who found \vs\ based on the line ratios to be $\gtrsim$ 5\,FWHM. We note however that as stated earlier these values are less certain for regions 5 and 6 given the likely contribution from X-ray radiation.

\subsection{ISM Density estimates} \label{sub:density_est}

A key physical quantity necessary to estimate the energetics of the bubble (see Section \ref{sec:bubble_discussion}) and that of the SNRs (Section~\ref{sub:mec_snrs}) is the pre-shock density of the ISM (\nism). This quantity can be estimated using Equation 3.4. from \cite{dopita_spectral_1996}, who provide a relationship between the surface brightness of the \hb\ line in the shock, \vs,\ and \nism:
\begin{equation}\label{eq:hb_shock}
f_{H\beta\text{shock}} = 7.44 \times 10^{-6} ~v_{100}^{2.41} \times \frac{n_\text{ISM}}{cm^{-3}} ~\text{erg}~\text{s}^{-1} \text{cm}^{-2}
.\end{equation}
Here $v_\text{100}$ is the shock velocity in units of 100 km/s. In cases where \vs\ $>$ 150 km/s, the shock will create a photoionised precursor that needs to be taken into account \citep{dopita_spectral_1996}. The same relationship as above but for the precursor term reads (their Equation 4.4.)
\begin{equation}\label{eq:hb_precursor}
f_{H\beta\text{precursor}} = 9.85 \times 10^{-6} ~v_{100}^{2.28} \times \frac{n_\text{ISM}}{\text{cm}^{-3}} ~\text{erg}~\text{s}^{-1} \text{cm}^{-2}
;\end{equation}
combining the two equations, $n_\text{ISM}$ can be expressed as
\begin{equation} \label{eq:n_ism}
 \frac{n_\text{ISM}}{\text{cm}^{-3}} = \frac{10^{6} L_{\text{H}\beta}}{A (7.44 v_{100}^{2.41} + 9.85 v_{100}^{2.28}) }
,\end{equation}
where $A$ is the surface area of the emitting region. The surface area of each region was estimated assuming a prolate ellipsoid (i.e. we assumed the shorter of the semi-axes for the third dimension of the ellipsoid). We estimated the density of the ISM for the two SNRs and the bubble, and for each of the elliptical regions shown in the right panel of Figure~\ref{fig:disp_maps}, where we found clear signs of shock excitation, as argued in Sections~\ref{sec:camel} and \ref{sub:bpt_diagrams}. We caution again that these estimates might be less robust for regions 5 and 6 given that a fraction of the \hb\ luminosity might be produced by X-ray excitation. We return to this point in Section~\ref{sub:mec_power}.

%We took into account the precursor term for all regions, except for regions 4 and 7, which both have \vs\ $<$ 150 km/s (we elaborate on this below).
   
The integrated flux in the \hb\ line map in each of the mentioned regions was corrected for reddening using the observed Balmer decrement map. To this end, we followed \cite{momcheva_nebular_2013}, who provide a relationship between the observed Balmer decrement and the colour excess:
\begin{equation} \label{eq:extinction}
    E(B-V) = \frac{-2.5}{k(H\beta) - k(H\alpha)} \times \log_{10} \left (\frac{(H\alpha/H\beta)_{int}}{(H\alpha/H\beta)_{obs}} \right ) 
.\end{equation}
% python ~/scripts/pythonscripts/maxime_project/muse_project/read_mapping.py [PQ]*sp_*.txt --> 2.95--3.0
Here we used the extinction curve of \cite{calzetti_dust_2000} with $R_v$ = 4.05 to obtain $k(H\beta) = 4.6$ and $k(H\alpha) = 3.3$. For the intrinsic \ha/\hb\ ratio, the \mappings\ libraries suggested values of $\approx$ 2.95--3.0 for SMC and LMC abundances (see Sections~\ref{sec:mappings} and \ref{sub:metalliticy}) and for shock velocities \vs\ = 175--225 km/s. We therefore assumed an intrinsic ratio \ha/\hb~= 3.0 for the extinction correction.

%python ~/scripts/pythonscripts/maxime_project/muse_project/deredden_momcheva.py -a camel_1_n2ha/cleaned_images -b camel_1_hb/cleaned_images -i 3. ;
%python ~/scripts/pythonscripts/maxime_project/muse_project/image_stats.py ./camel_1_hb/cleaned_images/cleancamel_1_hb_ssmooth_flux_indep_HBETAderedden.fits ./camel_1_hb/cleaned_images/cleancamel_1_hb_ssmooth_eflux_indep_HBETAderedden.fits -r density_estimates/regions.reg
%lines=$(ls camel_*/cleaned_images/cleancamel*flux*deredden.fits)
%python ~/scripts/pythonscripts/maxime_project/muse_project/image_stats.py $lines -r density_estimates/regions.reg
% There was a bug in Camel so the correct errors for Ha are in acamel_n2ha_nohb (but have been copied to the right folder) the uncorrect file is in /home/agurpide/optical_data/NGC1313/nancleancubes/coordadjusted/camel_1_n2ha/cleaned_images/wrong_cleancamel_1_n2ha_ssmooth_eflux_indep_HALPHA.fits
\begin{table*}
    \centering
\caption{Measured and estimated quantities of interest for the regions shown in the right panel of Figure~\ref{fig:disp_maps}. All uncertainties are at the 1$\sigma$ level.}
\label{tab:pre_shock}
    \resizebox{\textwidth}{!} {
    \begin{tabular}{ccccccccccccc}
    \hline
        Region & $a$\tablefootmark{a} & $b$\tablefootmark{b} & Area\tablefootmark{c} & $R$\tablefootmark{d} & $\Sigma L_{\text{H}\beta}$\tablefootmark{e} & $<$ E(B-V)$>$\tablefootmark{f} & FWHM\tablefootmark{g} & \vs\tablefootmark{h} & $n_\text{ISM}$ & $t$\tablefootmark{i} & $P_\text{mec}$\tablefootmark{j} & $E$\tablefootmark{k} \\ 
        & arcsec & arcsec & 10$^{40}$ cm$^{2}$ & pc & 10$^{36}$ erg/s & & km/s & km/s & cm$^{-3}$ & 10$^5$ yr & 10 $^{39}$ erg/s & 10$^{51}$ erg \\\hline\hline
        \noalign{\smallskip}
        SNR 1 & 1.51 & 1.51 & 11.6 & 257$\pm$6 & 6.37$\pm$0.02 & 0.16$\pm$0.06 & 166$\pm$27 & 220$\pm$25  & 0.5$\pm$0.1 & 0.24--0.27 & -- & 0.3--0.8 \\
        SNR 2 & 1.72 & 1.06 & 8.1 & 211$\pm$5 & 4.64$\pm$0.02 &  0.20$\pm$0.08 & 136$\pm$20 & 200$\pm$25 &0.7$\pm$0.2 & 0.27--0.34 & -- &0.3--0.8\\ 
        \hline
        \noalign{\smallskip}
        3 & 3.12 & 1.81 & 25.1 & 164$\pm$9 & 8.35$\pm$0.02 &  0.12$\pm$0.08 & 119$\pm$13 &165$\pm$25 &0.6$\pm$0.2 & -- & -- & --\\
        4 & 4.68 & 1.26 & 24.2 & 116$\pm$ 1& 15.25$\pm$0.03 & 0.18$\pm$0.08 & 110$\pm$12  & 170$\pm$25 & 1.1$\pm$0.4 & -- & -- &  --\\ 
        5 & 3.30 & 2.33 & 35.5 & 69$\pm$3 & 10.66$\pm$0.03 & 0.16$\pm$0.07 & 110$\pm$13 & 170$\pm$25& 0.5$\pm$0.2  & -- & -- &  --\\
        6 & 3.80 & 1.70 & 27.6 & 137$\pm$5 & 11.16$\pm$0.04  & 0.2$\pm$0.1 & 130$\pm$15 & 180$\pm$25 &  0.6$\pm$0.2 & -- & -- & --\\
        7 & 2.03 & 1.96 & 19.9 & 199$\pm$8 & 7.90$\pm$0.04  & 0.2$\pm$0.1 &  100$\pm$17 & 160$\pm$25 & 0.8$\pm$0.3 & -- & -- & --\\
        \hline
        \noalign{\smallskip}
        Bubble & 11 & 6.5 & 318.6 & 134--227 & 105.6$\pm$0.12 & 0.16$\pm$0.1 & 130 & 170$\pm$25 & 0.6$\pm$0.2 & 4.6$\pm$0.7--7.8$\pm$1.2 &  19$\pm$10--55$\pm$30 & $\sim$188--990 \\
        \hline\hline
    \end{tabular}
    }
    \tablefoot{\tablefoottext{a}{Semi-major axis of the elliptical region.}\tablefoottext{b}{Semi-minor axis.}\tablefoottext{c}{Surface area of the ellipsoidal region, for which we  assumed a radius equal to $b$ for the third dimension (i.e. a prolate spheroid).} \tablefoottext{d}{Average distance from the centre of the region to the average between the centre of the bubble and the position of the ULX. For the bubble we quote instead the semi-major and semi-minor axes.} \tablefoottext{e}{Integrated extinction-corrected luminosity in the region. The error is propagated from the error flux map and takes into account the uncertainty on E(B-V). We   assumed a distance to NGC 1313 X--1 of 4.25\,Mpc in the calculations of the luminosities.}\tablefoottext{f}{Mean and standard deviation of E(B-V) in each region to provide some reference values, although the extinction correction was  applied on a pixel-by-pixel basis.}\tablefoottext{g}{Maximum FWHM measured in each region. The error is the  typical spread in the region.}\tablefoottext{h}{Shock velocity estimated from the observed line ratios in each region after comparison with the \mappings\ libraries (see text for details).}\tablefoottext{i}{Estimated age of the bubble or, in the case of the SNRs, time since the initial explosion.} \tablefoottext{j}{Estimated average mechanical power required to inflate the bubble.} \tablefoottext{k}{Estimated explosion energy in the case of an evolved SNR in the pressure-driven snowplough phase (see Section~\ref{sub:mec_snrs} for details).}}
\end{table*}
The estimated \nism\ values are in overall good agreement with the 0.1--1 cm$^{-3}$ values typically found in other ULX bubbles \citep[e.g.][]{pakull_ultraluminous_2005}. In general the values found for all regions across the bubble are in good agreement,  except for region 4 where the slightly higher $n_\text{ISM}$ value might explain its higher surface brightness.

\subsection{Metallicity}\label{sub:metalliticy}
%https://irsa.ipac.caltech.edu/applications/DUST/
The presence of shocks heavily affects abundances in nebulae \citep[e.g.][]{1991PASP..103..815P} and requires specific modelling \citep[][]{allen_mappings_2008}. Consequently, empirical metallicity diagnostics remain restricted to  photoionisation zones, and are built with careful samples of reliable HII (or similar) regions. The most prevalent tools currently in use \citep[e.g.][]{perez-montero_ionized_2017} are often based on temperature and density estimations originating from weak recombination lines \citep[][]{osterbrock_astrophysics_2006}. However, these lines are either outside the observed bandpass of MUSE (e.g. the [O III]$\lambda$4363 and [S III]$\lambda$9532 lines) or are too weak to be detected above the continuum (e.g. [N II]$\lambda$5755). We therefore used the strong line method, through the \textit{S} (empirical) calibration provided by \cite{pilyugin_new_2016}, which only relies on strong lines (\hb\, [O~III]$\lambda$4959,5007, [N~II]$\lambda$6548,6584, and [S~II]$\lambda$6717,6371), clearly detected in our data cube. More specifically, as suggested by the authors, we used two different calibrations (their Equations 6 and 7, respectively) for the upper ($\log N_2 \geqslant$ --0.6) and lower ($\log N_2 <$ --0.6) branches, where $N_2$ = ([N~II]$\lambda$6548 + [N~II]$\lambda$6584) / \hb\  \citep[see][for more details]{pilyugin_new_2016}, to calculate \met.

Because this method is only calibrated for HII regions, we restricted our analysis to the dynamically cold regions. We followed the principle used in \cite{2017MNRAS.466..726D}, who considered the cold--warm delimitation of the [S II]/\ha\ BPT diagram of \cite{kewley_host_2006} (see  \citealt{lara-lopez_metal-things_2021} for a similar approach). Here we also excluded the new intermediate delimitation of \cite{law_sdss-iv_2021}, as we have seen that this delimitation region is affected by the X-ray ionisation of the source (Section~\ref{sub:bpt_diagrams}). Hence we restricted the analysis to the regions strictly classified as HII. 

\begin{figure}
    \centering
    \includegraphics[width=0.49\textwidth]{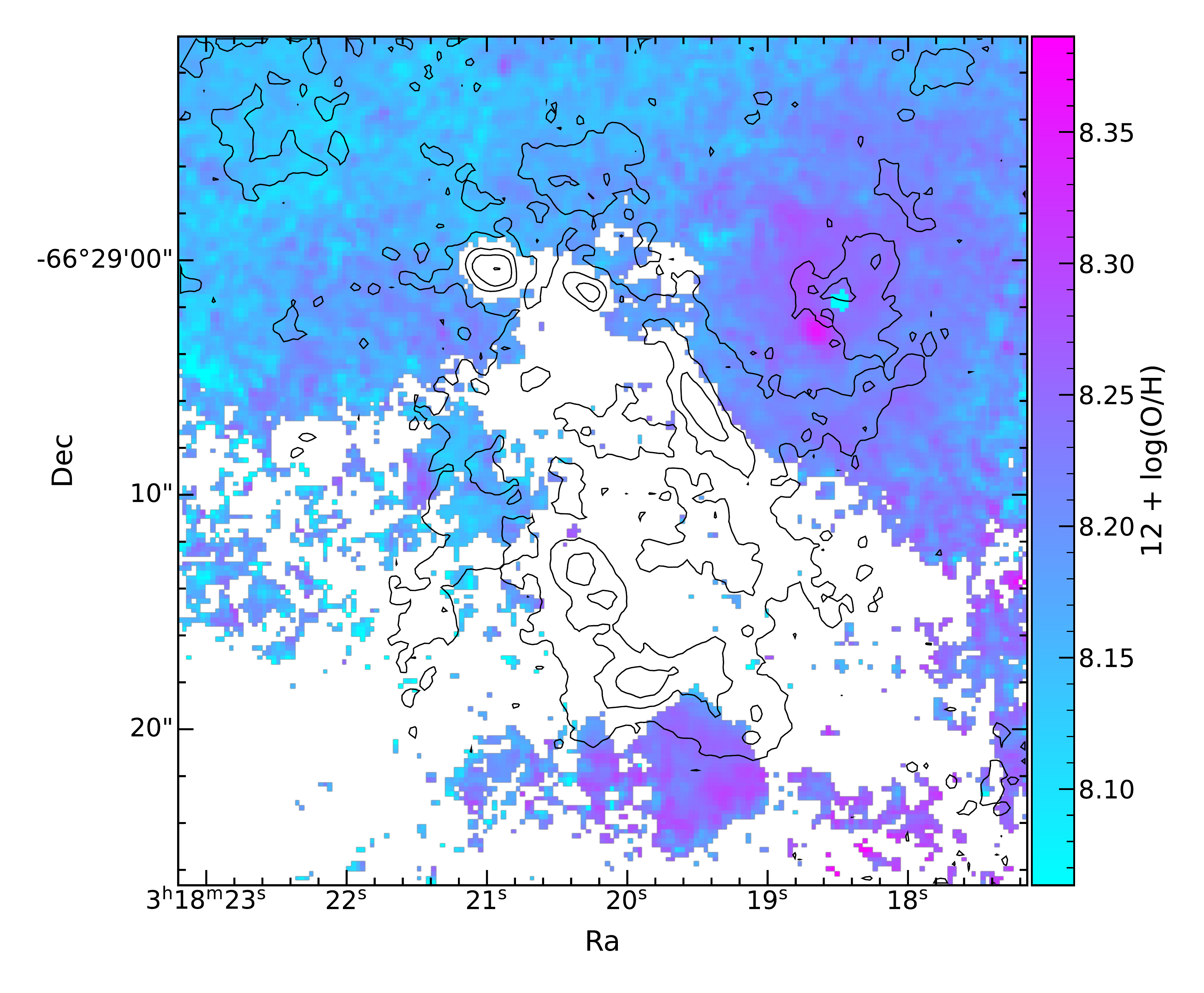}
    \caption{Metallicity map around \ulx\ only for those regions classified as HII regions (see text for details) based on the [S~II]/\ha\ vs [O~III]$\lambda$5007/\hb\ BPT diagram (Figure \ref{fig:bpt_diagrams}). Contours as per Figure~\ref{fig:astro}.}
    \label{fig:metal_ion}
\end{figure}

Given that some of the line ratios involved in the metallicity computation are rather distant in wavelength, we corrected the corresponding fluxes for extinction using the methodology outlined in Section \ref{sub:density_est}, this time adopting the theoretical value for Case B recombination of \ha/\hb~ $= 2.86$, more suitable for HII-like regions \citep{osterbrock_astrophysics_2006}. The values obtained for E(B-V) in the regions of interest in this case are in the 0.15--0.8 range. The resulting metallicity map is shown in Figure~\ref{fig:metal_ion}. 

%python ~/scripts/pythonscripts/maxime_project/muse_project/deredden_momcheva.py -a camel_1_n2ha/cleaned_images -b camel_1_hb/cleaned_images -i 2.86; python ~/scripts/pythonscripts/maxime_project/muse_project/H2diags.py -bpt lineratios/bpt_diagrams_v2/BPT_2.fits
We find a 3$\sigma$-clipped median and standard deviation of \met\ = 8.18$\pm$0.04. These estimates are slightly below the values reported by \cite{walsh_oh_1997}, who found that the metallicity in NGC~1313 followed a nearly flat profile with \met\ $\approx$8.4$\pm$0.1. We note that in the bright HII regions created by the stellar cluster north-west and south of the ULX the metallicity is closer to the values reported by \cite{walsh_oh_1997} (purple areas in the map). Therefore, while some of these differences in metallicity are likely due to the difference in the reddening correction and the metallicity calibration used, it may be possible that the method we employed is less suitable for these diffuse regions. Nevertheless, a plausible range for the metallicity around \theulx\ seems 8.1--8.4, which is in line with the estimates from the \mappings\ libraries (Section~\ref{sec:mappings}). We can also confidently say that NGC~1313~X--1 was formed in a subsolar environment. 

\subsection{Optical counterpart} \label{sub:optical_counterpart}
As stated earlier, the optical counterpart is not detected in the ACS/F555W equivalent MUSE image (Figure \ref{fig:astro}), nor in the white light MUSE image. The lowest Vega magnitude in any of the $U$, $V$, $I$ filters reported by \cite{yang_optical_2011} from analyses of the 2004 \hst\ data was $\sim$ 22.6 in the $U$ filter. The MUSE cube has a limiting AB magnitude of $\sim$20.6, as determined by the automatic pipeline \citep{weilbacher_data_2020}, which translates to a $\sim$ 20 Vega magnitude. Thus, our sensitivity is too low to assess whether the optical counterpart has varied with respect to the 2004 observations. The counterpart might be additionally blurred due to the limited spatial resolution (PSF FWHM $\sim$ 1") of the MUSE data cube.

\begin{figure*}
    \centering
    \includegraphics[width=0.49\textwidth]{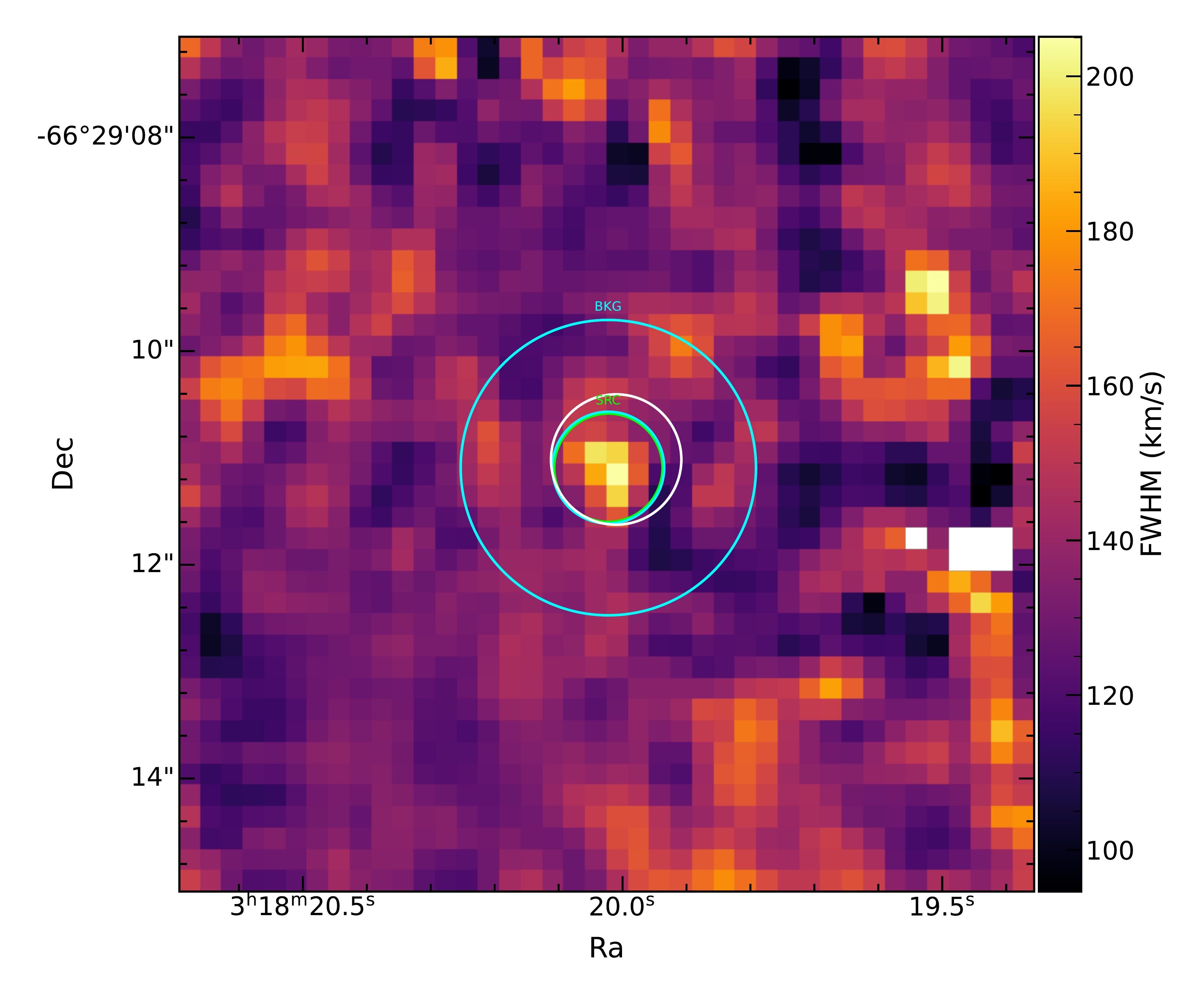}
        \includegraphics[width=0.49\textwidth]{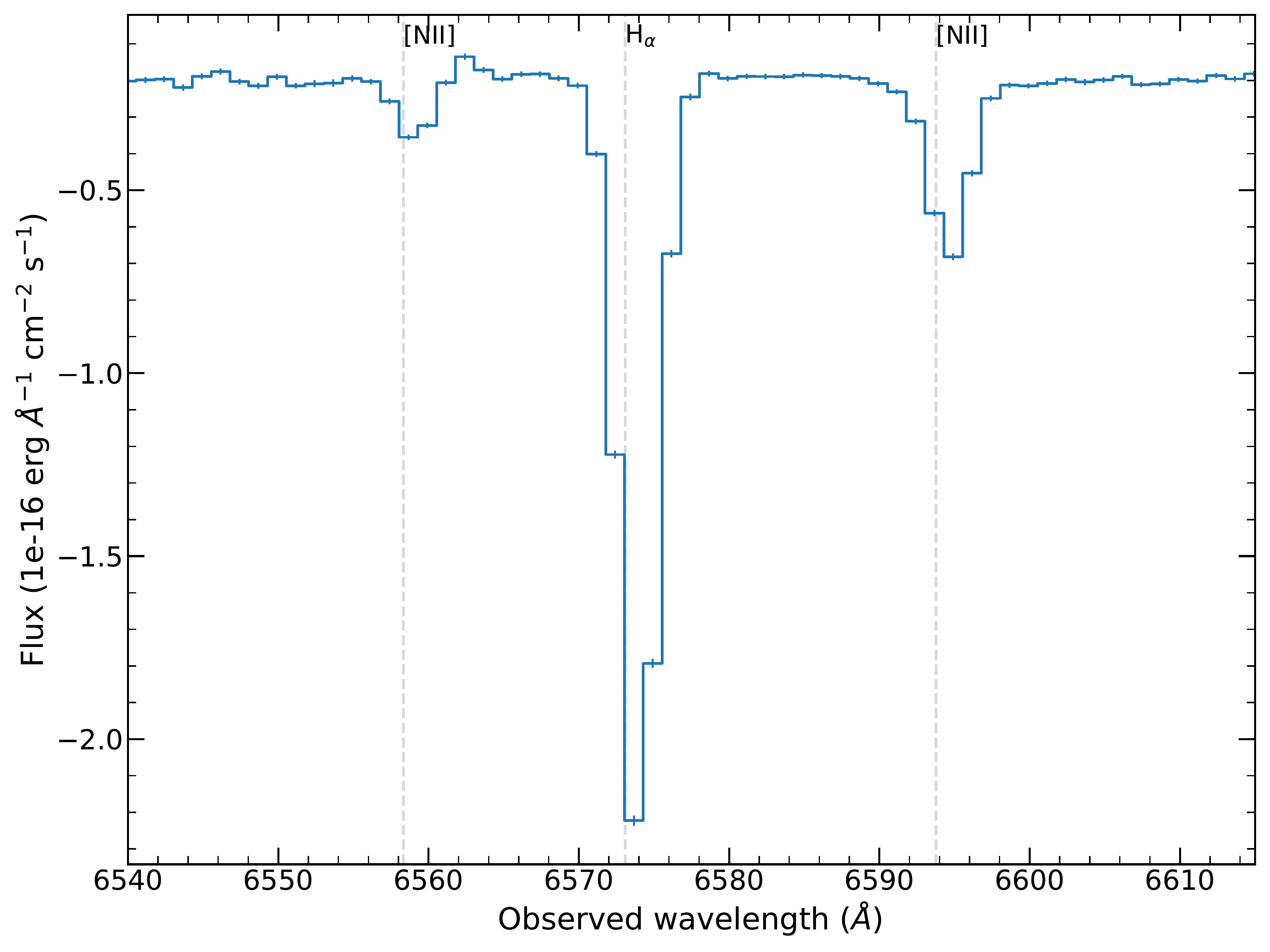}
        %saved to optical_counterpart_correct
    \caption{Extraction regions and optical spectrum of the counterpart. (left) Dispersion map for the [O~I]$\lambda$6300 line showing the enhancement around the ULX position and the extraction regions used. The extraction regions for the source and background are shown in green and cyan, respectively, whereas the white circle shows the ULX positions as determined in Section~\ref{sec:bubble_data_reduction}. (right) MUSE background-subtracted spectrum of the optical counterpart around the \ha-[N~II] complex. The main nebular lines are labelled and their expected positions based on the redshift of the galaxy are indicated with grey dashed lines.}
    \label{fig:extraction_region}
\end{figure*}

However, while the counterpart is not detected, as stated in Section~\ref{sec:camel}, we observed an enhancement of the width of the [O~I]$\lambda$6300 line at the position of \ulx, with the FWHM reaching a peak value of 200 km/s, as stated in Section~\ref{sec:camel} (see Figure~\ref{fig:extraction_region}). We determined the peak and the FWHM of this enhancement by fitting a 2D Moffat profile to the region around the peak emission to select a suitable extraction region for the spectrum of the optical counterpart. We used a circular region centred at the peak with a $\sim$0.5" radius for both semi-axes, as given by the estimated FWHM. We chose the background from an annulus centred at the source position to encompass some of the nebular emission. We set the inner radius to be equal to the source radius and the outer radius to 1.5" in order to avoid the significantly brighter south-east rim of the nebula.

Figure~\ref{fig:extraction_region} shows the background-subtracted spectrum around the \ha-[N~II] complex, which shows that the resulting spectrum is dominated by noise and is oversubtracted. There were no obvious features in the spectrum of the optical counterpart besides the sky residuals, suggesting that the emission is dominated by the nebular emission, which precluded us from characterising the optical counterpart. 
%python ~/scripts/pythonscripts/muse/get_image_peak.py cleancamel_1_hb_ssmooth_wave_indep_HBETA.fits -r determine_peak.reg
%ds9 cleancamel_1_oI_ssmooth_wavedisp_indep_OI6300corrected.fits -region ../../lineratios/ngc1313_muse555W_hst_img.reg
%python ~/scripts/pythonscripts/muse/get_image_peak.py -r ../../lineratios/determine_peak.reg cleancamel_1_oI_ssmooth_disp_indep_OI6300.fits

\section{Discussion} \label{sec:bubble_discussion}

The simultaneous spectral and imaging capabilities of MUSE  allowed us to analyse in detail the complex morphology of the bubble around \theulx, first noticed by \cite{pakull_optical_2002}. The nebula shows an indication of excitation by both shocks and EUV/X-rays. Shock excitation is seen at outer edges of the nebula close to the \ulx, revealed by the strong [O I]$\lambda$6300/\ha\ and [S II]/\ha\ ratios, together with the strong broadening (FWHM $\gtrsim$ 100 km/s) of the \ha, [N II], and [S II] lines near those regions (see Figures~\ref{fig:flux_maps}, \ref{fig:disp_maps} and \ref{fig:bpt_fwhm_regions}). This strong broadening at the edges of the bubble is also seen in other ULX bubbles where the kinematics have been studied, such as those around NGC~1313~X--2 \citep{ramsey_optical_2006} and NGC~5585~X--1 \citep{soria_ultraluminous_2021}. We have also seen that the gas in a region of $\sim 7.5$" around the ULX shows a high line-of-sight velocity compared to the surrounding gas (Figure~\ref{fig:vel_maps}). In particular, the mean \ha[N II][S II] and [O III] velocity maps show two high-velocity regions with a trough at the source position (Figure~\ref{fig:vel_profile}), which argues against a radial expansion, which could be the case for a SNR. This strongly suggests that the gas inside the bubble is being dynamically perturbed by winds emanating from the ULX, creating the cavity around it. The strong broadening at the edges of the bubble is then thought to be an indication of the gas being decelerated at the shock front, strongly supporting that the bubble has been (and is being) inflated by a wind and/or jets. Therefore, these results provide direct and supporting evidence for the presence of powerful winds in \ulx, so far only revealed through X-ray spectroscopy \citep{walton_iron_2016, pinto_resolved_2016, pinto_xmmnewton_2020}. We show below that the estimated mechanical power of the wind in these previous works is in remarkable agreement with that estimated from the bubble. The elongated shape and the roughly diametrically opposite regions 3 and 7 (see in particular the [O~I]$\lambda$6300 contours in Figure~\ref{fig:disp_maps}) might suggest that we are witnessing the activity of a jet. Follow-up radio observations would therefore be of great interest to confirm this. Given the clear evidence of shocks, below we compute the age and mechanical power of the bubble following previous works on shock-ionised bubbles \citep[e.g.][]{pakull_300-parsec-long_2010, soria_ultraluminous_2021}. 

Interestingly, the inner parts of the nebula also show a clear indication of strong  EUV/X-ray excitation. This is revealed by the high [O III]$\lambda$5007/\hb\ $>$ 5 close to the source and south of it (Figure \ref{fig:bpt_diagrams}), surrounded by a region of weakly ionised gas (Figure~\ref{fig:xin}), indicative of hard ionising radiation, in good agreement with the findings from \cite{pakull_optical_2002}. This morphology, a cavity with an X-ray ionised nebula in its interior, is in many respects very similar to the bubble around IC~342~X--1 reported by \cite{2003MNRAS.342..709R}. The morphology of the X-ray ionised region is also reminiscent of the prototypical photoionised nebula around Holmberg II~X--1, which shows an area of high excitation close to the source (traced by the He II $\lambda$4686 line), and strong [O I]$\lambda$6300 emission in the outer pars \citep{pakull_optical_2002, kaaret_high-resolution_2004, lehmann_integral_2005}. We have also  seen  that some of these regions fall in the AGN--LINER classification locus according to the BPT diagrams (Figure \ref{fig:bpt_diagrams}). Similar classification was found by \cite{berghea_spitzerobservations_2012} using infrared observations of Holmberg~II X--1 and NGC~6946 X--1, which are typical photoionised dominated nebulae \citep[e.g.][]{abolmasov_optical_2006}. EUV/X-ray excitation is also supported by the line ratios predicted by the \mappings\ libraries, which cannot account for the combined high [O I]$\lambda$6300/\ha\ or [O III]$\lambda$5007/\hb\ ratios, ruling out shocks as the main source of ionisation. Thus, \ulx\ shows signs of ionising the surrounding medium by means of both shocks and EUV/X-ray radiation, which highlights the complex morphology of the nebula and enormous power of ULXs in ionising the ISM. This has been recently highlighted in the work of \cite{simmonds_can_2021}, which have shown that a combination of realistic ULX broadband and stellar population spectra can reproduce the nebular He II $\lambda$4686 emission seen in low-metallicity galaxies, whose origin remains unclear.

Our study also suggest that ULX bubbles might be more complex than revealed by long-slit spectroscopy alone. In this regard, it is surprising that the nebula around NGC~5408 X--1 seems to show no signs of shock excitation \citep{cseh_black_2012} and displays a much smaller bubble \citep[$\sim$ 60\,pc;][]{grise_optical_2012}, given the detection of winds in X-rays \citep{pinto_resolved_2016} with similar outflow velocities ($\sim$ 0.2\,$c$). We note that \cite{cseh_black_2012} probed a small (2" $\times$ 1") region around NGC~5408~X--1 based on the [O III]$\lambda$5007 line map, whereas here we  show that the [O III]$\lambda$5007 emission is concentrated close to the source in the high-excitation regions. In fact, a similar conclusion was reached by \cite{mirioni_sources_2002} after performing long-slit spectroscopy of the high-excitation region around \theulx, thereby missing the shock-excited regions at further distances from the source. We believe that a more thorough study using IFU spectroscopy is needed to completely rule out the presence of shocks in sources such as NGC~5408~X--1.

The 452 $\times$ 266\,pc bubble around \theulx\ sits at the high end of the largest optical nebulae ever detected to date, comparable in size to the shocked bubble around the ULX in NGC~1313 X--2 \citep{pakull_ultraluminous_2005} or the bubble around Holmberg~IX~X--1 \citep{grise_optical_2011}. The detailed analysis of the nebula around NGC 5585 X--1 presented by \cite{soria_ultraluminous_2021} allows us to compare the X-ray properties and bubble characteristics of these two sources. The bubble around NGC~5585~X--1 is also similar in size and morphology with respect to that of \theulx\ and, as stated above, shows similar strong broadening towards the edges of the bubble with similar FWHM values and [S II]/\ha\ and [O I]$\lambda$6300/\ha\ ratios, typical of shock-dominated bubbles. However, it is interesting to observe that the bubble around NGC~5585~X--1 shows no signs of EUV/X-ray excitation, with [O III]$\lambda$5007/\hb\ only reaching values of $\sim$ 1.6 (log([O~III]/\hb) $\sim$ 0.2) well within the HII region classification in the [N II]/\ha\ diagram (cf. Figure~\ref{fig:bpt_diagrams}). While \theulx\ reaches unabsorbed luminosities of $\sim$ 1.8 $\times$ 10$^{40}$ erg/s \citep{gurpide_long-term_2021}, NGC~5585~X--1 only reaches $\sim$ 4~$\times$~10$^{39}$ erg/s \citep{soria_ultraluminous_2021}. The difference in ionising photons between these two sources may therefore account for the different [O III]$\lambda$5007/\hb\ ratios. This may suggest that the dimmer luminosity of NGC~5585~X-1 is intrinsic and not due to a geometrical effect due to an unfavourable viewing angle in which the funnel axis could have been strongly misaligned with respect to our line of sight. Interestingly, \theulx\ switches between soft and hard ultraluminous states \citep{sutton_ultraluminous_2013}, thought to belong to the ULXs for the which the intense accretion rate leads to the formation of a supercritical disk-wind funnel \citep{shakura_black_1973, poutanen_supercritically_2007}, whereas NGC~5585~X-1 belongs to the so-called broadened disk type \citep{soria_ultraluminous_2021}, which are thought to be sources close or at the Eddington limit \citep[e.g.][]{middleton_bright_2013}. This may be consistent with the implied higher accretion rate in \ulx. Alternatively, the gas in the cavity created by NGC~5585~X--1 might more rarefied than that of NGC~1313~X--1. We show below that the mechanical power of \ulx\ is comparable to that of NGC~5585~X--1. If the outflow rate depends on the Eddington accretion rate \citep[e.g.][]{shakura_black_1973, poutanen_supercritically_2007}, this may instead imply that the Eddington accretion rates of the two sources are similar, and that NGC~5585~X--1 might be viewed through an unfavourable viewing angle compared to NGC~1313~X--1. Another possibility, if indeed both sources have similar Eddington accretion rates, is that NGC~5585~X--1 hosts a lighter compact object, which could also account for its dimmer luminosity.

In this respect, it is unclear whether  we can derive some conclusions from the morphology of the X-ray ionised
nebula with regard to the level of anisotropy of the X-ray emission in \theulx. The offset between the peak of the [O III]$\lambda$5007/\hb\ ratio and the ULX position might be expected \citep[e.g.][]{halpern_x-ray_1980}, given the rather low ionisation potential to doubly ionise oxygen (35.1\,eV). The fact that the X-ray ionised region is only seen south of the ULX is likely due to the nebula being density-bounded in the north direction, given the absence of significant \ha\ emission there (Figure~\ref{fig:flux_maps}). Therefore, the nebula might not directly probe  the degree of anisotropy of the X-ray emission. Observations in the [O IV] 25.89\,$\mu$m or the He II $\lambda$4686 lines would be of interest as they should trace more closely the high-excitation regions closer to the ULX. The flux of the He II $\lambda$4686 photon-counting line \citep[e.g.][]{pakull_detection_1986} or photoionisation modelling of these regions might provide useful constraints on the EUV spectra of the source, degree of anisotropy, and/or inclination \citep[e.g.][]{kaaret_high-resolution_2004,abolmasov_optical_2008, abolmasov_optically_2009, berghea_first_2010}. MUSE observations of the nebulae around the Galactic supercritically accreting source SS433 \citep{dubner_high-resolution_1998}, long speculated to be an edge-on ULX \citep{king_brightest_2002,fabrika_jets_2004, middleton_nustar_2021}, and ULXs for which the inclination is suspected to be higher than for \ulx, such as NGC 5408 X--1 \citep{middleton_challenging_2011}, might also offer interesting comparisons with that of \ulx\ to constrain the inclination of these systems and the relationship between the bubble morphology and the accretion flow geometry.

\subsection{Age and mechanical power of the ULX bubble} \label{sub:mec_power}
The characteristic age of the bubble can be estimated assuming that a spherically expanding bubble has been carved by a steady wind with mechanical power $P_\text{mec}$ over the bubble lifetime ($t$) \citep{weaver_interstellar_1977}. The evolution of the bubble radius ($R$) over time as a function of $P_\text{mec}$ and the mass density of the unshocked ISM ($\rho$) is given by
\begin{equation}
    R = \left (\frac{125}{154\pi}\right)^{1/5} \left (P_\text{mec}  / \rho \right)^{1/5} t^{3/5}.
\end{equation}
We assume that the bubble expands into a neutral ISM, which has a mean molecular weight $\mu = 1.38$, and thus $\rho = \mu m_p n_\text{ISM}$, where $m_p$ is the proton mass and $n_\text{ISM}$ is the hydrogen number density estimated in Section~\ref{sub:density_est}. The expansion velocity is therefore
\begin{equation}
    v_\text{exp} = \frac{dR}{dt} = \frac{3}{5} R / t  
,\end{equation}
which we take as \vs. The age of the bubble is then $t = \frac{3 R}{5 v_\text{exp}}$. From these expressions the mechanical power can be estimated as
\begin{equation}
    P_\text{mec} = \left (\frac{154 \pi}{125} \right) R^2 v_\text{shock}^{3} (\mu m_p n_\text{ISM})
.\end{equation}
With these formulae we estimate the age and the mechanical power considering the major and minor semi-axes of the bubble for a typical \vs\ = 170\,km/s. The results are given in Table~\ref{tab:pre_shock}. The estimated ages are found to be (4.6$\pm$0.7) $\times$ 10$^5$\,yr and (7.8$\pm$1.2) $\times$ 10$^5$\,yr for the minor and major axes, respectively, so a typical age would be $\sim$ 6.2 $\times$ 10$^5$\,yr. For the mechanical power, we find (1.9$\pm$1) $\times$10$^{40}$\,erg/s and (5.5$\pm$3) $\times$10$^{40}$\,erg/s, essentially consistent within the uncertainties for both semi-axes. 

As stated in Section~\ref{sub:density_est}, the exact factor relating \vs\ and FWHM is rather uncertain. Assuming the perhaps more conservative case in which \vs = FWHM, which we take as $\sim$110\,km/s, the average FWHM from all regions (Table~\ref{tab:pre_shock}), we find slightly greater  mechanical power for the larger axis, but in good agreement within the uncertainties with the estimates given above. This is in part because the estimated \nism\ values are higher for lower shock velocities (cf. Equation~\ref{eq:n_ism}) and offset the lower shock velocity when estimating the mechanical power. We thus consider the estimates given above as a good figure for the mechanical power of the source. The age of the bubble would obviously increase by a factor of $\sim$ 1.47, thus a reasonable range may be $t$ $\sim$ (6.2--8.7) $\times$ 10$^5$\,yr. 

% For the mechanical- radiative ratio Siwek+2017 reports 50% but for 0.5 Myr it seems like the radiative luminosity exceeds the mechanical power (?) (Their Figure 3). They also report optical luminosity of 10^39 erg/s yet I measure 10^37 erg at most (at 6000A as quoted and correcting for absorption with E(B-V) = 0.2) using the flux emitted in the whole bubble

\begin{figure}
    \centering
    \includegraphics[width=0.49\textwidth]{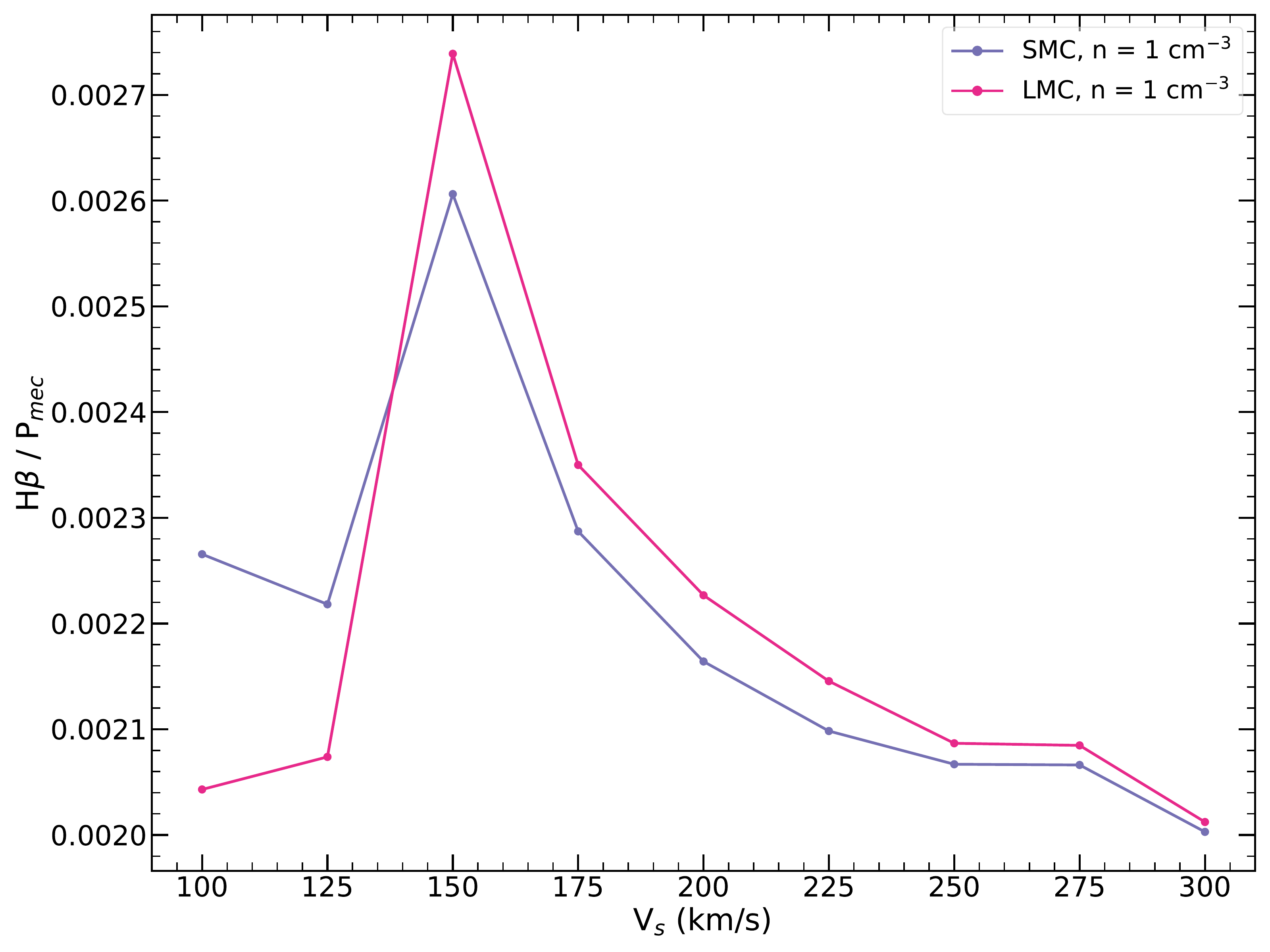}
    \caption{Ratio of the \hb\ luminosity to the mechanical power as a function of \vs\ for a plausible range of metallicities (SMC and LMC as derived in Section~\ref{sec:mappings}) assuming equipartition for the magnetic field value. }\label{fig:hb_pjet}
    %python ~/scripts/pythonscripts/maxime_project/muse_project/read_mapping.py P*b[e]*sp_lines.txt Q*b[e]*sp_lines.txt
\end{figure}

Another method commonly used to estimate the mechanical power is to use some of the optical lines as a proxy \citep[e.g.][]{pakull_300-parsec-long_2010, soria_super-eddington_2014}. To this end, we used the \hb\ line as it has a weak dependence on the shock velocity, whereas we have already determined the pre-shock gas density of the ISM (which we take as \nism = 1\,cm$^{-3}$) and the metallicity around NGC~1313~X--1. From standard bubble theory \citep{weaver_interstellar_1977} we expect the total radiative luminosity of the shock to be $L_{\text{T}}$~=~ 27/77\,$P_{\text{mec}}$. The total radiative luminosity of the shock as a function of \vs\ can be estimated from Equation
3.3 in \cite{dopita_spectral_1996}  \citep[see also][]{soria_super-eddington_2014},

\begin{equation}
\begin{split}
    L_\text{T} = \frac{1}{2} (\mu m_p n_\text{ISM}) \,v_s^3 = \\ 
    1.14 \times 10^{-3} \left(\frac{v_s}{100 \,\text{km}\text{s}^{-1}}\right)^3n_\text{ISM} ~\text{erg}\,\text{cm}^{-2}\,\text{s}^{-1}
    \end{split}
,\end{equation}
where we have assumed $\mu$ = 1.38 as above. The total radiative luminosity of the \hb\ line as a function of \vs\ can be obtained from the \mappings\ libraries \citep{allen_mappings_2008}\footnote{Here we use the \cite{allen_mappings_2008} tables as those compiled by \cite{alarie_extensive_2019} do not include these values.}. We therefore calculated the fraction of power radiated in the \hb\ line as a function of \vs\ for the best set of abundances  derived in Section~\ref{sec:mappings} (LMC and SMC). The resulting calculation is shown in Figure~\ref{fig:hb_pjet}. We  assume equipartition magnetic field values in the calculation, but the results are equivalent for ISM magnetic field values in the 1--10\,$\mu$G range. From the shock velocities derived in Table~\ref{tab:pre_shock}, we find $L_{\text{H}\beta}$ $\simeq$ (2.3--2.6) $\times$ 10$^{-3}$\,$P_\text{mec}$. 

We now need to estimate the total luminosity in the \hb\ emitted by shocks in the bubble. For this we simply add the $L_{\text{H}\beta}$ of each of the five shocked regions discussed in Figure~\ref{fig:disp_maps} (Table~\ref{tab:pre_shock}) which gives $\Sigma L_{\text{H}\beta}$ = (5.33 $\pm$ 0.01) $\times$ 10$^{37}$\,erg/s. We note that while there might be some contribution to the \hb\ luminosity coming from EUV/X-ray excitation, notably for region 5, this is likely compensated by the fact that we might be missing some \hb\ flux around regions 3, 4, and 6 given the coarser regions employed (Figure~\ref{fig:disp_maps}). We thus consider these regions adequate in order to provide an order of magnitude estimate on $P_\text{mec}$. With these values we obtained $P_\text{mec}$ $\simeq$ (3.8--4.3) $\times$ 10$^2$\,$\Sigma L_{\text{H} \beta} \simeq$ (2.2$\pm$0.3) $\times$ 10$^{40}$ erg/s. An approximate upper limit on the mechanical power using this method may be obtained by considering all the flux emitted in the whole bubble,  approximated as the cyan ellipse in Figure~\ref{fig:disp_maps},  and subtracting the flux emitted due to EUV/X-ray excitation, which we approximate as the polygon classified as intermediate and AGN in the [N II]/\ha\ BPT diagram (Figure~\ref{fig:bpt_diagrams}) south of the ULX. We measure $\Sigma L_{\text{H} \beta}$ = 1 $\times$ 10$^{38}$ erg/s, and therefore we estimate $P_\text{mec} \lesssim$ 4.0--4.5 $\times$ 10$^{40}$\,erg/s. If we additionally exclude the flux from region 5, which we saw is likely dominated by X-ray ionisation, we obtain $P_\text{mec} \lesssim$ 3.5--4.0 $\times$ 10$^{40}$\,erg/s.

We thus conclude that the estimated mechanical power is comparable to or higher than the X-ray luminosity of the source. This is in agreement with the estimates obtained from wind detections using X-ray spectroscopy \citep{pinto_resolved_2016, walton_iron_2016}, which also revealed that the wind dominates the energy output of the source. We note that this agreement is remarkable given that these estimates are completely independent. The mechanical power of the \ulx\ is comparable to that found for microquasars such as NGC\,7793--S26 or M83 Q1 \citep{soria_super-eddington_2014}, which also display mechanical powers exceeding their radiative luminosities, whereas the estimated age is commensurate with other ULX bubbles such as the one around the ULX in NGC 5585 \citep{soria_ultraluminous_2021} or IC~342~X--1 \citep{cseh_black_2012}.

\subsection{Age and explosion energy of the supernova remnants}\label{sub:mec_snrs}

We first consider the case in which these SNRs are in the free-adiabatic expanding phase, where  the expansion can be described by the well-known Sedov-Taylor self-similar solution. The radius of the SNR since the time of explosion ($t$) is defined solely by the initial explosion energy $E$ and the density of the ambient medium ($\rho$),
\begin{equation}
    R = \left( \frac{2.026 E}{\rho} \right ) = \left( \frac{2.026 E}{\mu n_\text{ISM} m_p} \right )^{1/5} t^{2/5}  
,\end{equation}
for which we have assumed neutral hydrogen, as in the previous section. The expansion velocity is therefore
\begin{equation}
    V = \frac{2R}{5t}
.\end{equation}
Using our estimates from Section~\ref{sec:mappings} for the shock velocity
and our estimates for \nism\ from Section~\ref{tab:pre_shock} for a radius of 17.5~pc, we estimate $t_\text{SNR1} = (16\pm2) \times 10^4$\,yr and $t_\text{SNR2} = (17\pm2) \times 10^4$\,yr. However, the energies obtained under this assumption are $\sim$ 10$^{49}$\,erg, two orders of magnitude below the canonical value of 10$^{51}$\,erg. Assuming a canonical energy of 10$^{51}$\,erg and using the analytical expressions from \cite{cioffi_dynamics_1988}, we find that the radius at which the Sedov-Taylor expansion stage ends is $\sim$ 19~pc and $\sim$17~pc for SNR 1 and SNR 2, respectively. It is thus possible that these are evolved SNRs that have already entered the so-called pressure-driven snowplough phase, in which radiative losses become important \citep{cioffi_dynamics_1988}. This might also be supported by the presence of several strong forbidden lines in their spectra (Figures~\ref{fig:flux_maps} and ~\ref{fig:snr_spectra}). 

\begin{figure}
    \centering
    \includegraphics[width=0.49\textwidth]{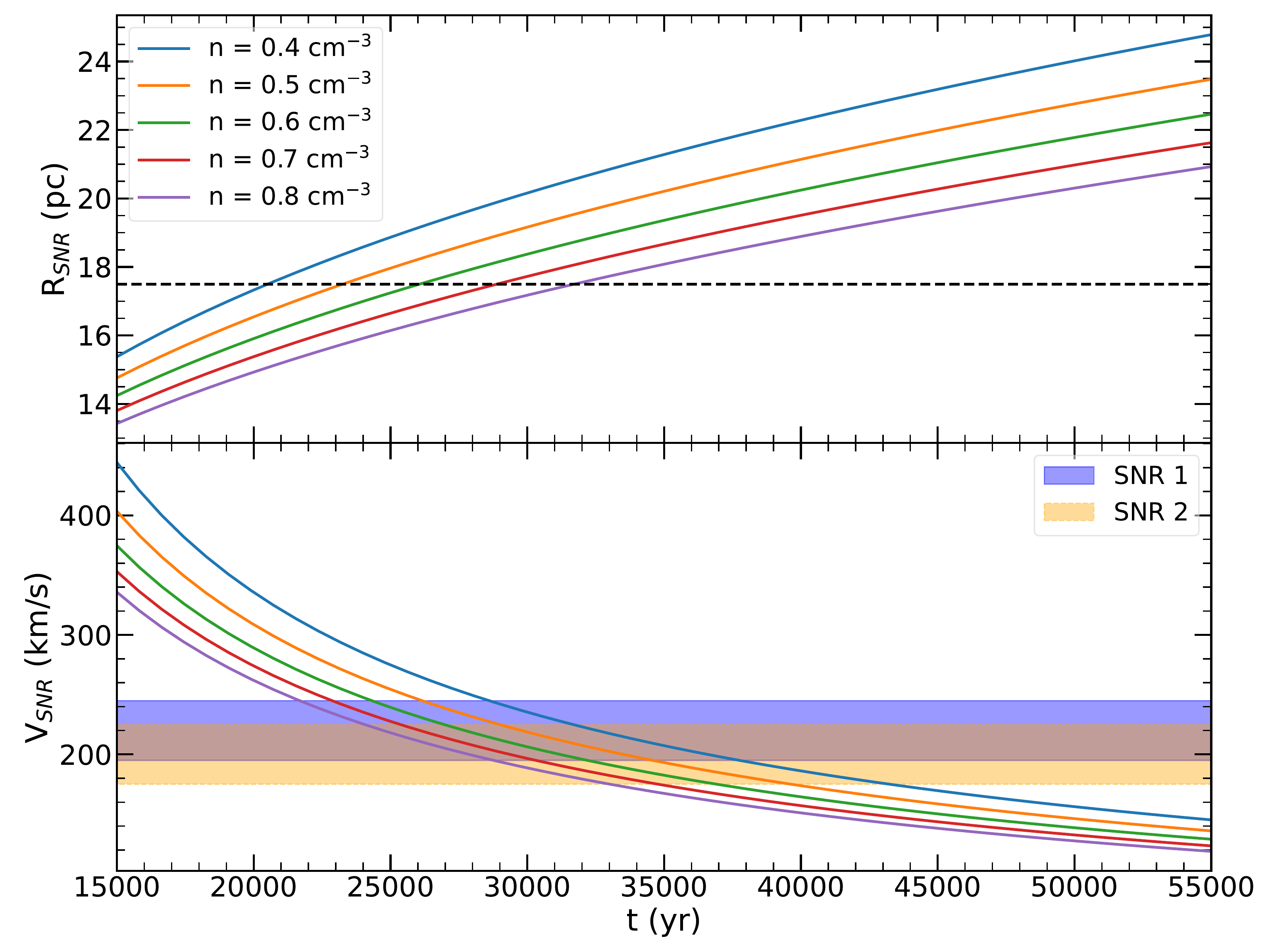}
    \caption{Evolution of the radius and expansion velocity of a SNR (green shaded areas) as a function of time for the pressure-drive snowplough phase using the analytical expressions from \cite{cioffi_dynamics_1988} for an initial explosion energy of 0.5 $\times$ 10$^{51}$. The evolution is calculated for the typical range of values found for \nism\ in Section~\ref{sub:density_est} (Table~\ref{tab:pre_shock}). The dashed line indicates the radii of the SNRs measured from the \hst\ images (Figure~\ref{fig:F567N_chandra}) and the blue and yellow shaded areas represent the range of shock velocities determined in Section~\ref{sec:mappings} (Table~\ref{tab:pre_shock}) for SNR 1 and 2, respectively.}
    \label{fig:snr_age}
\end{figure}

We therefore refine our calculation by using the analytical expressions from \cite[][see also \citealt{fesen_g107090_2020}]{cioffi_dynamics_1988}. Figure~\ref{fig:snr_age} shows the evolution of the radius and the SNR expansion velocity as a function of time. We find good agreement with the calculations for  the radii and for the velocities assuming an initial explosion energy of 0.5 $\times$ 10$^{51}$\,erg. From the figure we can estimate $t_\text{SNR1} = 2.4-2.7 \times 10^4$\,yr and $t_\text{SNR2} = 2.5-3.4 \times 10^4$\,yr (for a fixed $E$ = 0.5 $\times$ 10$^{51}$ erg). We find 0.3 < $E$ < 0.8 $\times$ 10$^{51}$~erg as a plausible range of energies to match the observed radii and derived velocities.

We can do a similar exercise for the ULX bubble, assuming again that, given its large size ($R$ $>$ 130\,pc; Table~\ref{tab:pre_shock}), it is  an evolved remnant in the snowplough phase. The expression for the energy explosion in the radiative phase is \citep{cioffi_dynamics_1988}
\begin{equation}
    E_\text{51} = 6.8 \times 10^{43} \left (\frac{R}{\text{pc}} \right)^{3.16} \left (\frac{v_\text{exp}}{\text{km} \text{s}^{-1}} \right)^{1.35} \left (\frac{n_\text{ISM}}{\text{cm}^{-3}} \right)^{1.16} \text{erg}
    ,\end{equation}
where following \cite{cseh_black_2012} we  also adopted the metallicity factor $\xi_\text{m}$ as unity for simplicity. Taking the values from Table~\ref{tab:pre_shock} we arrive at $E$ = 188--990 $\times$ 10$^{51}$\,erg for the minor and major axes, respectively, more than two orders of magnitude above the typical SNR explosion energy and in the range of estimates of other ULX bubbles \citep[e.g.][]{pakull_optical_2002, berghea_detection_2020}. This is perhaps not surprising as SNRs are not expected to remain optically bright once expanded beyond $\sim$ 100\,pc for an input canonical energy of 10$^{51}$\,erg \citep[][]{1989ApJ...340..355B}, and indeed the largest and oldest known SNRs have radii of just $\simeq$ 50\,pc \citep[see e.g.][and references therein]{leahy_x-ray_1986, williams_supernova_2004, 2020MNRAS.498.5194F}. Therefore these results are in good agreement with previous works \citep[e.g.][]{pakull_optical_2002, cseh_black_2012, urquhart_newly_2019, berghea_detection_2020} which suggest instead that a continuous injection of energy is more likely to explain the cavities found around some ULXs. However, we note that we cannot exclude the possibility that the ULX perturbed the SNR that gave birth to the compact object. This would be similar to the case of the SS433/W50 nebula \citep[][although see also \citealt{ohmura_continuous_2021}]{goodall_when_2011} or its potentially younger version Circinus X--1 \citep{heinz_lord_2015, coriat_twisted_2019}.

\section{Conclusions} \label{sec:bubble_conclusions}
We have shown that \ulx\ ionises a region of 452 $\times$ 266\,pc via EUV/X-ray--X-ray excitation and shocks. Thanks to the simultaneous spectro-imaging capabilities of MUSE, we were   able to diagnose the main sources of ionisation of the bubble in a spatially resolved manner: the inner regions close to the ULX show all the hallmarks of an X-ray ionised region, whereas the outer edges shows clear signs of shock-excitation. The latter is revealed by the strong broadening of the nebular lines at the edges of the bubble combined with the high line-of-sight velocity measured inside the bubble cavity close to the source, strongly suggesting that the bubble is being inflated by a wind. The elongated shape of the nebula may point to the presence of jet activity, and thus radio observations would be of great interest here. Future observations probing high-excitation lines of the X-ray ionised nebula inside the bubble might be of interest to constrain the degree of beaming of the source.

The mechanical power is found to be in excess of 10$^{40}$\,erg/s, suggesting that the mechanical power of the source is comparable or even higher than the radiative throughput. This is in line with other estimates made by \cite{pakull_300-parsec-long_2010}, \cite{soria_super-eddington_2014} and \cite{soria_ultraluminous_2021}, suggesting that the mechanical feedback in the super-Eddington regime is superior to the radiative one. Obtaining a statistical sample of ULXs with estimated mechanical and radiative powers will be of interest to constrain the energy that goes into powering the wind, its dependency with the mass-accretion rate, and its link with the class of less-luminous microquasars with   equally strong super-Eddington mechanical powers, such as SS433 or NGC 300 S10 \citep{urquhart_newly_2019, mcleod_optical_2019}. The high mechanical power of ULXs make them good sources for cosmic-ray acceleration \citep{inoue_high_2017, abeysekara_very-high-energy_2018}, which could be targeted in the future with the Cherenkov Telescope Array \citep{consortium_design_2011}, the next-generation of ground-based gamma-ray observatories.

With this work we hope to have highlighted the promising avenue that IFU spectroscopy provides in our understanding of the ULX bubble nebulae and the energetic feedback of the super-Eddington regime.

\begin{acknowledgements}
The authors are deeply grateful to M. Pakull and R. Soria for their careful read and constructive comments on an earlier version of this manuscript. We would also like to thank M. Coriat for his helpful comments on the discussion related to the presence of any jet-like signature. We finally thank I. Pastor-Marazuela for her comments which lead us to investigate and discover the two SNRs. This work has made use of data obtained from the Chandra Data Archive as well as observations made with the NASA/ESA Hubble Space Telescope and obtained from the Hubble Legacy Archive, which is a collaboration between the Space Telescope Science Institute (STScI/NASA), the Space Telescope European Coordinating Facility (ST-ECF/ESA), and the Canadian Astronomy Data Centre (CADC/NRC/CSA). Software used: Python v3.8, mpdaf \citep{piqueras_mpdaf_2017}. 
       
\end{acknowledgements}

\bibliographystyle{aa}
\bibliography{biblio,biblio2,biblio3}
\begin{appendix} %First appendix
\onecolumn
\section{Line ratios of the shocked regions}
\begin{figure*}[h]
    \centering
    \includegraphics[width=0.40\textwidth]{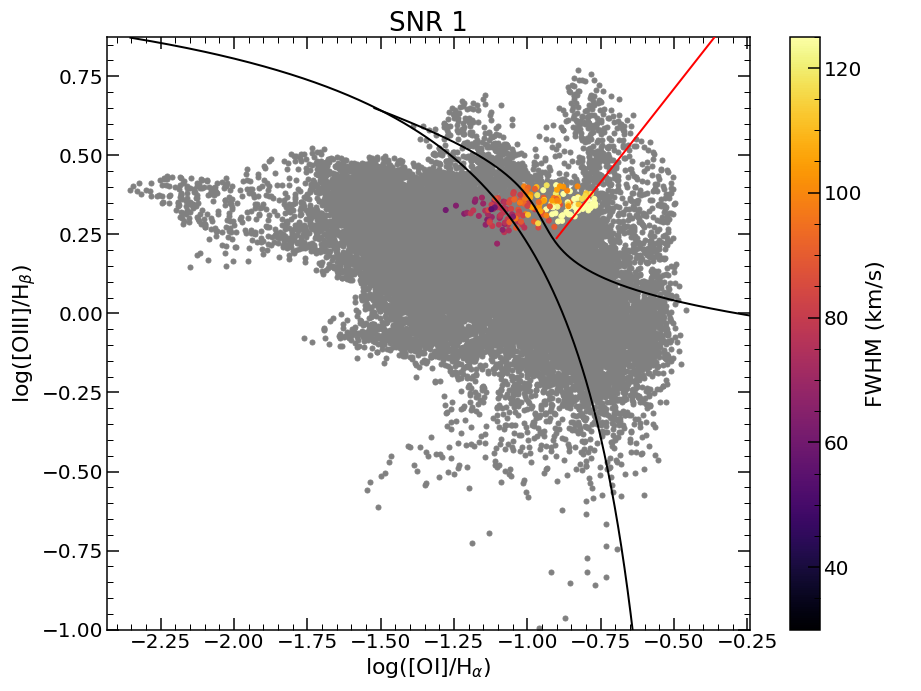}
    \includegraphics[width=0.40\textwidth]{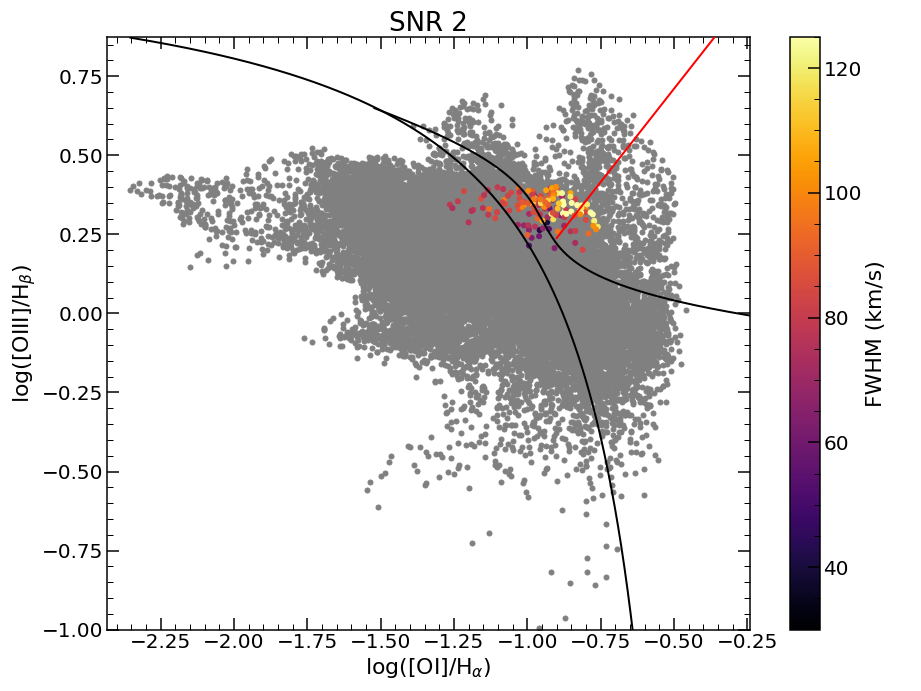}
    \includegraphics[width=0.40\textwidth]{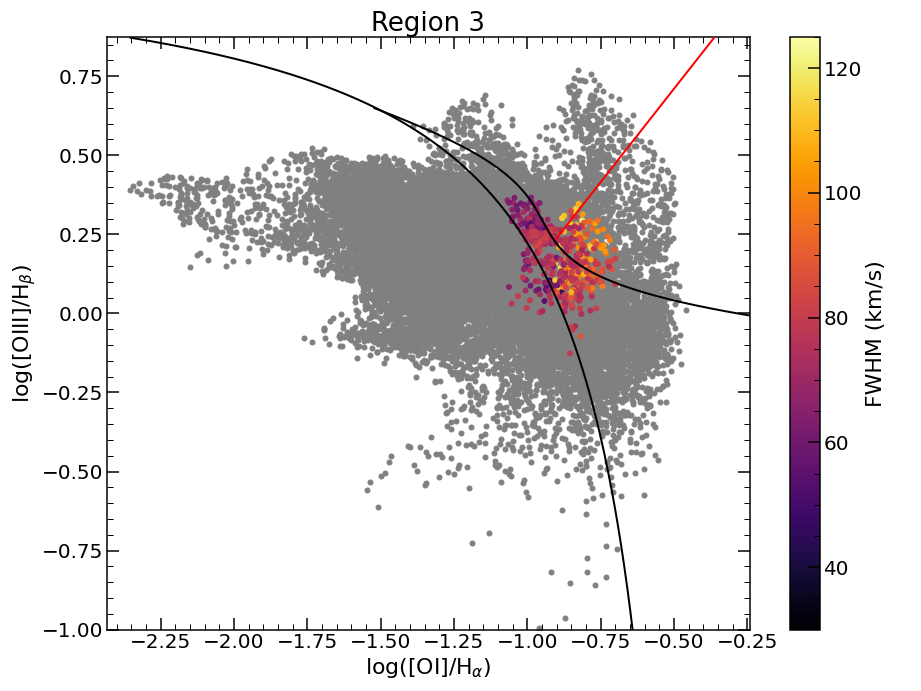}
    \includegraphics[width=0.40\textwidth]{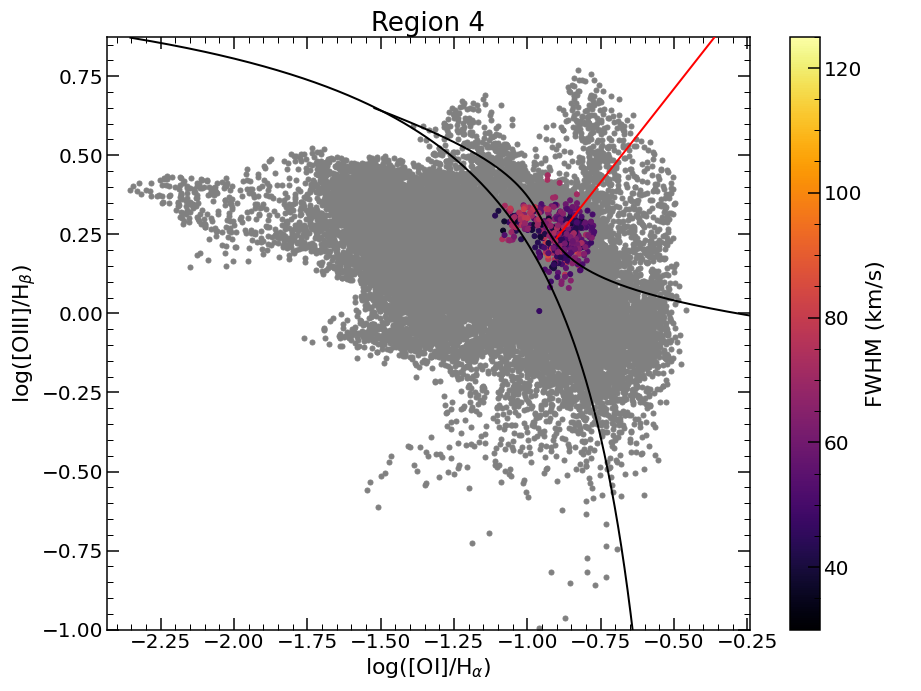}
    \includegraphics[width=0.40\textwidth]{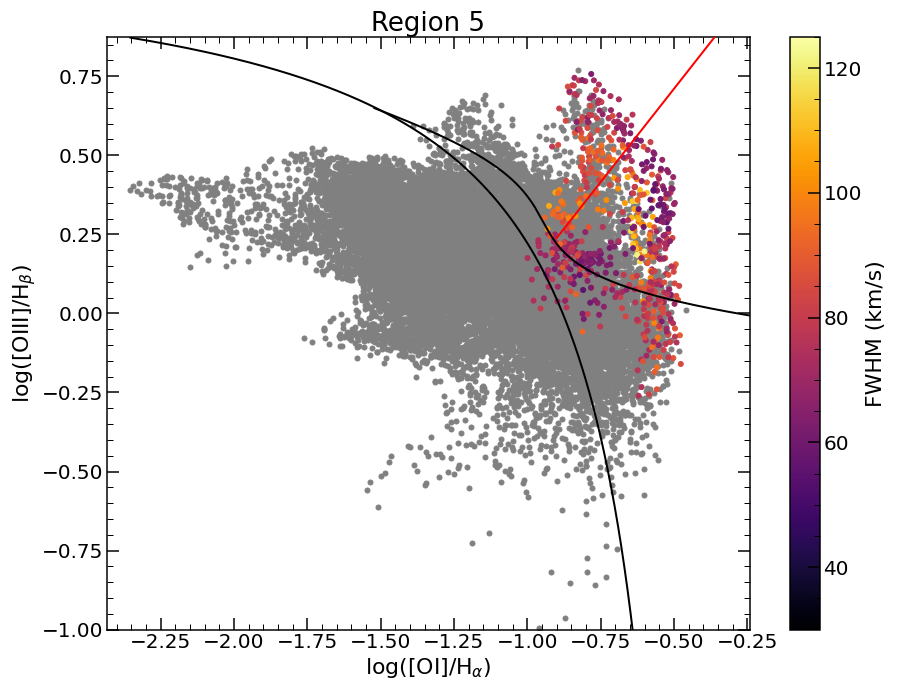}
    \includegraphics[width=0.40\textwidth]{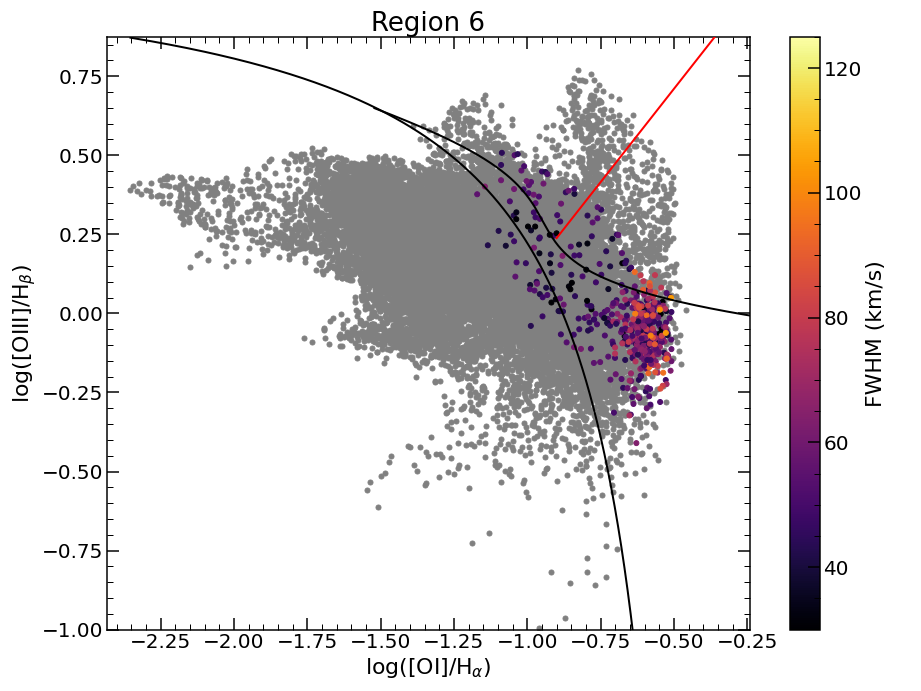}
    \includegraphics[width=0.40\textwidth]{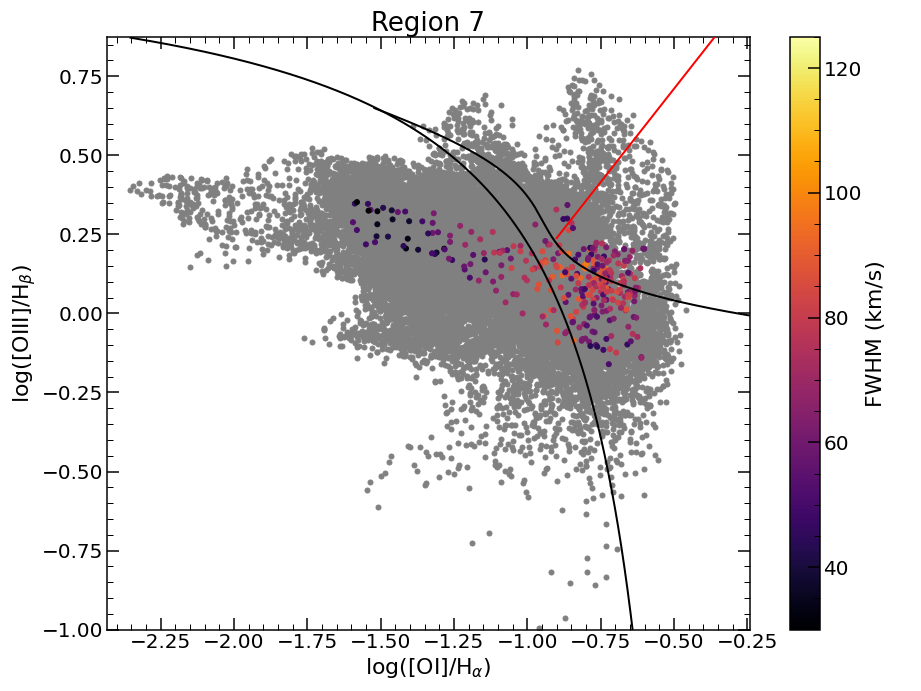}
    \caption{As in Figure~\ref{fig:bpt_fwhm}, but showing only the different regions from Figure~\ref{fig:disp_maps}.}\label{fig:bpt_fwhm_regions}
\end{figure*}
\end{appendix}
\end{document}